\documentclass[notitlepage,11pt,floatfix,aps,nofootinbib,superscriptaddress]{revtex4-1} 
\pdfoutput=1
\usepackage{setspace}

\usepackage{amssymb,amsfonts,multirow}

\usepackage{epsfig}

\usepackage{amssymb,url,bm}
\usepackage{graphicx}
\usepackage{hyperref}

\usepackage{color}

\def\di{\displaystyle}
\usepackage{array}
\usepackage{amsmath}
\usepackage{slashed}

\long\def\del #1 \enddel { }

\usepackage{epsfig}

\usepackage{xcolor,colortbl}

\definecolor{Black}{gray}{0}
\definecolor{Gray}{gray}{0.85}
\definecolor{LightGray}{gray}{0.93}
\definecolor{LightGreen}{rgb}{0.88, 1, 0.88}
\definecolor{LightCyan}{rgb}{0.88,1,1}
\definecolor{LightRed}{rgb}{1, 0.85, 0.85}
\definecolor{LightYellow}{rgb}{1, 1, 0.85}
\definecolor{LightBlue}{rgb}{0.87, 0.94, 1}
\definecolor{white}{gray}{1}
\usepackage{epsfig}

\usepackage{array,mathtools,amssymb,booktabs}
\newcolumntype{C}{>{$}c<{$}}
\AtBeginDocument{
\heavyrulewidth=.16em
\lightrulewidth=.1em
\cmidrulewidth=.03em
\belowrulesep=.4ex
\belowbottomsep=0pt
\aboverulesep=.4ex
\abovetopsep=0pt
\cmidrulesep=\doublerulesep
\cmidrulekern=.5em
\defaultaddspace=.5em
}

\usepackage{hyperref}

\newcolumntype{G}{>{\columncolor{LightGray}}c}

\renewcommand{\thesection}{{\bf \Roman{section}}}

\usepackage{hyperref}

\def\beq{\begin{equation}}
\def\eeq{\end{equation}}

\newcommand*{\fp}[1]{{FP}${}_{#1}$}
\newcommand*{\FP}[1]{$\bm{{\rm FP}_{#1}}$}
\newcommand*{\Fp}[1]{{{\rm \bf FP}_{\bm{#1}}}}

\def\bea{\arraycolsep .1em \begin{eqnarray}}
\def\eea{\end{eqnarray}}
\def\Tr{{\rm Tr}}

\def\eps{\epsilon}

\def\al#1{\alpha_{\rm {#1}}}
\def\eq#1{(\ref{#1})}

\def\s0#1#2{\mbox{\small{$ \frac{#1}{#2} $}}}
\def\0#1#2{\frac{#1}{#2}}

\def\al#1{\alpha_#1}
\def\grgl{\:\hbox to -0.2pt{\lower2.5pt\hbox{$\sim$}\hss}{\raise3pt\hbox{$>$}}\:}
\def\klgl{\:\hbox to -0.2pt{\lower2.5pt\hbox{$\sim$}\hss}{\raise3pt\hbox{$<$}}\:}

\def\lsim{\mathrel{\rlap{\lower4pt\hbox{\hskip1pt$\sim$}}
    \raise1pt\hbox{$<$}}}                % less than or approx. symbol
\def\gsim{\mathrel{\rlap{\lower4pt\hbox{\hskip1pt$\sim$}}
    \raise1pt\hbox{$>$}}}                % greater than or approx. symbol

\newcommand{\xalt}{\widetilde{X}}
\newcommand{\yalt}{\widetilde{Y}}

\makeatletter
    \def\CT@@do@color{%
      \global\let\CT@do@color\relax
            \@tempdima\wd\z@
            \advance\@tempdima\@tempdimb
            \advance\@tempdima\@tempdimc
    \advance\@tempdimb\tabcolsep
    \advance\@tempdimc\tabcolsep
    \advance\@tempdima2\tabcolsep
            \kern-\@tempdimb
            \leaders\vrule
                    \hskip\@tempdima\@plus  1fill
            \kern-\@tempdimc
            \hskip-\wd\z@ \@plus -1fill }
    \makeatother
\begin{document}
${}$\vskip1cm

\title{More asymptotic safety guaranteed}
\author{Andrew~D.~Bond}
\email{a.bond@sussex.ac.uk}
\author{Daniel F.~Litim}
\email{d.litim@sussex.ac.uk}
\affiliation{\mbox{Department of Physics and Astronomy, U Sussex, Brighton, BN1 9QH, U.K.}}

\begin{abstract}

We study interacting fixed points and phase diagrams of  simple and semi-simple
quantum field theories in four dimensions  involving non-abelian gauge fields, fermions and  scalars in the Veneziano limit.
Particular emphasis is put on 
new phenomena which arise due to the semi-simple nature of the theory.
Using matter field  multiplicities as free parameters, we find  a large variety of interacting conformal fixed points with stable vacua and crossovers inbetween. Highlights include
semi-simple gauge theories with exact asymptotic safety,
theories with one or several 
interacting fixed points in the IR,  
 theories where one of the gauge sectors is both UV free and IR free,
and theories with weakly  interacting fixed points in  the UV and the IR limits.
The phase diagrams for various simple and semi-simple settings are also given. 
Further aspects such as perturbativity beyond the Veneziano limit, conformal windows, and implications for model building are  discussed.

\end{abstract}
\vskip-1cm

\maketitle

\newpage
\begin{spacing}{.83}
\tableofcontents
\end{spacing}

\section{\bf Introduction}

Asymptotic freedom is a key feature of non-Abelian gauge theories \cite{Gross:1973id,Politzer:1973fx}. It predicts that  interactions weaken with growing energy due to quantum effects, thereby reaching a free ultraviolet (UV) fixed point under the renormalisation group.  
Asymptotic safety, on the other hand, stipulates that running couplings may very well asymptote into an interacting  UV fixed point at highest energies  \cite{Wilson:1971bg,Weinberg:1980gg}. 
The most striking difference between asymptotically free and asymptotically safe theories relates to residual interactions in the UV. Canonical power counting is modified, whence
establishing asymptotic safety  in a reliable manner becomes a challenging  task  \cite{Falls:2013bv}. 

Rigorous results for asymptotic safety at weak coupling have been known since long for models including either scalars, fermions, gauge fields or gravitons, and  away from their respective critical dimensionality
\cite{Gastmans:1977ad,Tomboulis:1977jk,Christensen:1978sc,Tomboulis:1980bs,
Weinberg:1980gg,Peskin:1980ay,Smolin:1981rm,Bardeen:1983rv,Gawedzki:1985uq,
Gawedzki:1985ed,deCalan:1991km,Kazakov:2007su}. 
In these toy models asymptotic safety  arises through the cancellation of tree level and leading order quantum terms.  
Progress has also been made to substantiate the asymptotic safety conjecture beyond weak coupling \cite{Falls:2013bv}. This is of particular relevance for quantum gravity where  good evidence has arisen in a variety of different settings~\cite{Reuter:1996cp,Litim:2003vp,Litim:2006dx,Niedermaier:2006ns,Litim:2008tt,
Codello:2008vh,Litim:2011cp,Falls:2014tra,Biemans:2016rvp,Folkerts:2011jz,Christiansen:2012rx,Dona:2013qba,Meibohm:2015twa,Christiansen:2017cxa,Falls:2017lst}.

An important new development in the understanding of asymptotic safety has been initiated in \cite{Litim:2014uca} where it was shown that  certain  four-dimensional quantum field theories involving $SU(N)$ gluons, quarks, and scalars can develop weakly coupled UV fixed points.
Results have been extended beyond classically marginal interactions \cite{Buyukbese:2017ehm}. Structural insights into the renormalisation of general gauge theories have led to necessary and sufficient conditions  for asymptotic safety, alongside strict no go theorems
 \cite{Bond:2016dvk,Bond:2017sem}.  
Asymptotic safety invariably arises as a quantum critical phenomenon through  cancellations at loop level for which
 all three types of elementary degrees of freedom --- scalars, fermions and gauge fields --- are required. 
Findings have also been extended to cover supersymmetry \cite{Bond:2017suy} and UV conformal windows \cite{Bond:2017tbw}.
 Throughout, it is found that suitable Yukawa interactions are pivotal  \cite{Bond:2016dvk,Bond:2017sem}. 

In this paper,  we are interested in fixed points of
semi-simple gauge theories. Our primary motivation is the semi-simple nature of the Standard Model, and the prospect for asymptotically safe extensions thereof \cite{Bond:2017wut}. We are particularly interested in  semi-simple theories where interacting fixed points and asymptotic safety can be established rigorously \cite{Bond:2016dvk}.
More generally, we also wish to 
understand 
how  low- and high-energy fixed points are generated dynamically, what their features are, and whether novel phenomena arise owing to  the semi-simple nature of the underlying gauge symmetry. 
Understanding the stability of a Higgs-like ground state at interacting fixed points is also of interest in view of the ``near-criticality" of the Standard Model vacuum \cite{EliasMiro:2011aa,Buttazzo:2013uya}.

We investigate these questions  for    quantum field theories  with  
$SU(N_C)\times SU(N_c)$ local gauge symmetry coupled to  massless fermionic and singlet scalar  matter.  Our models  also have a  global $U(N_{\rm F})_L \times U(N_{\rm F})_R\times U(N_{\rm f})_L \times U(N_{\rm f})_R$  flavor symmetry, and are characterised by up to nine independent  couplings. Matter field multiplicities serve as free parameters.   We obtain rigorous results from the leading orders in perturbation theory by adopting a Veneziano limit. We then provide a comprehensive classification of  quantum field theories according to their UV and IR limits, their fixed points, and eigenvalue spectra. 
Amongst these, we find semi-simple gauge theories with exact asymptotic safety in the UV. 
We also find a large variety of theories with  crossover- and low-energy fixed points. Further novelties include theories with inequivalent yet fully attractive IR conformal fixed points,  theories with weakly  interacting fixed points in both the UV and the IR, and massless theories with a non-trivial gauge sector which is UV free and IR free. 
We  illustrate our results by providing general phase diagrams for simple and semi-simple gauge theories with and without Yukawa interactions.

The paper is organised as follows. General aspects of weakly interacting fixed points in $4d$  gauge theories are laid out in Sec.~\ref{FPgeneral}, together with  first results and expressions for universal exponents. In Sec.~\ref{MM} we introduce concrete families of semi-simple gauge theories coupled to elementary singlet ``mesons'' and suitably charged massless fermions. Perturbative RG equations for all gauge, Yukawa and scalar couplings and masses in a Veneziano limit are provided to the leading non-trivial orders in perturbation theory. Sec.~\ref{FPs} presents  our results for all interacting perturbative fixed points and their universal scaling exponents. Particular attention is paid to new effects which arise due to the semi-simple nature of the models.  Sec.~\ref{stab}   provides the corresponding fixed points in the scalar sector. It also establishes stability of the quantum vacuum whenever a physical fixed point arises in the gauge sector.  Using field multiplicities as free parameters, Sec.~\ref{class} provides  a complete classification of distinct models with asymptotic freedom or asymptotic safety in the UV, or without UV completions, together with their scaling in the deep IR.  In Sec.~\ref{PD}, the generic phase diagrams for simple and semi-simple gauge theories with and without Yukawas are discussed. The phase diagrams, UV -- IR transitions, and aspects of IR conformality   are analysed in more depth  for sample theories with asymptotic freedom and asymptotic safety. Further reaching topics such as exact perturbativity, extensions beyond the Veneziano limit, and conformal windows are discussed in Sec.~\ref{disc}. Sec.~\ref{sum} closes with a brief summary.

\section{\bf Fixed points of gauge theories}\label{FPgeneral}
In this section, we discuss general aspects of interacting fixed points in semi-simple gauge theories which are weakly coupled to matter, with or without Yukawa interactions, following \cite{Bond:2016dvk,Bond:2017sem}.  We also introduce some notation and conventions. 

\subsection{Fixed points in perturbation theory}
We are interested in the renormalisation of general gauge theories coupled to matter fields, with or without Yukawa couplings. The running of the gauge couplings $\alpha_i=g_i^2/(4\pi)^2$ with the renormalisation group scale $\mu$ is determined by the beta functions of the theory. Expanding them perturbatively up to two loop we have
\begin{align}\label{eq:RGE}
	\mu \partial_\mu \alpha_i \equiv \beta_i= \alpha_i^2(-B_i + C_{ij}\alpha_j
	- 2\,Y_{4,i}) + O(\alpha^4)\,,
\end{align}
where a sum over gauge group factors $j$ is implied.  
The one- and two-loop  gauge contributions $B_i$ and $C_{ij}$  and the two-loop Yukawa contributions $Y_{4,i}$ are known for general gauge theories, see~\cite{Machacek:1983tz,Machacek:1983fi,Machacek:1984zw,Luo:2002ti,Bond:2016dvk} for explicit expressions. While $B_i$ and $C_{ii}$ may take either sign, depending on the matter content, the Yukawa contribution  $Y_{4,i}$ and the off-diagonal gauge contributions $C_{ij}$ $(i\neq j)$ are strictly positive in any quantum field theory.
Scalar couplings do not play any role at this order in perturbation theory.
The effect of Yukawa couplings is incorporated by projecting the gauge beta functions \eq{eq:RGE} onto the Yukawa nullclines $(\beta_Y = 0)$, leading to explicit expressions for $Y_{4,i}$ in terms of the gauge couplings $g_j$. 
Moreover, for many theories  the Yukawa contribution along nullclines can be written as $Y_{4,i}=D_{ij}\,\alpha_j$ with $D_{ij} \ge 0$ \cite{Bond:2016dvk}. We  can then go one step further  and express the net effect of Yukawa couplings as a shift of the two loop gauge contribution,    
	$C_{ij} \rightarrow C_{ij}' = C_{ij} - 2D_{ij} \leq C_{ij}$. Notice that the shift will always be by some negative amount  provided at least one of the Yukawa couplings is non-vanishing.   It leads to the reduced gauge beta functions
\begin{align}\label{eq:RGE2}
	\beta_i = \alpha_i^2(-B_i + C'_{ij}\alpha_j) + O(\alpha^4)\,.
\end{align}
 Fixed points solutions of \eq{eq:RGE2} are either free or interacting and $\alpha^*=0$ for some or all gauge factors is always a self-consistent solution. 
Consequently, interacting fixed points are solutions to
\begin{align}\label{eq:matFP}
	B_i &= C_{ij}'\,\alpha_j^*\,,\quad\text{subject to}\quad\alpha_i^*> 0\,,
\end{align}
where only those rows and columns are retained where gauge couplings are interacting. 

Next we discuss the role of Yukawa couplings for the fixed point structure. In the absence of Yukawa couplings, the two-loop coefficients remain unshifted $C_{ij}' = C_{ij}$. An immediate consequence of this is that any interacting fixed point  must necessarily be IR. The reason is as follows: for an interacting fixed point to be UV, asymptotic freedom cannot be maintained for all gauge factors, meaning that some $B_i < 0$. However, as has been established in  \cite{Bond:2016dvk}, $B_i \leq 0$ necessarily entails $C_{ij} \geq 0$ in any $4d$ quantum gauge theory. If the left hand side of \eqref{eq:matFP} is negative, if only for a single row, positivity of $C_{ij}$ requires that some $\alpha_j^*$ must take negative values for  a fixed point solution to arise. This, however,  is unphysical \cite{Dyson:1952tj} and we are left with $B_i > 0$ for each $i$, implying that  asymptotic freedom remains intact in all gauge sectors. Besides the Gaussian, the theory may have weakly interacting infrared Banks-Zaks  fixed points in each gauge sector, as well as products thereof, which arise as solutions to \eq{eq:matFP} with the unshifted coefficients. 

In the presence of Yukawa couplings, the coefficients $C_{ij}'$ can in general take either sign. This has far reaching implications. Firstly, the theory can additionally display gauge-Yukawa fixed points where both the gauge and the Yukawa couplings take interacting values. 
Most importantly, solutions to \eqref{eq:matFP} are then no longer limited to theories with asymptotic freedom. Instead, interacting fixed points can be infrared, ultraviolet, or of the crossover type. In general we may expect gauge-Yukawa fixed points for each independent Yukawa nullcline. In summary, perturbative fixed points are either $(i)$ free and given by the Gaussian, or $(ii)$ free in the Yukawa but interacting in the gauge sector (Banks-Zaks fixed points), or $(iii)$ simultaneously interacting  in the gauge and the Yukawa sector (gauge-Yukawa fixed points), or $(iv)$ combinations and products of $(i)$, $(ii)$ and $(iii)$. Banks-Zaks fixed points are always IR, while the Gaussian and gauge-Yukawa fixed points can be either UV or IR. Depending on the details of the theory and its Yukawa structure, either the Gaussian or one of the interacting gauge-Yukawa fixed points will arise as the  ``ultimate'' UV fixed point of the theory and may serve to define the theory fundamentally \cite{Bond:2017sem}.

The effect of scalar quartic self-couplings on the fixed point is strictly sub-leading in terms of the values of the fixed points, as they do not affect the running of gauge couplings at this order of perturbation theory. However, as to have a true fixed point we must acquire one in all couplings, they provide additional constraints on the physicality of candidate gauge-Yukawa fixed points, as we additionally require that the quartic couplings take fixed points which are both real-valued, and lead to a bounded potential which leads to a stable vacuum state.

\subsection{Gauge couplings}

Let us now consider a semi-simple gauge-Yukawa theory with non-Abelian gauge fields under the semi-simple gauge group ${\cal G}_1\otimes {\cal G}_2$ coupled to fermions and scalars.  We  have two non-Abelian gauge couplings $\al 1$ and  $\al 2$, which are related to the fundamental gauge couplings via $\alpha_i=g_i^2/(4\pi)^2$. 
The running of gauge couplings within perturbation theory is given by
\beq\label{beta12}
\begin{array}{rcl}
\beta_1&=&
\displaystyle
-B_1\, \alpha_1^2+C_1\,\alpha_1^3 +G_1\,\alpha_1^2\,\alpha_2\,,\\[1ex]
\beta_2&=&
\displaystyle
 -B_2\, \alpha_2^2+C_2\,\alpha_2^3 +G_2\,\alpha_2^2\,\alpha_1\,.
\end{array}
\eeq
Here, $B_i$ are the well known one-loop coefficients.  In theories without Yukawa interactions, or where Yukawa interactions take Gaussian values, the numbers  $C_i$ and $G_i$ are the two-loop coefficients which arise owing to the gauge loops and owing to the mixing between gauge groups, meaning $C_i\equiv C_{ii}$ (no sum), and $G_{1}\equiv C_{12}$, $G_{2}\equiv C_{12}$, see \eq{eq:RGE}. In this case, we also have that $C_i, G_i\ge 0$ as soon as $B_i<0$.\footnote{General formal expressions of loop coefficients in the conventions used here are given in  \cite{Bond:2016dvk}.} For theories where Yukawa couplings take interacting fixed points
the numbers  $C_i$ and $G_i$ receive corrections due to the Yukawas, $C_i\equiv C'_{ii}$ (no sum), and $G_{1}\equiv C'_{12}$, $G_{2}\equiv C'_{12}$, see \eq{eq:RGE2}. Most notably, strict positivity of $C_i$ and $G_i$ is then no longer guaranteed \cite{Bond:2016dvk}.

\begin{table*}
\begin{center}
\begin{tabular}{cccc}
\toprule
\rowcolor{LightGreen}
&&&\\[-4mm]
\rowcolor{LightGreen}
\multicolumn{2}{c}{\bf fixed point} 
&${}\quad\bm\alpha_{\rm \bf gauge}{}\quad$
&${}\quad\bm\alpha_{\rm \bf Yukawa}{}\quad$
\\
\midrule
\rowcolor{LightGray}
Gauss&G&$=0$&$=0$\\
Banks-Zaks&BZ& $\neq 0$&$=0$\\
\rowcolor{LightGray}
gauge-Yukawa\ \  &GY& $\neq 0$ & $\neq 0$\\
\bottomrule
\end{tabular}
\caption{Conventions to denote the basic fixed points (Gaussian, Banks-Zaks, or gauge-Yukawa) of  simple gauge theories weakly coupled to matter.}
 \label{tFPdef}
 \end{center}
\end{table*}

In either case, the fixed points of the combined system are determined by the vanishing of \eq{beta12}. For a general semi-simple gauge theory with two gauge factors, one finds four different types of fixed points. The Gaussian  fixed point
\beq
\label{1}
(\al1^*,\al2^*)=(0,0)
\eeq
always exist (see Tab.~\ref{tFPdef} for our conventions). It is the UV fixed point of the theory as long as the one-loop coefficients obey $B_i>0$. The theory may also develop partially interacting fixed points, 
\bea
\label{2}
&&(\al1^*,\al2^*)=\left(0,\frac{B_2}{C_2}\right)\,,\\
\label{3}
&&(\al1^*,\al2^*)=\left(\frac{B_1}{C_1},0\right)\,.
\eea
Here, one of the gauge coupling is taking Gaussian values whereas the other one is interacting. The interacting fixed point is of the Banks-Zaks type  \cite{Caswell:1974gg,Banks:1981nn}, provided Yukawa interactions are absent. This then also implies that the gauge coupling is asymptotically free. Alternatively, the interacting fixed point can  be of the gauge-Yukawa type, provided that Yukawa couplings take an interacting fixed point themselves. In this case, and depending on the details of the Yukawa sector, the fixed point can be either IR or UV. 
Finally, we also observe fully interacting fixed points
\beq
\label{4}
(\al1^*,\al2^*)=\left(\frac{C_2B_1-B_2G_1}{C_1C_2-G_1G_2},\frac{C_1B_2-B_1G_2}{C_1C_2-G_1G_2}\right)\,.
\eeq
As such, fully interacting fixed points \eq{4} can be either UV or IR, depending on the specific field content of the theory.   In all cases we will additionally require that the couplings obey
\beq\label{pos}
\begin{array}{rcl}
\alpha_1&\ge& 0\,,\\[1ex]
\alpha_2&\ge& 0\,.
\end{array}
\eeq
to ensure they reside in the physical regime of the theory \cite{Dyson:1952tj}.

\begin{center}
\begin{table}
\begin{tabular}{ccccc}
 \toprule
 \rowcolor{LightBlue}
 {}\ \ \ \bf coupling\ \ \ &\multicolumn{4}{c}{\ \ order in perturbation theory\ \ }\\
\midrule
\rowcolor{LightGray}     
 $\beta_{\rm gauge}$&1&2&2&$n+1$ \\
\rowcolor{white}
  $\beta_{\rm Yukawa}$ &0&1&1&$n$\\
\rowcolor{LightGray}
   $\beta_{\rm scalar}$ &0&0&1&$n$\\
\midrule
\rowcolor{LightGreen}
    \ \ \bf approximation\ \ &{}\ \ \ \ \ LO\ \ \ \ \  &\ \ NLO\ \  &\ NLO${}^\prime$\ &\ $n$NLO${}^\prime$\   \\
\bottomrule
\end{tabular}
\caption{\label{Tab} Relation between approximation level and the loop order up to which couplings are retained in perturbation theory, following the terminology of \cite{Litim:2014uca,Litim:2015iea}.}\label{tNLO}
\end{table} 
\end{center}

\subsection{Yukawa couplings}
In order to proceed, we must specify the Yukawa sector. 
We assume three types of non-trivially charged fermions 
with charges under ${\cal G}_1$ and ${\cal G}_2$.  Some or all of the fermions which are only charged under ${\cal G}_1$ (${\cal G}_2$)
also couple to scalar fields via Yukawa couplings $\al Y$ ($\al y$), respectively. 
The scalars may or may not be charged under the gauge symmetries. 
They will have quartic self couplings which play no primary role for the fixed point analysis at weak coupling \cite{Bond:2016dvk}. 
Within perturbation theory, the beta functions for the gauge and Yukawa couplings are of the form
\beq\label{12Yy}
\begin{array}{rcl}
\beta_1&=&
\displaystyle
 -B_1\, \alpha_1^2+C_1\,\alpha_1^3 -D_1\,\alpha_1^2\,\al{Y}+G_1\,\al1^2\,\al2\,,\\[.5ex]
\beta_Y&=&
\displaystyle
 \ \ E_1\,\al{Y}^2-F_1\,\al{Y}\,\al 1\,,\\[.5ex]
\beta_2&=&
\displaystyle
 -B_2\, \al2^2+C_2\,\al2^3 -D_2\,\al2^2\,\al y +G_2\,\al2^2\,\al1\,,\\[.5ex]
\beta_y&=&
\displaystyle
 \ \ E_2\,\al{y}^2-F_2\,\al{y}\,\al 2\,.
\end{array}
\eeq
The RG flow is given up to two-loop in the gauge couplings, and up to one-loop in the Yukawa couplings. 
We refer to this as the NLO approximation, see Tab.~\ref{Tab} for the terminology.

We are interested in the fixed points of the theory, defined implicitly via the vanishing of the beta functions for all couplings.  The Yukawa couplings can display either a Gaussian or an interacting  fixed point
\beq\label{Y}
\begin{array}{rcl}
\displaystyle
\al Y^*&=&
\displaystyle
0\,,\quad\al Y^*=\frac{F_1}{E_1}\,\al 1^*\,,\\[2ex]
\al y^*&=&
\displaystyle
0\,,\quad\al y^*=\frac{F_2}{E_2}\,\al 2^*\,.
\end{array}
\eeq
Depending on whether none, one, or both  of the Yukawa couplings take an interacting fixed point, the system \eq{12Yy} reduces to \eq{beta12} whereby the two-loop coefficients $C_i$ of the gauge beta functions are shifted according to
\beq \label{C'}
\begin{array}{rcl}
\alpha_Y^*\neq 0:\quad C_1\to C_1'&=&
\displaystyle
C_1-D_1\,\0{F_1}{E_1}\le C_1\,,\\[2ex]
\displaystyle
\alpha_y^*\neq 0:\quad C_2\to C_2'&=&
\displaystyle
C_2-D_2\0{F_2}{E_2}\le C_2\,.
\end{array}
\eeq
Notice also that in this model the values for the mixing terms $G_i$ do not depend on whether the corresponding Yukawa couplings vanish, or not, due to the fact that no fermions charged under both groups are involved in Yukawa interactions. Owing to the fixed point structure of the Yukawa sector \eq{Y}, the formal fixed points \eq{1}, \eq{2}, \eq{3} and \eq{4} have the multiplicity $1, 2, 2$ and  $4$, respectively. In total, we  end up with nine qualitatively different fixed points FP${}_1$ -- FP${}_9$, summarised in Tab.~\ref{tFPs}: FP${}_1$ denotes the unique Gaussian fixed point. FP${}_2$ and FP${}_3$ correspond to a Banks-Zaks fixed point in one of the gauge couplings, and a Gaussian in the other. They can therefore be interpreted effectively as a ``product'' of a Banks-Zaks with a Gaussian fixed point. Similarly, at FP${}_4$ and FP${}_5$, one of the Yukawa couplings remains interacting, and they can therefore effectively be viewed as the  product of a gauge-Yukawa (GY) type fixed point in one gauge coupling with a Gaussian fixed point in the other. The remaining fixed points FP${}_6$ -- FP${}_9$ are interacting in both gauge couplings. These fixed points are the only ones which are sensitive to the two-loop mixing coefficients $G_1$ and $G_2$.  At FP${}_6$, both Yukawa couplings vanish meaning that it is effectively a product of two Banks-Zaks type fixed points.
At FP${}_7$ and FP${}_8$, only one of the Yukawa couplings vanish, implying that these are products of a gauge-Yukawa with a Banks-Zaks fixed point. Finally, at FP${}_9$, both Yukawa couplings are non-vanishing meaning that this is effectively the product of two gauge-Yukawa fixed points. 
\begin{table*}
\begin{center}
\begin{tabular}{cccccc}
\toprule
\rowcolor{LightBlue}
&
 \multicolumn{2}{c}{\cellcolor{LightRed}  \bf gauge couplings}
&
 \multicolumn{2}{c}{\cellcolor{LightYellow}  \bf Yukawa couplings}
&\cellcolor{LightGreen} \bf \  fixed point \
\\
\rowcolor{LightBlue}
\multirow{-2}{*}{\bf \ fixed point\  }
&\cellcolor{LightRed}  
${}\quad  \bm{\al 1^*}\quad$
&\cellcolor{LightRed}  
${}\quad \bm{\al 2^*}\quad$
&\cellcolor{LightYellow} 
${}\quad\quad  \bm{\al Y^*}\quad \quad$
&\cellcolor{LightYellow} 
${}\quad \quad \bm{\al y^*}\quad \quad$
&\cellcolor{LightGreen}  
{\bf \ \ \ type\ \ \ }\\
\midrule
\rowcolor{LightGray}
&&&&& \\[-2.5ex]
\rowcolor{LightGray}
\bf FP${}_{\bf 1}$
&0&0&0&0
&\bf G $\bm \cdot$ G\\[.5ex]
&&&&& \\[-2.5ex]
\bf FP${}_{\bf 2}$
&$\displaystyle\0{B_1}{C_1}$&0&0&0
&\bf BZ $\bm \cdot$  G\\[1.5ex]
\rowcolor{LightGray}
&&&&& \\[-2.5ex]
\rowcolor{LightGray}
\bf FP${}_{\bf 3}$
&0&$\displaystyle\0{B_2}{C_2}$&0&0
&\bf  G $\bm \cdot$  BZ\\[1.5ex]
&&&&& \\[-2.5ex]
\bf FP${}_{\bf 4}$
&$\displaystyle\0{B_1}{{C'_1}}$&0&$\displaystyle\0{F_1}{E_1}\,\al 1$&0
&\bf  GY $\bm \cdot$ G\\[1.5ex]
\rowcolor{LightGray}
&&&&& \\[-2.5ex]
\rowcolor{LightGray}
\bf FP${}_{\bf 5}$
&0&$\displaystyle\0{B_2}{{C'_2}}$&0&$\displaystyle\0{F_2}{E_2}\,\al 2$
&\quad\bf  G $\bm \cdot$ GY\\[1.5ex]
&&&&& \\[-2.5ex]
\bf FP${}_{\bf 6}$
&$\ \displaystyle\frac{C_2B_1-B_2G_1}{C_1C_2-G_1G_2}\ $
&$\ \displaystyle\frac{C_1B_2-B_1G_2}{C_1C_2-G_1G_2}\ $&0&0
&\bf  BZ $\bm \cdot$ BZ\\[1.5ex]
\rowcolor{LightGray}&&&&& \\[-2.5ex]
\rowcolor{LightGray}\bf FP${}_{\bf 7}$
&$\displaystyle\frac{C_2B_1-B_2G_1}{{C'_1}C_2-G_1G_2}$
&$\displaystyle\frac{{C'_1}B_2-B_1G_2}{{C'_1}C_2-G_1G_2}$&
$\displaystyle\frac{F_1}{E_1}\,\al 1$&0
&\bf  GY $\bm \cdot$ BZ\\[1.5ex]
&&&&& \\[-2.5ex]
\bf FP${}_{\bf 8}$
&$\displaystyle\frac{{C'_2}B_1-B_2G_1}{C_1{C'_2}-G_1G_2}$
&$\displaystyle\frac{C_1B_2-B_1G_2}{C_1{C'_2}-G_1G_2}$&0&
$\displaystyle\frac{F_2}{E_2}\,\al 2$
&\bf  BZ $\bm \cdot$ GY\\[1.5ex]
\rowcolor{LightGray}
&&&&& \\[-2.5ex]
\rowcolor{LightGray}
\bf FP${}_{\bf 9}$
&$\displaystyle\frac{{C'_2}B_1-B_2G_1}{{C'_1C'_2}-G_1G_2}$
&$\displaystyle\frac{{C'_1}B_2-B_1G_2}{{C'_1C'_2}-G_1G_2}$
&$\displaystyle\frac{F_1}{E_1}\,\al 1$
&$\displaystyle\frac{F_2}{E_2}\,\al 2$
&\bf  GY $\bm \cdot$  GY\\[1.5ex]
\bottomrule
\end{tabular}
\caption{The various types of  fixed points in gauge-Yukawa theories with semi-simple gauge group ${\cal G}_1\otimes {\cal G}_2$ and \eq{12Yy}, \eq{C'}. 
We also indicate how the nine qualitatively  different fixed points can be interpreted as products of the Gaussian (G), Banks-Zaks (BZ) and gauge-Yukawa (GY) fixed points as seen from the individual gauge group factors (see main text).}
 \label{tFPs}
\end{center}
\end{table*}

In theories where none of the fermions carries gauge charges under both gauge groups, we have that $G_1= 0=G_2$. In this limit, 
and at the present level of approximation, 
the gauge sectors do not  communicate with each other and the ``direct product'' interpretation of the fixed points as detailed above becomes exact. For the purpose of this work
we will find it useful to refer to the effective ``product'' structure of interacting fixed points even in settings with $G_1,G_2\neq 0$. 
Whether any of the fixed points is factually realised in a given theory crucially depends on the explicit values of the various loop coefficients. We defer an explicit investigation for certain ``minimal models'' to Sec.~\ref{MM}.

\subsection{Scalar couplings}

In  \cite{Bond:2016dvk}, it has been established that scalar self-interactions play no role for the primary occurrence of weakly interacting fixed points in the gauge- or gauge-Yukawa sector.  On the other hand, for consistency, scalar couplings must nevertheless take free or interacting fixed points on their own. The necessary and sufficent conditions for this to arise have been given in \cite{Bond:2016dvk}. Firstly, scalar couplings must take physical (real) fixed points. Secondly, the theory must display a stable ground state at the fixed point in the scalar sector. Below, we will analyse concrete models and show that both of these conditions are non-empty.

\subsection{Universal scaling exponents}
We briefly comment on the universal behaviour and scaling exponents at the interacting fixed points of Tab.~\ref{tFPs}. Scaling exponents 
arise as the eigenvalues $\vartheta_i$ 
of the stability matrix 
\beq\label{M} M_{ij}=\partial\beta_i/\partial\alpha_j|_*\eeq 
at fixed points. Negative or positive eigenvalues correspond to relevant or irrelevant couplings respectively. They imply that couplings approach the fixed point following a power-law behaviour in RG momentum scale, 
\beq \alpha_i(\mu)-\alpha_i^*=\sum_n c_n\, V^n_i \,\left(\frac{\mu}{\Lambda}\right)^{\vartheta_n}+{\rm subleading}\,.\eeq  
Classically, we have that $\vartheta\equiv  0$. Quantum-mechanically, and at a Gaussian fixed point, eigenvalues continue to vanish and the behaviour of couplings is determined by higher order effects. Then couplings are either exactly marginal $\vartheta\equiv  0$ or marginally relevant $\vartheta\to 0^-$ or marginally irrelevant $\vartheta\to 0^+$. In a slight abuse of language we will from now on denote  relevant and marginally relevant ones as $\vartheta\le 0$, and vice versa for irrelevant ones.   

Given that the scalar couplings do not feed back to the gauge-Yukawa sector at the leading non-trivial order in perturbation theory, we may neglect them for a discussion of the eigenvalue spectrum
\beq\{\vartheta_i,i=1,\cdots 4\}\,,
\eeq
 related to the two gauge and Yukawa couplings.
The fixed point FP${}_1$ is Gaussian in all couplings, and the scaling of couplings are either marginally relevant or marginally irrelevant. Only if $B_i>0$ trajectories can emanate from the Gaussian, meaning that it is a UV fixed point iff the theory is asymptotically free in both couplings. Furthermore, asymptotic freedom in the gauge couplings entails asymptotic freedom in the Yukawa couplings leading to four marginally relevant couplings with eigenvalues
\beq\label{4R}
\vartheta_1, \vartheta_2,\vartheta_3,\vartheta_4 \le 0
\eeq
The fixed points FP${}_2$ and FP${}_3$ are products of a Banks-Zaks in one gauge sector with a Gaussian fixed point in the other. Scaling exponents are then of the form
\beq\label{3R}
\vartheta_1, \vartheta_2,\vartheta_3\le 0< \vartheta_4 
\eeq
provided the gauge sector with Gaussian fixed point is asymptotically free. For IR free gauge coupling, we instead have the pattern
\beq\label{1R}
\vartheta_1<0\le \vartheta_2,\vartheta_3,\vartheta_4\,.
\eeq
At the fixed points FP${}_4$ and FP${}_5$, the theory is the product of a Gaussian and  a gauge-Yukawa fixed point.  Consequently, four possibilities arise: Provided that  the theory is asymptotically safe at the gauge-Yukawa fixed point and asymptotically or infrared free at the Gaussian, scaling exponents are of the form \eq{3R} or \eq{1R}, respectively. Conversely, if the gauge Yukawa fixed point is IR, the eigenvalue spectrum reads 
\beq\label{2R}
\vartheta_1, \vartheta_2\le0\le \vartheta_3,\vartheta_4 
\eeq
if the Gaussian is asymptotically free. Finally, if the Gaussian is IR free and the gauge-Yukawa fixed point IR, all couplings are UV irrelevant and
\beq\label{0R}
0\le \vartheta_1, \vartheta_2,\vartheta_3,\vartheta_4\,.
\eeq
More work is required  to determine the scaling exponents at the  fully interacting fixed points FP${}_6$ -- FP${}_9$. To that end, we write the characteristic polynomial of the stability matrix as
\beq
\sum_{n=0}^4 T_n\,\vartheta^n=0\,.
\eeq
The coefficients $T_n$ are functions of the loop coefficients. Introducing  $B=|B_1|$ and $B_2=P\,B_1$, with $P$ some free parameter, we can make a scaling analysis in the limit $B\ll 1$. Normalising the coefficient $T_4$ to unity, $T_4=1$, it then follows from the structure of the beta functions that $T_0={\cal O}(B^6), T_1={\cal O}(B^4), T_2={\cal O}(B^2)$ and  $T_3={\cal O}(B)$ to leading order in $B$. In the limit where $B\ll 1$ we can deduce exact closed expressions for the leading order behaviour of the eigenvalues from solutions to two quadratic equations,
\beq\label{dtheta}
\begin{array}{rcl}
0&=&\ \ \ \,\vartheta^2+T_3\,\vartheta+T_2\\[1ex]
0&=&T_2\, \vartheta^2+T_1\,\vartheta+T_0\,.
\end{array}
\eeq
 The general expressions are quite lengthy and shall not be given here explicitly. We note that the four eigenvalues of the four couplings at the four fully interacting fixed points FP${}_6$ -- FP${}_9$ are the four solutions to \eq{dtheta}. Irrespective of their signs, and barring exceptional numerical cancellations, we conclude that
two scaling exponents are quadratic and two are linear in  $B$,
\beq\label{theta}
\begin{array}{rcl}
\vartheta_{1,2}&=&
\displaystyle
-\frac{1}{2}\left(T_3\pm\sqrt{T^2_3-4T_2}\right)={\cal O}(B^2)\\[2ex]
\vartheta_{3,4}&=&
\displaystyle
-\frac{1}{2 T_2}\left(T_1\pm\sqrt{T^2_1-4T_0\,T_2}\right)={\cal O}(B)\,.
\end{array}
\eeq
This is reminiscent of fixed points in gauge-Yukawa theories with a simple gauge group. The main reason for the appearance of two eigenvalues of order ${\cal O}(B^2)$ relates to the gauge sector, where the interacting fixed point arises through the cancellation at two-loop level.  Conversely, two eigenvalues of order ${\cal O}(B)$ relate to the Yukawa couplings, as they arise from a cancellation at one-loop level.
This completes the discussion of fixed points in general weakly coupled semi-simple gauge theories.

\section{\bf Minimal models}\label{MM}
In this section we introduce in concrete terms a family of semi simple gauge theories whose interacting fixed points will be analysed exactly within perturbation theory in the Veneziano limit.

\subsection{Semi-simple gauge theory}
We consider  families of massless four-dimensional quantum field theories with a semi-simple gauge group
\beq\label{gaugegroup}
SU(N_{\rm C}) \times SU(N_{\rm c})
\eeq
for general non-Abelian factors with $N_{\rm C}\ge 2$ and $N_{\rm c}\ge 2$. 
Specifically, our models contains $SU(N_{\rm C})$ gauge fields $A_\mu$ with field strength $F_{\mu\nu}$,
and $SU(N_{\rm c})$ gauge fields $a_\mu$ with field strength $f_{\mu\nu}$. 
The gauge fields are coupled to $N_{\rm F}$ flavors of fermions $Q_i$, $N_{\rm f}$ flavors of fermions $q_i$, and $N_\psi$ flavors of fermions $\psi_i$. The fermions $(Q,q,\psi)$ transform in the fundamental representation 
of the first, the second, and both gauge group(s) \eq{gaugegroup}, respectively, as summarised in Tab.~\ref{tCharges}. The Dirac fermions $\psi$ are responsible for the semi-simple character of the theory and
serve as messengers to communicate between gauge sectors. All fermions are Dirac  to guarantee anomaly cancellation. 
The fermions $(Q,q)$ additionally couple via Yukawa interactions to an $N_{\rm F}\times N_{\rm F}$ matrix scalar field $H$ and an $N_{\rm f}\times N_{\rm f}$ matrix scalar field $h$, respectively. The scalars $H$ and $h$ are invariant under $U(N_{\rm F})_L \times U(N_{\rm F})_R$ and $U(N_{\rm f})_L \times U(N_{\rm f})_R$ global flavor rotations, respectively, and
 singlets under the gauge symmetry. They can be viewed as elementary mesons in that they carry the same global quantum numbers as the singlet scalar bound states $\sim \langle Q\bar Q\rangle$ and $\sim \langle q\bar q\rangle$. The fermions $\psi$ are not furnished with Yukawa interactions.

The fundamental action is taken to be the sum of the individual Yang-Mills actions, the fermion kinetic terms, the Yukawa interactions, and the scalar kinetic and self-interaction Lagrangeans $L=L_{\rm YM}+L_F+L_Y+L_S+L_{\rm pot}$, with
\beq \label{L}
\begin{array}{rcl}
L_{\rm YM}&=&
\displaystyle
 - \frac{1}{2} \Tr \,F^{\mu \nu} F_{\mu \nu}
 - \frac{1}{2} \Tr \,f^{\mu \nu} f_{\mu \nu}\\[1.5ex]
L_F&=& 
\displaystyle
\Tr\left(\overline{Q}\,  i\slashed{D}\, Q \right)
+\Tr\left(\overline{q}\,  i\slashed{D}\, q\right)
+\Tr\left(\overline{\psi}\,  i\slashed{D}\, \psi\right)
\\[1.5ex]
L_Y&=&
\displaystyle
Y \,\Tr\left(\overline{Q}_L H Q_R + \overline{Q}_R H^\dagger Q_L\right)
+y \,\Tr\left(\overline{q}_L h q_R + \overline{q}_R h^\dagger q_L\right)
\\[1.5ex]
L_S&=&
\displaystyle
\Tr\,(\partial_\mu H ^\dagger\, \partial^\mu H) +\Tr\,(\partial_\mu h ^\dagger\, \partial^\mu h) \\[1.5ex]
L_{\rm pot}&=&
\displaystyle
-U\,\Tr\,(H ^\dagger H )^2  -V\,(\Tr\,H ^\dagger H )^2 \\[1.5ex]
&&
-u\,\Tr\,(h ^\dagger h )^2  -v\,(\Tr\,h ^\dagger h )^2 
-w\,\Tr\,H ^\dagger H   \,\Tr\,h ^\dagger h  \,.
\end{array}
\eeq
The trace $\Tr$ denotes the trace over both color and flavor indices, and the decomposition $Q=Q_L+Q_R$ with $Q_{L/R}=\frac 12(1\pm \gamma_5)Q$ is understood for all fermions $Q$ and $q$. 
Mass terms are neglected at the present stage as their effect is subleading to the main features developed below.
In four dimensions, the theory is renormalisable in perturbation theory.

\begin{table}[t]
\begin{center}
    \begin{tabular}{ c  c  c  c cccc}
\toprule
\rowcolor{LightBlue}
& \multicolumn{3}{c}{\cellcolor{LightYellow}\bf fermions}
& \multicolumn{2}{c}{\cellcolor{LightGreen} \bf scalars}
& \multicolumn{2}{c}{\cellcolor{LightRed}  \bf gauge fields}\\
    \rowcolor{LightBlue}\multirow{-2}{*}{\bf\  representation\ } & 
\cellcolor{LightYellow} ${}\quad Q{}\quad$  & 
\cellcolor{LightYellow}${}\quad q{}\quad$  & 
\cellcolor{LightYellow} ${}\quad\psi{}\quad$  & 
\cellcolor{LightGreen}  ${}\quad H{}\quad$  &  
\cellcolor{LightGreen}  ${}\quad h{}\quad$  &  
\cellcolor{LightRed}      ${}\quad\ \ A_\mu{}\ \ \quad$ & 
\cellcolor{LightRed}    ${}\quad \ \ a_\mu{}\ \ \quad$\\
\midrule
\rowcolor{LightGray}     
under~$SU(N_{C})$& $N_{\rm C}$ & 1 & $N_{\rm C}$ & 1 & 1 & $N_{\rm C}^2-1$ & 1\\
under~$ SU(N_{c})$& 1 & $N_{\rm c}$ & $N_{\rm c}$ & 1 & 1 & 1 & $N_{\rm c}^2-1$ \\
\rowcolor{LightGray}     
multiplicity
& $N_{\rm F}$ & $N_{\rm f}$ & $N_\psi$ & $N_{\rm F}^2$ & $N_{\rm f}^2$ & 1 & 1\\
\bottomrule
\end{tabular}
\caption{Representation under the semi-simple gauge symmetry \eq{gaugegroup} together with  flavour multiplicities of all fields. Gauge (fermion) fields are either in the adjoint (fundamental) or trivial representation.}
 \label{tCharges}
 \end{center}
\end{table}

 The theory has nine classically marginal coupling constants given by the two gauge couplings, the two Yukawa couplings,  and five quartic scalar couplings. We write them as
\beq
\label{couplings}
\begin{array}{lcr}
\displaystyle
\al 1=\frac{g_1^2\,N_{\rm C}}{(4\pi)^2}\,,\quad
\al 2=\frac{g_2^2\,N_{\rm c}}{(4\pi)^2}\,,\quad
\al Y=\frac{Y^{2}\,N_{\rm C}}{(4\pi)^2}\,,\quad
\al y=\frac{y^{2}\,N_{\rm c}}{(4\pi)^2}\,,
&&\\[2ex]
\displaystyle
\al U=\frac{{u}\,N_{\rm F}}{(4\pi)^2}\,,\quad
\al V=\frac{{v}\,N^2_{\rm F}}{(4\pi)^2}\,,\quad
\al u=\frac{{u}\,N_{\rm f}}{(4\pi)^2}\,,\quad
\al v=\frac{{v}\,N^2_{\rm f}}{(4\pi)^2}\,,
\end{array}
\eeq
where we have normalized the couplings with the appropriate loop factor and powers of $N_C, N_c, N_{\rm F}$ and $N_{\rm f}$ in view of the Veneziano limit to be adopted below. Notice the additional power of $N_{\rm F}$ and $N_{\rm f}$ in the definitions of the scalar double-trace couplings. We normalise the quartic ``portal''  coupling as
\beq\label{w} 
\al w=\frac{{w}\,N_{\rm F}N_{\rm f}}{(4\pi)^2}\,.
\eeq
It is responsible for a mixing amongst the scalar sectors starting at tree level. 
Below, we use  the shorthand notation $\beta_i\equiv\partial_t\alpha_i$ with $i=(1,2,Y,y,U,u,V,v,w)$ to indicate the $\beta$-functions for the couplings \eq{couplings}. To obtain explicit expressions for these, we exploit the formal results summarised in \cite{Machacek:1983tz,Machacek:1983fi,Machacek:1984zw}. 
The semi-simple character of the theory is switched off if the $N_\psi$ messenger fermions  (which carry charges under both gauge groups) are replaced by 
$N_1$ and $N_2$ Yukawa-less fermions in the fundamental of $SU(N_C)$ and $SU(N_c)$, respectively,
with
\beq\label{Relax}
\begin{array}{l}
N_1=N_c\,N_\psi\,,
\\
N_2=N_C\,N_\psi\,.
\end{array}
\eeq
If in addition $\alpha_w=0$, the theories \eq{L} reduce to a ``direct product'' of simple gauge Yukawa theories with \eq{Relax}.
Also, in the limit where one of the gauge groups is switched off, $\alpha_1\equiv 0$ (or $\alpha_2\equiv 0$), one gauge sector and the scalars decouples straightaway, and we are left with a simple gauge theory.  Finally, if  $N_1=0=N_2$, we recover the  models of  \cite{Litim:2014uca} in each gauge sector (displaying asymptotic safety for certain field multiplicities). Below, we will find it useful to contrast results with those from the ``direct product'' limit.

\subsection{Free parameters and Veneziano limit}
We now discuss the set of fundamentally free parameters of our models. On the level of the Lagrangean, the free parameters of the theory are the matter field multiplicities 
\beq\label{Ns}
N_{\rm C},\quad N_{\rm c},\quad N_{\rm F},\quad N_{\rm f},\quad N_\psi\,. 
\eeq
Notice that the $N_\psi$ fermions $\psi$ are centrally responsible for interactions between the gauge sectors. In the limit
\beq\label{N=0}
N_\psi=0
\eeq
the interaction between gauge sectors reduces to effects mediated by the portal coupling $\alpha_w\neq 0$, which are strongly loop-suppressed. In this limit, results for fixed points and running couplings fall back to those for the individual gauge sectors \cite{Litim:2014uca}. 
Results for fixed points for general $N_\psi$ are deferred to App.~\ref{AppN}. Here, we will set $N_\psi$ to a finite value,
\beq\label{N=1}
N_\psi=1\,.
\eeq
This leaves us with four free parameters. In order to achieve exact perturbativity, we perform a Veneziano limit \cite{Veneziano:1979ec} by sending the number of colors and the number of flavors $(N_{\rm C}, N_{\rm c}, N_{\rm F},N_{\rm f})$ to infinity but keeping their ratios fixed. This reduces the set of free parameters of the model down to three, which we chose to be
\beq
\label{RST}
R=\frac{N_{\rm c}}{N_{\rm C}}\,,\quad 
S=\frac{N_{\rm F}}{N_{\rm C}}\,,\quad
T=\frac{N_{\rm f}}{N_{\rm c}}\,.
\eeq
The ratio 
\beq
\label{F}
F=\frac{N_{\rm f}}{N_{\rm F}}
\eeq
is then no longer a free  parameter, but fixed as $F=R\,T/S$ from \eq{RST}. By their very definiton, the parameters \eq{RST}  are positive semi-definite and can take values $0\le F,R,S,T\le \infty$.
However, we will see below that their values are further constrained if we impose perturbativity for all couplings.\subsection{Perturbativity to leading order}
The RG evolution of couplings is analysed within the perturbative loop expansion. To leading order (LO), the running of the gauge couplings reads $\beta_i=-B_i\,\al i^2$ (no sum), with $B_i$ the one-loop gauge coefficients for the gauge coupling $\alpha_i$. In the Veneziano limit, the one-loop coefficients  take the form \beq\label{Bi}
B_i=-\frac43 \eps_i\,.
\eeq
In terms of \eq{RST} and in the Veneziano limit, the parameters $\eps_i$ are given by
\beq \label{eps12}
\begin{array}{rcl}
	\epsilon_1 &=&\displaystyle S+ R- \frac{11}{2}\,,\\[2ex] 
	\epsilon_2 &= &\displaystyle T+ \frac{1}{R}- \frac{11}{2}\,.
\end{array}
\eeq
We can therefore trade the free parameters $(S,T)$ defined in \eq{RST} for $(\eps_1,\eps_2)$ and consider the set \beq\label{eps12R}
(\eps_1,\eps_2,R)
\eeq 
as free parameters which characterise the matter content of the theory. Under the exchange of gauge groups we have
\beq\label{eps21}
(\eps_1,\eps_2,R)\to (\eps_2,\eps_1,R^{-1})\,.
\eeq 
For fixed $R$, we observe that $R-\frac{11}{2}\le \eps_1<\infty$ and $1/R-\frac{11}{2}\le \eps_2<\infty$. 
Perturbativity in either of the gauge couplings requires that both one-loop coefficients $B_i$ are parametrically small compared to unity. Therefore we impose
\beq\label{eps}
0<|\eps_i|\ll1\,.
\eeq
This requirement of exact perturbativity in both gauge sectors entails the important constraint
\beq\label{R}
\frac{2}{11}< R<\frac{11}{2}\,.
\eeq
Outside of this range, no physical values for $S$ and $T$ can be found such that \eq{eps} holds true. Inside this range, physical values are constrained within $0\le S,T\le \0{11}2-\02{11}$. 
The  parameters \eq{eps12R} have a simple interpretation. The small paramaters $\eps_i$  control the perturbativity within each of the gauge sectors, whereas the parameter $R$ controls the ``interactions'' between the two gauge sectors. It is the presence of $R$ which makes these theories intrinsically semi-simple, rather than being the direct product of two simple gauge theories.  Perturbativity is no longer required in the limit where one of the gauge sectors is switched off, and the constraint \eq{R} is relaxed into
\beq\label{Relax2}
\begin{array}{rcc}
\displaystyle
0\le R<\frac{11}{2}&\quad\text{if}&\quad \alpha^*_2\equiv 0\,,\\[1ex]
\displaystyle
\frac{2}{11}< R<\infty&\quad\text{if}&\quad \alpha^*_1\equiv 0\,.
\end{array}
\eeq
The parametrisation \eq{eps12R} is most convenient for expressing the relevant RG beta functions for all couplings.

Finally, for some of the subsequent considerations we replace the two small parameters $(\eps_1,\eps_2)$ by  $(\eps,P)$, a single small parameter $\eps$ proportional to $\eps_1$ together with a parameter $P$ related to the ratio between $\eps_1$ and $\eps_2$. Specifically, we introduce
\beq 
\label{P} 
\begin{array}{rcl}
\eps_1&=&\, R\,\eps\,,\\[1ex]
\eps_2&=&\displaystyle
P\,\0{\eps}{R}\,.
\end{array}
\eeq
which is equivalent to $P= R^2\,{\eps_2}/{\eps_1}$ together with $\eps=\eps_1/R$ and $\eps=P\,R\,\eps_2$.\footnote{The choice \eq{P} can be motivated by dimensional analysis  of \eq{eps12} which shows that $\eps_1$ and $\eps_2$ formally scale as $\sim R$ and $\sim1/R$ for large or small $R$, respectively, whereby their ratio $\eps_1/\eps_1$ scales as $\sim R^2$. The large-$R$ behaviour is factored-out by our parametrisation.}   Since $R$ can only take finite positive values, the additional rescaling with $R$ does not affect the relative sign between $\eps_1$ and $\eps$. In this manner we have traded the free parameters $(\eps_1,\eps_2,R)$  for 
\beq
\label{epsPR}
(R,P,\eps)\,.
\eeq
Notice  that the parameter $P$ can be expressed as
\beq\label{Pexplicit}
P=\frac{1+(N_{\rm f}-\0{11}{2} N_{\rm c})/N_{\rm C}}{1+(N_{\rm F}-\0{11}{2}N_{\rm C})/N_{\rm c}}
\eeq
in terms of the field multiplicities \eq{Ns}. It thus may take any real value of either sign with  $-\infty<P<\infty$, whereas $R$ must take values within the range \eq{R}. Moreover, 
\beq
\eps=1+\frac{N_{\rm F}-\s0{11}{2}N_{\rm C}}{N_{\rm c}}\,.
\eeq
In this parametrisation, the ratio of fermion flavour multiplicities \eq{F} becomes
\beq\label{epsF}
F=\frac{11 R-2}{11-2R}
+\frac{2R}{11-2R}\left(\frac{P}{R}-\frac{11R-2}{11-2R}\right)\eps
+{\cal O}(\eps^2)\,.
\eeq
We also observe that the substitution
\beq\label{exchange}
(R,P,\eps)\to \left(R^{-1},P^{-1},P\eps\right)
\eeq
relates to the exchange of gauge groups. The parametrisation \eq{epsPR} is most convenient for analysing the various interacting fixed points and their scaling exponents (see below).
This completes the definition of our models.

\subsection{Anomalous dimensions}\label{anomalous}
We  provide results for the anomalous dimensions associated to the fermions and scalars. Furthermore,  if mass terms are present, their  renormalisation is induced through the RG flow of the gauge, Yukawa, and scalar couplings. Following \cite{Litim:2014uca}, we define the scalar anomalous dimensions as
$\Delta_S=1+\gamma_S$, where $\gamma_S\equiv  \frac{1}{2}{d\ln Z_S}/{d\ln\mu}$
and $S=H,h$. Within perturbation theory, the one and two loop contributions read
\beq
\label{gammaH}
\begin{array}{rcl}
\gamma_H&=&
\displaystyle
\alpha_Y
-\032\left(\0{11}{2}
-\eps_1-R\right)\alpha_Y^2
+\052\alpha_Y\,\alpha_1
+2\alpha_U^2 +{\cal O}(\alpha^3) \,,
\\[2ex]
\gamma_h&=&
\displaystyle
\,\alpha_y
-\032\left(\0{11}{2}+\eps_2-\01R\right)\alpha_y^2
+\052\alpha_y\,\alpha_2
+2\alpha_u^2 +{\cal O}(\alpha^3) \,.
\end{array}
\eeq
For the fermion anomalous dimensions $\gamma_F\equiv {d\ln Z_F}/{d\ln\mu}$
with $F=Q,q,\psi$, we find
\beq
\label{gammaH}
\begin{array}{rcl}
\gamma_Q&=&
\displaystyle
\left(\0{11}{2}+\eps_1-R\right)\alpha_Y
+\xi_1\,\alpha_1 +{\cal O}(\alpha^2) \,,
\\[2ex]
\gamma_q&=&
\displaystyle
\left(\0{11}{2}+\eps_2-\01R\right)\alpha_Y
+\xi_2\,\alpha_2 +{\cal O}(\alpha^2) \,,
\\[2ex]
\gamma_\psi&=&
\displaystyle
\xi_1\,\alpha_1
+\xi_2\,\alpha_2 +{\cal O}(\alpha^2) \,,
\end{array}
\eeq
where $\xi_1$ and $\xi_2$ denote the gauge fixing parameters for the first and second gauge group respectively.

The anomalous dimension for the scalar mass terms can be derived from the composite operator $\sim M^2\,\Tr\, H^\dagger H$ and $\sim m^2\,\Tr\, h^\dagger h$. Introducing the mass anomalous dimension  $\gamma_{M}=  d\ln M^2/d \ln \mu$, and similarly for $m$,  
one finds 
\beq\label{gammam}
\begin{array}{rcl}
\gamma_M&=&8\al U+4 \al V+2\al Y+{\cal O}(\alpha^2) \\[1ex]
\gamma_m&=&8\al u\,+4 \al v\; +2\al y\ +{\cal O}(\alpha^2) \,,
\end{array}
\eeq
to one-loop order.  We also compute the running of the mass terms for the scalars
\beq\label{massRunS}
\begin{array}{rcl}	\beta_{M^2} &=& \gamma_M\, M^2
+ 2 F \, m^2\, \alpha_w 
+ {\cal O}(\alpha^2,\alpha\,m^2_F) \,,\\[1ex]
	\beta_{m^2} &=& \gamma_m\,  m^2
+ 2F^{-1}\,M^2\,\alpha_w + {\cal O}(\alpha^2,\alpha\,m^2_F) \,,
\end{array}
\eeq
where the parameter $F\equiv N_{\rm f}/N_{\rm F}$ solely depends on $R$ to leading order in $\eps$, see \eq{epsF}. Notice that the coupling $\alpha_w$ induces a mixing amongst the different scalar masses already at one-loop level.

Analogously, the anomalous dimension for the fermion mass operator is defined as
$\Delta_F=3+\gamma_{M_F}$ with $\gamma_{M_F}\equiv  {d\ln M_F}/{d\ln\mu}$,
and $M_F$ stands for one of the fermion masses with $F=Q,q$ or $\psi$. Within perturbation theory, the one loop contributions read
\beq
\label{gammaF}
\begin{array}{rcl}
\gamma_{M_Q}&=&
\displaystyle
\alpha_Y\, \left(\0{13}{2}+\eps_1-R\right)-3\, \alpha_1-+{\cal O}(\alpha^2) \ ,\\[2ex]
\gamma_{M_q}&=&
\displaystyle
\alpha_y\, \left(\0{13}{2}+\eps_2-\01R\right)-3\, \alpha_2+{\cal O}(\alpha^2) \\[2.5ex]
\gamma_{M_\psi}&=&
-3\, (\alpha_1+\alpha_2)+{\cal O}(\alpha^2) \,.
\end{array}
\eeq
 For the fermion masses we have the running
\begin{align}
\nonumber
	\beta_{m_Q} &= \gamma_{M_Q}\,m_Q
\,,\\
\label{massRunF}
	\beta_{m_q} &= \gamma_{M_q}\,m_q
\,,\\
\nonumber
	\beta_{m_\psi} &= \gamma_{M_\psi}\,m_\psi
\,.
\end{align}
We note that $\gamma_{M_\psi}$ is manifestly negative. For $\gamma_{M_Q}$ and $\gamma_{M_q}$ we observe that the gauge and Yukawa contributions arise with manifestly opposite signs in the parameter  regime \eq{eps}, \eq{R}. Hence either of these  may take either sign, depending on whether the gauge or Yukawa contributions dominate.

\subsection{Running couplings beyond the leading order}
We now go beyond the leading order in perturbation theory and provide the complete, minimal set of RG equations which display exact and weakly interacting fixed points. To that end, we must  retain terms up to two loop order in the gauge coupling, or else an interacting fixed point cannot arise. At the same time, in order to explore the feasibility of asymptotically safe UV fixed points we must  retain the Yukawa couplings \cite{Bond:2016dvk}, minimally at the leading non-trivial order which is one loop. Following \cite{Litim:2014uca} we refer to this level of approximation in the gauge-Yukawa sector as next-to-leading order (NLO). In the presence of scalar fields, we also must retain the quartic scalar couplings at their leading non-trivial order. We refer to this approximation of the gauge-Yukawa-scalar sector as NLO${}^\prime$ \cite{Litim:2015iea}, see Tab.~\ref{Tab}. This is  the minimal order in perturbation theory at which a fully interacting fixed point can be determined in all couplings with canonically vanishing mass dimension. 

In general, the RG flow for the gauge and Yukawa couplings at NLO${}^\prime$ is strictly independent of the scalar couplings owing to the fact that scalar loops only arise starting from the two loop order in the Yukawa sector, and at three (four) loop order in the gauge sector, if the scalars are charged (uncharged). Furthermore, the scalar sector at NLO${}^\prime$ depends on the Yukawa couplings, but not on the gauge couplings owing to the fact that the scalars are uncharged. Consequently,  we observe a partial decoupling of the gauge-Yukawa sector 
$(\alpha_1,\alpha_2,\alpha_Y,\alpha_y)$  and the scalar sector $(\alpha_U,\alpha_V,\alpha_u,\alpha_v,\alpha_{w})$. This structure will be exploited systematically below to identify all interacting fixed points.

We begin with the gauge-Yukawa sector where we find the coupled beta functions \eq{12Yy} which are characterised by ten  loop coefficients $C_i, D_i, E_i, F_i$ and $G_i$ $(i=1,2)$, together with the coefficients $B_i$ given in \eq{Bi} or, equivalently, the perturbative control parameters \eq{eps12}. The one-loop coefficients arise in the Yukawa sector and take the values
\beq \label{EF}
\begin{array}{lcl}
E_1 = 13 + 2\left(\epsilon_1 - R\right)\,,
\quad \quad &&
F_1=6\,,
\\[1ex]
\displaystyle
E_2 = 13 + 2\left(\epsilon_2 - \frac{1}{R}\right)\,,\quad\quad &&
F_2=6\,.
\end{array}
\eeq
At the two-loop level we have six coefficients related to the gauge, Yukawa, and mixing contribution, which are found to be
\beq \label{CDG}
\begin{array}{lcl}
&&
C_1 = 
\displaystyle
25 + \frac{26}{3}\epsilon_1\,,\quad
D_1 = 2\left(\epsilon_1-R+ \frac{11}{2}\right)^2\,,\quad
G_1=2 R\\[2ex]
&&
\displaystyle
C_2 = 25 + \frac{26}{3}\epsilon_2\,,\quad
D_2 = 2\left(\epsilon_2- \frac{1}{R}+ \frac{11}{2}\right)^2\,,\quad
G_2=\frac{2}{R}
\end{array}
\eeq
A few comments are in order.
Firstly, the loop coefficients $D_i, E_i, F_i, G_i>0$ as they must for any quantum field theory.
Additionally we confirm that $C_i>0$  \cite{Bond:2016dvk}, provided the parameters $\eps_i$ are in the perturbative regime \eq{eps}. 
Secondly, provided that $R=0$ in the expressions for $\eps_1, E_1$ and $G_1$, 
and $1/R=0$ in those for 
$\eps_2, E_2$ and $G_2$,
the system \eq{12Yy} falls back onto a direct product of simple gauge-Yukawa theories, each of the type discussed in \cite{Litim:2014uca}. Notice that this limit cannot be achieved parametrically in $R$. The reason for this is the presence of $N_\psi$ fermions which are charged under both gauge groups. They contribute with reciprocal multiplicity $R\leftrightarrow 1/R$ to the Yukawa-induced loop terms $D_i$ and  $E_i$ as well as to the mixing terms $G_i$. Exact decoupling of the gauge sectors then becomes visible only in the parametric limit where $N_\psi\to 0$ whereby all terms involving $R$ or $1/R$ drop out.
Finally, we note that the exchange of gauge groups ${\cal G}_1\leftrightarrow {\cal G}_2$ corresponds to $R\leftrightarrow 1/R$ and $S\leftrightarrow T$, implying $\eps_1\leftrightarrow \eps_2$ and $P\leftrightarrow 1/P$, respectively. Evidently, at the symmetric point $R=1$ and $\eps_1=\eps_2$ (or $P=1$) we have exact exchange symmetry between gauge group factors. 

Inserting \eq{EF}, \eq{CDG} and \eq{Bi} into the general expression \eq{12Yy}, we  obtain the perturbative RG flow for the gauge-Yukawa system at NLO accuracy
\beq\label{12Yyexplicit}
\begin{array}{rcl}
\beta_1&=&
\displaystyle
 \043\eps_1\, \alpha_1^2+\left(25 + \frac{26}{3}\epsilon_1\right)\alpha_1^3 -2\left(\epsilon_1-R+ \frac{11}{2}\right)^2\alpha_1^2\,\al{Y}+2R\,\al1^2\,\al2\,,\\[1.5ex]
\beta_2&=&
\displaystyle
  \043\eps_2\, \al2^2+\left(25 + \frac{26}{3}\epsilon_2\right)\al2^3 -2\left(\epsilon_2- \frac{1}{R}+ \frac{11}{2}\right)^2\al2^2\,\al y +\02R\,\al2^2\,\al1\,,\\[2.5ex]
\beta_Y&=&
\displaystyle
 \left[13 + 2\left(\epsilon_1 - R\right)\left]\al{Y}^2\right.\right.-6\,\al{Y}\,\al 1\,,\\[2ex]
\beta_y&=&
\displaystyle
 \left[13 + 2\left(\epsilon_2 - \s0{1}{R}\right)\right]\al{y}^2-6\,\al{y}\,\al 2\,.
\end{array}
\eeq
We observe that the running of Yukawa couplings at one loop is determined by the fermion mass anomalous dimension \eq{gammaF},
\beq\label{YyQq}
\begin{array}{rcl}
\beta_Y&=&2\,\gamma_{M_Q}\,\alpha_Y\,,\\
\beta_y&=&2\,\gamma_{M_q}\,\alpha_y\,.
\end{array}
\eeq
The result for the mass anomalous dimensions \eq{gammaF} can also be derived diagrammatically from the flow of the Yukawa vertices \eq{12Yyexplicit}, thus  offering an independent confirmation for the link \eq{YyQq}.

Next, we turn to the scalar sector and the running of quartic couplings to leading order in perturbation theory, which is one loop. 
At NLO${}^\prime$ accuracy, we have \eq{12Yyexplicit} together with the beta functions for the quartic scalar couplings which are found to be
\beq\label{HVhv}
\begin{array}{rcl}
\beta_U&=&-[11+2(\eps_1-R)]\alpha_Y^2+4\alpha_U(\alpha_Y+2\alpha_U)\,,\\[1ex]
\beta_V&=&\ \ 12\alpha_U^2+4\alpha_V(\alpha_V+4 \alpha_U +\alpha_Y)+\alpha_w^2\,,\\[1ex]
\beta_u&=&-[11+2(\eps_2-\s01R)]\alpha_y^2+4\alpha_u(\alpha_y+2\alpha_u)\,,\\[1ex]
\beta_v&=&\ \ 12\alpha_u^2+4\alpha_v(\alpha_v+4 \alpha_u +\alpha_y)+\alpha_w^2\,,\\[1ex]
\beta_w&=&\ \ \alpha_w\left[8(\alpha_U+\alpha_u)+4(\alpha_V+\alpha_v)
+2(\alpha_Y+\alpha_y)\right]\,.\end{array}
\eeq
Their structure is worth a few remarks: Firstly,  in the Veneziano limit, $\beta_w$ contains no term quadratic in the coupling $\alpha_w$ as the coefficient is of the order ${\cal O}(N^{-1}_{\rm F} N^{-1}_{\rm f})$ and suppressed by inverse powers in flavour multiplicities.  Secondly, we notice that $\beta_w$ comes out proportional to $\alpha_w$. 
Consequently, $\alpha_w$ is a technically natural coupling according to the rationale of \cite{'tHooft:1979bh}, unlike all the other quartic interactions, implying that 
\beq\label{wFP}
\alpha_w^*=0
\eeq
constitutes an exact fixed point of the theory. Comparison with \eq{gammam} shows that the proportionality factor is the sum of the scalar mass anomalous dimensions, $\beta_w=\alpha_w(\gamma_M+\gamma_m)$.   The quartic coupling $\alpha_w$ would be promoted to a free parameter characterising a line of fixed points with exactly marginal scaling provided that its  beta function vanishes identically at one loop. This would require the vanishing of the sum of scalar anomalous mass dimensions at the fixed point,
\beq
\gamma^*_M+\gamma^*_m=0\,.
\eeq
Below, however, we will establish that such scenarios are incompatible with vacuum stability (see Sect.~\ref{stab}). 
Moreover, at interacting fixed points we invariably find that 
\beq\label{gammaPos}
\gamma^*_M+\gamma^*_m>0
\eeq
as a consequence of vacuum stability. This implies that  
$\alpha_w$ 
constitutes an infrared free coupling at any interacting fixed point with a stable ground state.
For the purpose of the present study, we therefore limit ourselves to fixed points with \eq{wFP}.
We then observe that the running of the remaining scalar couplings is solely fuelled by the Yukawa couplings. Furthermore, the scalar subsectors associated to  the different gauge groups are disentangled in our approximation.\footnote{The degeneracy is lifted as soon as the quartic coupling $\alpha_w\neq 0$, see \eq{L}, \eq{couplings}.} Interestingly, the beta functions for  $(\alpha_U,\alpha_V)$ and  $(\alpha_u,\alpha_v)$ are related by  the substitution $R\leftrightarrow 1/R$ and $\eps_1\leftrightarrow \eps_2$. Moreover, the double trace scalar couplings do not couple back into any of the other couplings and their fixed points are entirely dictated by the corresponding single trace scalar and the Yukawa coupling \cite{Litim:2014uca}. This structure allows for a straightforward systematic analysis of all weakly coupled fixed points of the theory to which we turn next.

\section{\bf Interacting fixed points}\label{FPs}
In this section, we present our results for exact  fixed points in the Veneziano limit, corresponding to interacting conformal field theories, and the universal scaling exponents in their vicinity. 

\subsection{Parameter space}\label{AppX}

\begin{figure}[t]
\begin{center}
\includegraphics[scale=.25]{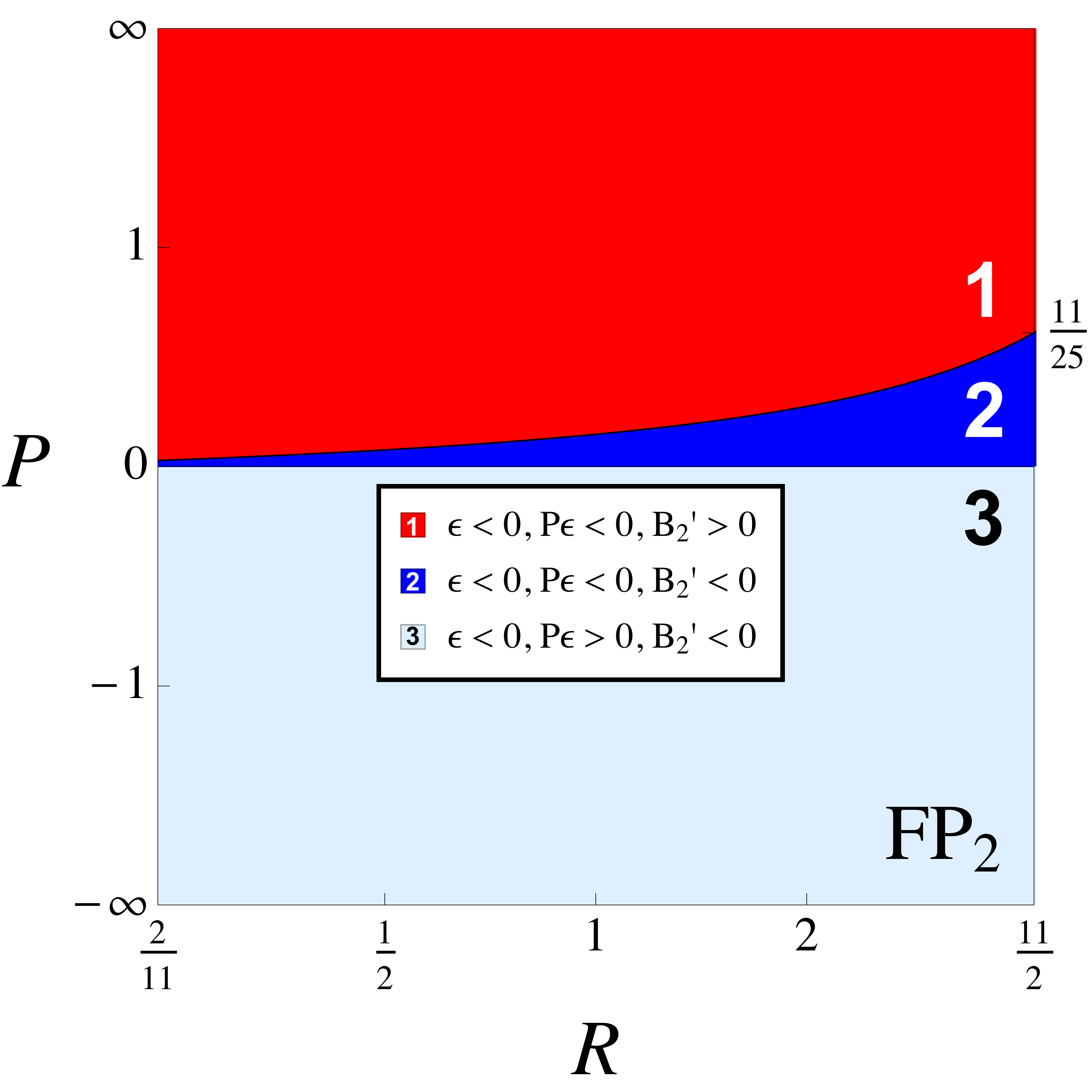}
\\
\vskip.3cm
\hskip-.2cm
\includegraphics[scale=.25]{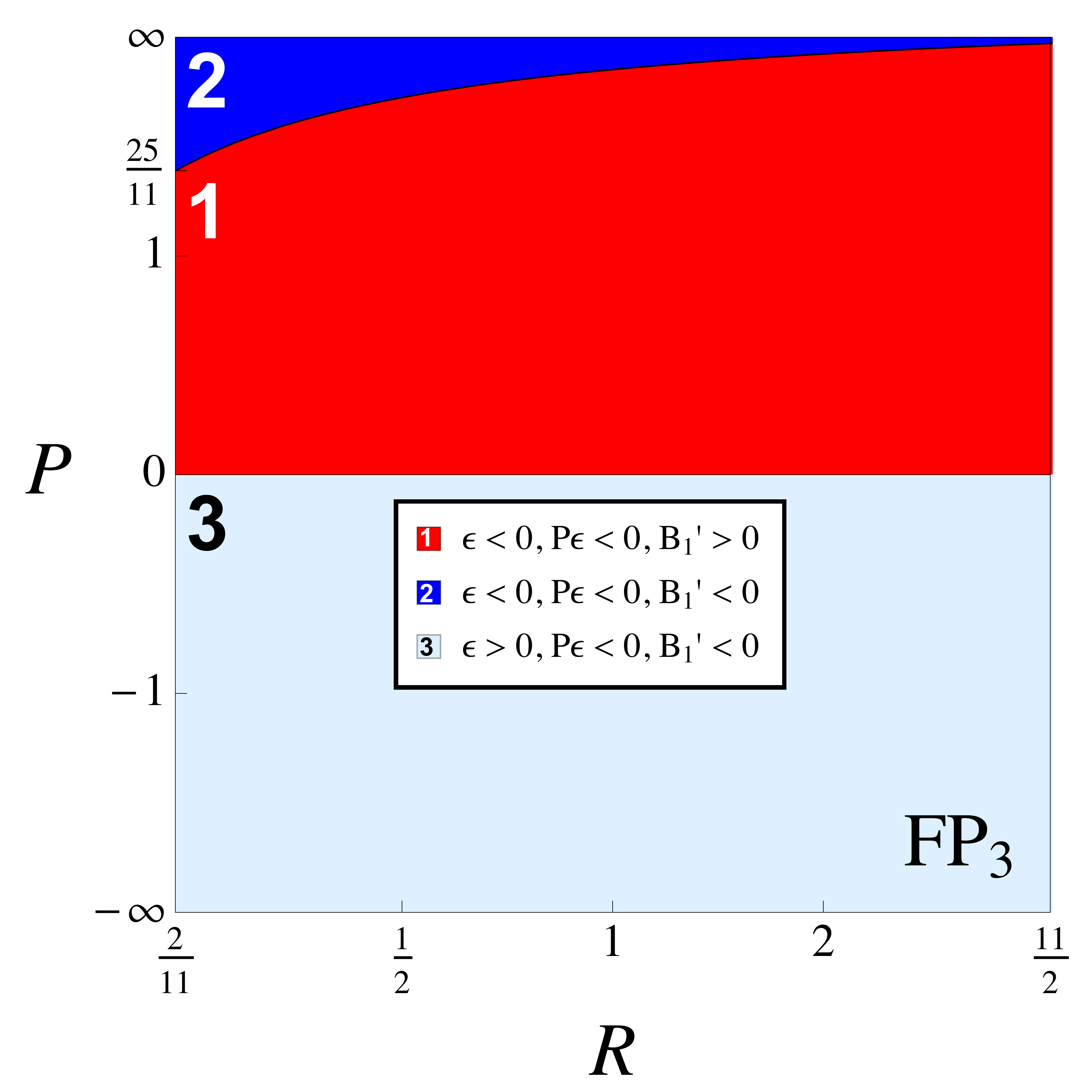}
\caption{The phase space of parameters \eq{PRepsN} for the partially interacting fixed points FP${}_2$ (upper panel) and FP${}_3$ (lower panel) where one of the two gauge sectors remains interacting and all Yukawa couplings vanish. The inset indicates the different parameter regions and conditions for existence, including whether the non-interacting gauge sector is asymptotically free $(B'>0)$ or infrared free $(B'<0)$, see Tab.~\ref{tBprime},\ref{tPara15}. 
}\label{pFP23} 
\end{center}
\end{figure}

In Tab.~\ref{tFPeps} we state our results for the gauge and Yukawa couplings to leading order in \eq{eps} at all fixed points,  following the nomenclature of Tab.~\ref{tFPs}.  Expressions are given as functions of the parameters $(P,R,\eps)$, 
\beq\label{PRepsN}
\begin{array}{rcl}
P&=&
\displaystyle
\frac{1+(N_{\rm f}-\0{11}{2} N_{\rm c})/N_{\rm C}}{1+(N_{\rm F}-\0{11}{2}N_{\rm C})/N_{\rm c}}\\[2.5ex]
R&=&
\displaystyle
\frac{N_{\rm c}}{N_{\rm C}}\\[2.5ex]
{\rm sgn}\,\eps&=&{\rm sgn}\left(N_{\rm c}+N_{\rm F}-\frac{11}{2}N_{\rm C}\right)\,,\\[.5ex]
\end{array}
\eeq
which only depend on the matter and gauge field multiplicities \eq{Ns},  and $N_\psi=1$. Results for general $N_\psi$ are given in App.~\ref{AppN}.  We also observe \eq{R}, unless stated otherwise. 
 Constraints on the parameters $(R,P,\eps)$ 
and other information is summarised   Figs.~\ref{pFP23},~\ref{pFP45},~\ref{pFP6},~\ref{pFP78} and~\ref{pFP9}  and in Tabs.~\ref{tBprime}~\ref{tPara15},~\ref{tPara69} for the various fixed points.   
Below, certain characteristic values $\frac{2}{11}<R_1<R_2<R_3<R_4<\frac{11}2$ for the parameter $R$ are of particular interest, namely
\beq\label{R1234}
\begin{array}{rcl}
R_1 &= &
\displaystyle
\frac{343 - 3\sqrt{9361}}{100} \approx 0.53\,,	\\	[1.2ex]	
				R_2 &=& \displaystyle
\frac{43 - 9\sqrt{5}}{38} \approx 0.60\,,\\[1.2ex]	
	R_3 &=&\displaystyle
\frac{43 + 9\sqrt{5}}{38} \approx 1.66\,,	\\[1.2ex]	
 R_4 &=&\displaystyle
\frac{343 + 3\sqrt{9361}}{100}\approx 1.90\,.
\end{array}
\eeq
Their origin is explained in App.~\ref{AppX}. After these preliminaries we are in a position to analyse the fixed point spectra.

 \begin{table*}[b]
\begin{center}
\begin{tabular}{cllcc}
\toprule
\rowcolor{LightBlue}
&
\cellcolor{LightRed}  
&\cellcolor{LightYellow}
&\cellcolor{LightGreen} 
\\[-1mm]
\cellcolor{LightBlue}
\multirow{-2}{*}{$\bm{\#}$}&
\multirow{-2}{*}{\cellcolor{LightRed}  \quad\quad\quad\quad\quad\bf gauge couplings}
&\multirow{-2}{*}{\cellcolor{LightYellow}  \quad\quad\bf Yukawa couplings}
&
\multirow{-2}{*}{\cellcolor{LightGreen} \bf type} 
\\

\midrule

\rowcolor{LightGray}
\bf \ FP${}_{\bf 1}\ $
&$\alpha^*_1=0\,,\quad\quad\quad\alpha^*_2=0\,,$
&$\alpha^*_Y=0\,,\quad\quad\quad\alpha^*_y=0\,,$
& \bf  G $\bm \cdot$  G
\\[1ex]
&&& \\[-2.5ex]
\bf \ FP${}_{\bf 2}\ $
&
$\alpha^*_1=-\s0{4}{75}R\eps\,,\ \, \alpha^*_2=0\,,$
&
$\alpha^*_Y=0\,,\quad\quad\quad \alpha^*_y=0\,,$&
 \bf  BZ $\bm \cdot$  G
 \\[1ex]
\rowcolor{LightGray}
&&& \\[-2.5ex]
\rowcolor{LightGray}
\bf FP${}_{\bf 3}$
&$\alpha^*_1=0\,,\quad\quad\quad\alpha^*_2=-\s0{4}{75}\frac{P\eps}{R}\,,$
&
$\alpha^*_Y=0\,,\quad\quad\quad \alpha^*_y=0\,,$
& \bf  G $\bm \cdot$  BZ
\\[1ex]
&&& \\[-2.5ex]
\bf FP${}_{\bf 4}$&
$\alpha^*_1=\s0{2}3 \s0{(13-2R)R\eps}{(2R-1)(3R-19)}\,,\quad\quad\ \ \alpha^*_2=0\,,$
&
$\alpha^*_Y=\s0{4R\eps}{(2R-1)(3R-19)}\,,\quad  \alpha^*_y=0\,,$
& \bf  GY $\bm \cdot$  G
\\[1ex]
\rowcolor{LightGray}
&&& \\[-2.5ex]
\rowcolor{LightGray}
\bf FP${}_{\bf 5}$
&$ \alpha^*_1=0\,,\quad  \alpha^*_2=\0{2}3 \frac{(13-2/R)}{(2/R-1)(3/R-19)} \frac{P\eps}{R}\,,$
&
$\alpha^*_Y=0\,,\  \alpha^*_y=\frac{4\,P\eps/R}{(2/R-1)(3/R-19)}\,,$
&\bf  G $\bm \cdot$  GY
\\[1ex]
&&& \\[-2.5ex]
\bf FP${}_{\bf 6}$&
$\alpha^*_1= \0{-4(25 - 2P/R)}{1863} R\eps\,,\quad 
\alpha^*_2=\0{-4(25 - 2R/P)}{1863}\frac{P\eps}{R}$
&$\alpha^*_Y=0\,,\quad\quad\quad \alpha^*_y=0\,,$
&\bf  BZ $\bm \cdot$  BZ
\\[1ex]
\rowcolor{LightGray}&&& \\[-2.5ex]
\rowcolor{LightGray}
&
$\alpha^*_1=\0{2}9\frac{(13-2R)(25 - 2P/R)}{50R^2-343R+167} R\eps$
&
$\alpha^*_Y=\0{4}3 \frac{25 - 2P/R}{50R^2-343R+167} R\eps$
&
\\[1ex]
\rowcolor{LightGray}
\multirow{-2}{*}{\bf FP${}_{\bf 7}$}
&
$\alpha^*_2= \0{4}9\frac{(13-2R)R/P + (2R-1)(19-3R)}{50R^2 - 343R + 167} \frac{P\eps}{R}$
&$\alpha^*_y=0\quad\quad\quad\quad\quad\quad\quad\ \ $&
\multirow{-2}{*}{\bf  GY $\bm \cdot$  BZ}
\\[1ex]
&&& \\[-2.5ex]&
$\alpha^*_1=\0{4}9\frac{(13-2/R)P/R + (2/R-1)(19-3/R)}{50/R^2 - 343/R + 167} R\eps$
&$\alpha^*_Y=0\quad\quad\quad\quad\quad\quad\quad\ \ $
\\[1ex]
\multirow{-2}{*}{\bf FP${}_{\bf 8}$}
&
$\alpha^*_2=\0{2}9\frac{(13-2/R)(25 - 2R/P)}{50/R^2-343/R+167} \frac{P\eps}{R}
\quad\quad\ \ $
&
$\alpha^*_y=\0{4}3 \frac{25 - 2R/P}{50/R^2-343/R+167} \frac{P\eps}{R}$
&
\multirow{-2}{*}{\bf  BZ $\bm \cdot$  GY}
\\[1ex]
\rowcolor{LightGray}&&& \\[-2.5ex]
\rowcolor{LightGray}
&
$\alpha^*_1=\0{2}9\0{(13-2R)[(13-2/R)P/R+(\s02R-1)(3/R-19)]}{(19R^2 - 43R + 19)(2/R^2 -13/R + 2)}R\eps\  $
&
$\alpha^*_Y=\0{6\,\alpha_1^*}{13-2R}$
&\\[1ex]
\rowcolor{LightGray}
\multirow{-2}{*}{\bf FP${}_{\bf 9}$}
&
$\alpha^*_2=
\0{2}9\0{(13-2/R)[(13-2R)R/P+(2R-1)(3R-19)]}{(19/R^2 - 43/R + 19)(2R^2 -13R + 2)} \frac{P\eps}{R}$
&
$\alpha^*_y=
\0{6\,\alpha_2^*}{13-2/R}$
&
\multirow{-2}{*}{\bf   GY $\bm \cdot$  GY\ }\\[1ex]
\bottomrule
\end{tabular}
\caption{Gauge and Yukawa couplings at  interacting fixed points following Tab.~\ref{tFPs} to the leading order in $\eps$ and  in terms of $(R,P,\eps)$. Valid domains for $(\eps,P,R)$ in \eq{PRepsN} are detailed in Tab.~\ref{tPara15},~\ref{tPara69}.\\ }
 \label{tFPeps}
\end{center}
\end{table*}

\subsection{Partially and fully interacting fixed points}
\label{pifp}
Gauge theories with \eq{12Yyexplicit}, \eq{HVhv} can have two types of interacting fixed points: partially interacting ones where one gauge coupling takes the Gaussian fixed point  (\fp2,\fp3,\fp4,\fp5), and fully interacting ones where both gauge sectors remain interacting (\fp6,\fp7,\fp8,\fp9), see Tab.~\ref{tFPs}. At partially interacting fixed points, one gauge sector decouples and the semi-simple theory with \eq{12Yyexplicit}, \eq{HVhv} effectively reduces to a simple gauge theory. Simple gauge theories have three possible types of perturbative fixed points: the Gaussian (G), the Banks-Zaks (BZ), and gauge-Yukawa (GY) fixed points for each independent linear combination of the Yukawa couplings \cite{Bond:2016dvk}.  In our setting, at \fp2 and \fp4 we have that  $\alpha_2^*\equiv 0$, and the theory reduces to a simple gauge theory with
  \beq\label{Simple}
\begin{array}{rcl}
\beta_1&=&
\displaystyle
 \ \ \043\eps_1\, \alpha_1^2+\left(25 + \frac{26}{3}\epsilon_1\right)\alpha_1^3 -2\left(\epsilon_1-R+ \frac{11}{2}\right)^2\alpha_1^2\,\al{Y}\\[2ex]
\beta_Y&=&\displaystyle
 \ \ \left[13 + 2\left(\epsilon_1 - R\right)\left]\al{Y}^2\right.\right.-6\,\al{Y}\,\al 1\\[1.5ex]
\displaystyle
\beta_U&=&-[11+2(\eps_1-R)]\alpha_Y^2+4\alpha_U(\alpha_Y+2\alpha_U)\,,\\[1ex]
\beta_V&=&\ \ 12\alpha_U^2+4\alpha_V(\alpha_V+4 \alpha_U +\alpha_Y)\,,
\end{array}
\eeq
at NLO${}^\prime$ accuracy, where the parameter $R$ with
\beq\label{Rrelaxed}
0\le R=\frac{N_1}{N_{\rm C}}<\frac{11}{2}
\eeq
measures the number of Yukawa-less Dirac fermions $N_1$ in the fundamental representation   in units of $N_{\rm C}$. Notice that $N_1$ is related to $N_\psi$ via \eq{Relax} in the theories \eq{L}. 
On the other hand, $N_1$ can be viewed as an independent parameter (counting the Yukawa-less fermions in the fundamental representation of the gauge group) if one were to switch off the semi-simple character of the theory.
For $R=0$ the theory \eq{Simple} reduces to the  one investigated in \cite{Litim:2014uca}. The lower bound on $R$ \eq{R} is relaxed in \eq{Rrelaxed}, because the requirement of perturbativity for an interacting  fixed point in the other gauge sector has become redundant. 
We observe the $R$-independent Banks-Zaks (BZ) fixed point $\alpha_1^*=\frac{4}{75}\eps_1$ which is, invariably, IR.
To leading order in $\eps_1$ we also find a gauge-Yukawa (GY) fixed point 
\beq\label{GYsimple}
\begin{array}{rcl}
\alpha_g^*&=&
\di
\frac{26-4R}{57-9 R}\,\frac{\eps_1}{1-2 R}\\[2ex]
\alpha_Y^*&=&
\di
\frac{4}{19-3 R}\frac{\eps_1}{1-2 R}\\[2ex]
\alpha_U^*&=&
\di
\frac{\sqrt{23-4R}-1}{19-3 R}\,\frac{\eps_1}{1-2 R}\\[2ex]
\alpha_V^*&=&
\di
\frac{-2\sqrt{23-4R}+
\sqrt{20-4R+6\sqrt{23-4R}}}{19-3 R}\frac{\eps_1}{1-2 R}\,.
\end{array}
\eeq
For $\eps_1>0$, the GY fixed point is UV and physical as long as $0\le R<\frac12$. It can be interpreted as a ``deformation'' of the UV fixed point analysed in  \cite{Litim:2014uca} owing to the presence of charged Yukawa-less fermions. Once $R>\frac12$, however, the fixed point is physical iff $\eps_1<0$ where it becomes an IR fixed point. This new regime is entirely due to the Yukawa-less fermions  and does not arise in the model of  \cite{Litim:2014uca}. This pattern can also be read off from the scaling exponents, which, at the gauge Yukawa fixed point and to the leading non-trivial order in $\eps_1$,  are given by
 \beq\label{evGYsimple}
\begin{array}{rcl}
\vartheta_g&=&
\di
-\frac{104}{171}\,
\frac{1-\s02{13}R}{1-\s03{19}R}\,
\frac{\eps_1^2}{1-2 R}\\[2.5ex]
\vartheta_y&=&
\di
\ \ \frac{4}{19}\,\frac{1}{1-\s03{19}R}\frac{\eps_1}{1-2 R}\\[2.5ex]
\vartheta_u&=&\di
\ \ \frac{16\sqrt{23-4R}}{19-3 R}\,\frac{\eps_1}{1-2 R}\\[2.5ex]
\vartheta_v&=&\di
\ \ \frac{8\sqrt{20+6\sqrt{23-4R}-4R}}{19-3 R}\,\frac{\eps_1}{1-2 R}\,.
\end{array}
\eeq
 For $\eps_1>0$ and $R<\s012$ asymptotic safety is guaranteed with
$\vartheta_g<0<\vartheta_y, \vartheta_u, \vartheta_v\,,$
showing that the UV fixed point has one relevant direction.  The scaling exponents reduce to those in \cite{Litim:2014uca} for $R=0$. Conversely,  for $\eps_1<0$ and $R>\s012$ the theory is asymptotically free and the interacting fixed point is fully IR attractive with 
$0<\vartheta_g,\vartheta_y, \vartheta_u, \vartheta_v\,.$
Results straightforwardly translate to the partially interacting fixed points \fp3 and \fp5 where $\alpha_1^*\equiv 0$. The  explicit $\beta$-functions in the other gauge sector are found from \eq{Simple} -- \eq{evGYsimple} via the replacements $\eps_1\leftrightarrow\eps_2$ and $R\leftrightarrow 1/R$,  leading to
\beq\label{Simple2}
\begin{array}{rcl}
\beta_2&=&
\displaystyle
\ \ \; \043\eps_2\, \alpha_2^2+\left(25 + \frac{26}{3}\epsilon_2\right)\alpha_2^3 -2\left(\epsilon_2-\frac1R+ \frac{11}{2}\right)^2\alpha_2^2\,\al{y}\\[2ex]
\beta_y&=&\displaystyle
\ \ \;  \left[13 + 2\left(\epsilon_2 - \frac1R\right)\right]\al{y}^2-6\,\al{y}\,\al 2\\[2ex]
\displaystyle
\beta_u&=&
\displaystyle
-\left[11+2(\eps_2-\frac1R)\right]\alpha_y^2+4\alpha_u(\alpha_y+2\alpha_u)\,,\\[2.5ex]
\beta_v&=&\ \ \; 12\alpha_u^2+4\alpha_v(\alpha_v+4 \alpha_u +\alpha_y)\,.
\end{array}
\eeq
Evidently, the coordinates of the fully interacting gauge-Yukawa fixed point and the corresponding universal scaling exponents of \eq{Simple2} are given by \eq{GYsimple}, \eq{evGYsimple} after obvious replacements.
Moreover, in \eq{Simple2} the parameter $R$ with
\beq\label{Rrelaxed2}
0\le \frac1R=\frac{N_2}{N_{\rm c}}<\frac{11}{2}
\eeq
measures the number of Yukawa-less Dirac fermions $N_2$ in the fundamental representation   in units of $N_{\rm c}$, see \eq{Relax}. 
The only direct communication between the different gauge sectors in \eq{L} is through the off-diagonal two-loop gauge contributions $G_i$. Were it not for the fermions $\psi$ which are charged under both gauge groups, the theory \eq{L} with  \eq{12Yyexplicit}, \eq{HVhv} would be the ``direct product'' of the simple model \eq{Simple}, \eq{Rrelaxed} with its counterpart \eq{Simple2}, \eq{Rrelaxed2}. In this limit we will find nine ``direct product'' fixed points with scaling exponents from each pairing of the possibilities (G, BZ, GY) in each sector. 

Below,  we contrast findings for the full semi-simple setting \eq{12Yyexplicit}, \eq{HVhv} with those from the ``direct product'' limit in order to pin-point effects which uniquely arise from the semi-simple character of the theories \eq{L}.

At any of the partially interacting fixed points, the semi-simple character of the theory becomes visible in the non-interacting sector. In fact, 
contributions from the $\psi$ fermions modify the effective one-loop coefficient $B_i\to B_i'$ according to
\beq\label{Bprime}
\begin{array}{l}
\alpha_1^*=0:\quad B_1\to B_1'=B_1+G_1\,\alpha^*_2\\[1ex]
\alpha_2^*=0:\quad B_2\to B_2'=B_2+G_2\,\alpha^*_1\,.
\end{array}
\eeq
No such effects can materialize in a ``direct product'' limit. Moreover, these contributions always arise with a positive coefficient ($B'>B)$ and are absent if $N_\psi=0$ (where $G_i=0)$. For $N_\psi\neq 0$, asymptotic freedom can thereby be changed into infrared freedom, but not the other way around. 
This result is due to the fact that the Yukawa couplings are tied to individual gauge groups, and so by this structure we cannot have any Yukawa contributions to $B'$. 
In principle, the opposite effect can equally arise: it would require Yukawa couplings which contribute to both gauge coupling $\beta$-functions, and would therefore have to involve at least one field which is charged under both gauge groups \cite{Bond:2016dvk}.  
Tab.~\ref{tBprime} shows the effective one loop coefficients at partially interacting fixed points as a function of field multiplicities.

\begin{table}[t]
\begin{center}
\begin{tabular}{cl}
  \rowcolor{LightGreen}
\toprule
\rowcolor{LightGreen}
\cellcolor{LightBlue}
$\bm{\#}\ $
& ${}$
\quad\quad\quad$\bm{B'}$ \bf coefficient	
\\[.4mm]
\midrule
\rowcolor{LightGray}&\\[-4mm]
\rowcolor{LightGray}
\FP2 	& $\ \ \di B_2' = -\frac{4}{3}\left(1 - \frac{2}{25}R/P\right)\0{P\epsilon}R\ \ $
\\[.4mm]

\rowcolor{white}
&\\[-3mm]
\rowcolor{white}
\FP3 &  $\ \ B_1' =  -\frac{4}{3}\left(1 - \frac{2}{25}P/R\right)R\epsilon$
\\[.4mm]

\rowcolor{LightGray}
&\\[-3mm]
\rowcolor{LightGray}
\FP4 	& $\ \ \di B_2' = -\frac{4}{3}\left(1 - X(R)/{P}\right)\0{P\epsilon}R$
\\[.4mm]

\rowcolor{white}
&\\[-3mm]
\rowcolor{white}
\FP5 &  $\ \ B_1' = -\frac{4}{3}\left(1 - P/{\xalt(R)}\right)R \epsilon$
\\[.4mm]

\bottomrule
\end{tabular}
\end{center}
 \caption{Shown are the effective one-loop coefficients $B'$ for the non-interacting gauge coupling  at \fp2, \fp3, \fp4 and \fp5, and their dependence on model parameters. $B'>0$ corresponds to asymptotic freedom. Notice that $B'$ changes sign across the boundaries $P=2R/25, 25R/2,X(R)$, and $\xalt(R)$, respectively, with $X$ and $\xalt$ given in \eq{P1234}.}
\label{tBprime}
\end{table}

\begin{table}[b]
\begin{center}
\begin{tabular}{cccccc}
  \rowcolor{LightGreen}
\toprule
\rowcolor{LightGreen}
\cellcolor{LightBlue}
&\multicolumn{3}{c}{\cellcolor{LightGreen}\bf \ \ \ parameter range\ \ \ }
&
&
\\
\rowcolor{LightGreen}
\cellcolor{LightBlue}
\multirow{-2}{*}{$\bm{\#}\ $}	
	& \ sign$\ \epsilon\ $	
	& $R$	
	& $P$	
	&\multirow{-2}{*}{\cellcolor{LightGreen}\bf eigenvalue spectrum}
	&\multirow{-2}{*}{\cellcolor{LightGreen}\bf info}
\\[.4mm]
\rowcolor{white}
\midrule
\rowcolor{white}
\FP1 	
	& $\pm$  
	& $\ \left(\frac{2}{11},\frac{11}{2}\right)\ $ 
	&$\ \ (-\infty,+\infty)\ \ $ 
	& \eq{4R}, \eq{2R}, or \eq{0R}
	&\cellcolor{LightRed}  Gaussian
\\[.4mm]
\midrule
\rowcolor{LightGray} &&&&& \cellcolor{LightRed} Fig.~\ref{pFP23} (upper panel) \\
\rowcolor{LightGray}
  	&  $-$	
 	&$\left(\02{11},\frac{11}{2}\right)$
 	&$(\frac{2}{25}R,+\infty)$
 	&  	$\vartheta_{1,2,3}\le0<\vartheta_{4}$
 	& region 1
\\[.4mm]
\rowcolor{LightGray} 
	&  $-$	
	&$\left(\frac{2}{11},\frac{11}{2}\right)$
	&$(0,\frac{2}{25}R)$
	& $\quad\quad\  \vartheta_{1}<0\le\vartheta_{2,3,4}$
	& region 2
\\[.4mm]
\rowcolor{LightGray}
 \multirow{-4}{*}{\cellcolor{LightGray}\FP2}
	&  $-$	
	&$\left(\frac{2}{11},\frac{11}{2}\right)$
	&$(-\infty,0)$
	& $\quad\quad\  \vartheta_{1}<0\le\vartheta_{2,3,4}$
	& region 3
\\[.4mm]

\midrule
\rowcolor{white} &&&&&\cellcolor{LightRed} Fig.~\ref{pFP23} (lower panel)\\
 	&$-$
	&$\left(\frac{2}{11},\frac{11}{2}\right)$	
	& $\left(0,\frac{25}{2}R\right)$  
	&  $\vartheta_{1,2,3}\le0<\vartheta_{4}$
	& region 1
\\[.4mm]
 \rowcolor{white}
 	&$-$
 	&$\left(\frac{2}{11},\frac{11}{2}\right)$	
 	& $\left(\frac{25}{2}R,+\infty\right)$  
 	&   $\quad\quad\  \vartheta_{1}<0\le\vartheta_{2,3,4}$ 
	& region 2
\\[.4mm]
\multirow{-4}{*}{\FP3}
	&$+$	
	&$\left(\frac{2}{11},\frac{11}{2}\right)$	
	& $\left(-\infty,0\right)$  
	&$\quad\quad\  \vartheta_{1}<0\le\vartheta_{2,3,4}$ 
	& region 3
\\[.4mm]

\midrule
\rowcolor{LightGray} &&&&& \cellcolor{LightRed} Fig.~\ref{pFP45} (upper panel)\\

\rowcolor{LightGray}
 	&  $-$ 
 	& $\left(\frac{1}{2},\frac{11}{2}\right)$
 	&$(-\infty,X(R))$
 	& $\ \quad\quad\quad \quad\quad 0\le\vartheta_{1,2,3,4}$
	& region 1 \& 3
\\[.4mm]
\rowcolor{LightGray}
 	&  $-$ 
	& $\left(\frac{1}{2},\frac{11}{2}\right)$
	&$(X(R),+\infty)$
	& $\quad\, \vartheta_{1,2}\le0<\vartheta_{3,4}$
	& region 2
\\[.4mm]
\rowcolor{LightGray}
	&$+$ 
	& $\left(\frac{2}{11},\frac{1}{2}\right)$ 	
	&$(X(R),+\infty)$		 
	& 	  $\quad\quad\  \vartheta_{1}<0\le\vartheta_{2,3,4}$
	& region 4 \& 6
\\[.4mm]
\rowcolor{LightGray}
\multirow{-5}{*}{\cellcolor{LightGray}\FP4}	  
	&$+$ 
	& $\left(\frac{2}{11},\frac{1}{2}\right)$ 	
	&$(-\infty,X(R))$
	& $\vartheta_{1,2,3}\le0<\vartheta_{4}$
	& region 5
\\[.4mm]

\midrule &&&&& \cellcolor{LightRed} Fig.~\ref{pFP45} (lower panel) \\
	&$-$
 	&$\big(\frac{2}{11},2\big)$	
 	& $\big(\xalt(R),+\infty\big)$  
 	& $\quad\quad\  \quad\quad\quad  0\le\vartheta_{1,2,3,4}$
	&region 1
\\[.4mm]
	&$-$
	& $\big(\frac{2}{11},2\big)$  
	&$\big(0,\xalt(R)\big)$
	& $\quad\  \vartheta_{1,2}\le0<\vartheta_{3,4}$
	&region 2
\\[.4mm]
 	&$+$
	&$\left(\frac{2}{11},2\right)$  
	&$\big(-\infty,0\big)$ 
	&$\quad\quad\quad\quad\quad\ \;  0\le\vartheta_{1,2,3,4}$
	&region 3
\\[.4mm]
 	&$-$
 	&$\big(2,\frac{11}{2}\big)$
 	& $\big(-\infty,\xalt(R)\big)$  
 	&$\quad\quad\; \; \vartheta_{1}<0\le\vartheta_{2,3,4}$ 
	& region 4
\\[.4mm]
	 &$-$
	&$\big(2,\frac{11}{2}\big)$	
	& $\big(\xalt(R),0\big)$  
	&$\vartheta_{1,2,3}\le0<\vartheta_{4}$ 
	&region 5
\\[.4mm]
	 \multirow{-7}{*}{\FP5}
	&$+$
	&$\big(2,\frac{11}{2}\big)$	
	& $\big(0,+\infty\big)$  
	&$\quad\quad\  \vartheta_{1}<0\le\vartheta_{2,3,4}$ 
	&region 6
\\[.4mm]\bottomrule
\end{tabular}
\end{center}
 \caption{Parameter regions where the partially interacting fixed points \fp1 -- \fp5 exist, along with regions of relevancy for eigenvalues and effective one-loop terms, where applicable. The boundary functions $X(R)$ and $\xalt(R)$ are given in \eq{P1234}. The coefficient $B'$ for the gauge coupling at the Gaussian fixed point is given in Tab.~\ref{tBprime}.
}\label{tPara15}
\end{table}

\subsection{Gauss with Banks-Zaks} 

Next, we discuss all fixed points one-by-one, and determine the valid parameter regimes $(R,P,\eps)$ for each of them. We recall that $N_\psi=1$ in our models. Whenever appropriate, we also compare results with the ``direct product'' limit, whereby the diagonal contributions from the Yukawa-less $\psi$-fermions are retained but their  off-diagonal contributions to the other gauge sectors suppressed (see Sect.~\ref{pifp}). This comparison allows us to quantify the effect related to the semi-simple nature of the models \eq{L}. 

\begin{figure}[t]
\begin{center}
\vskip-.5cm
\hskip-.7cm
\includegraphics[scale=.275]{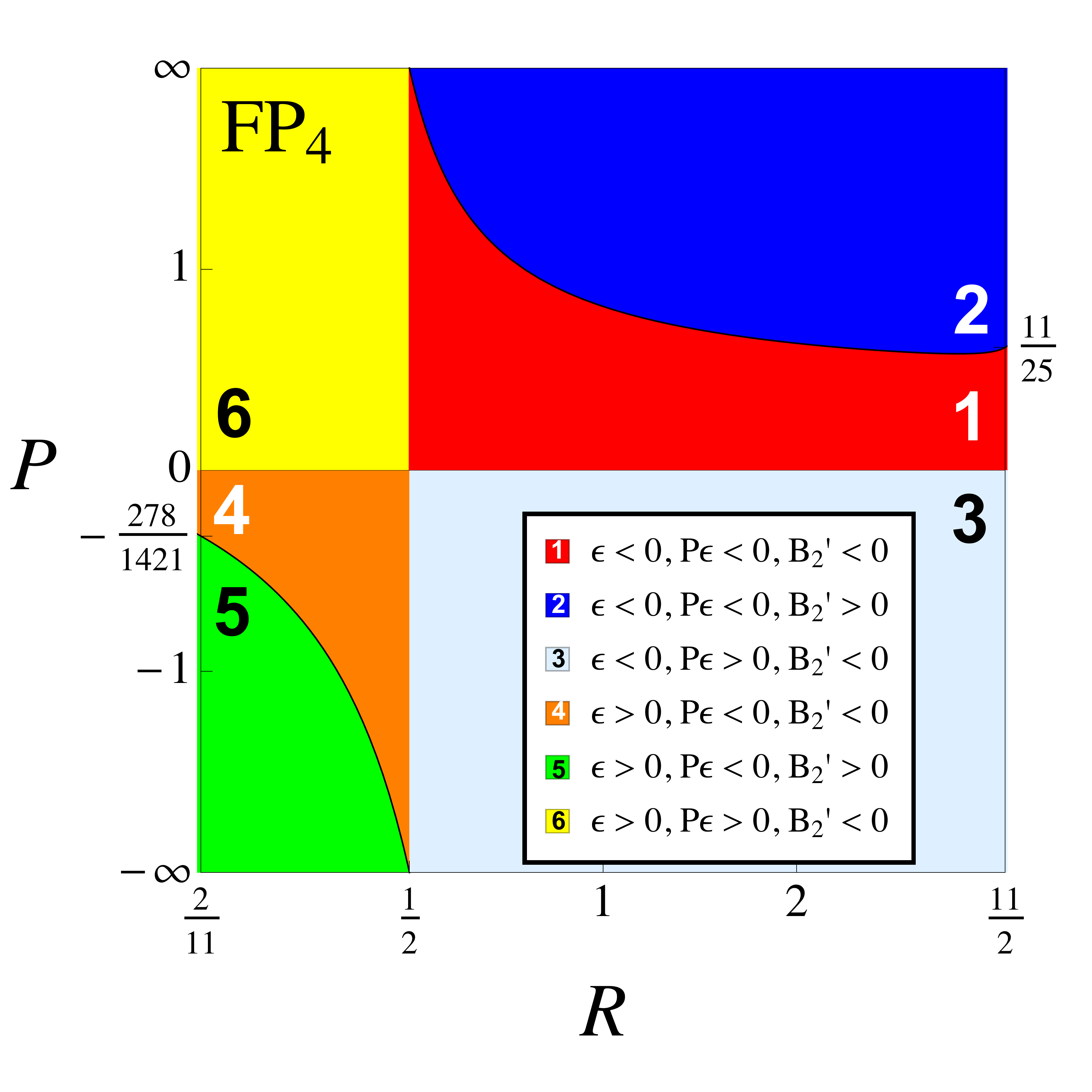}
\\
\vskip-.3cm
\includegraphics[scale=.275]{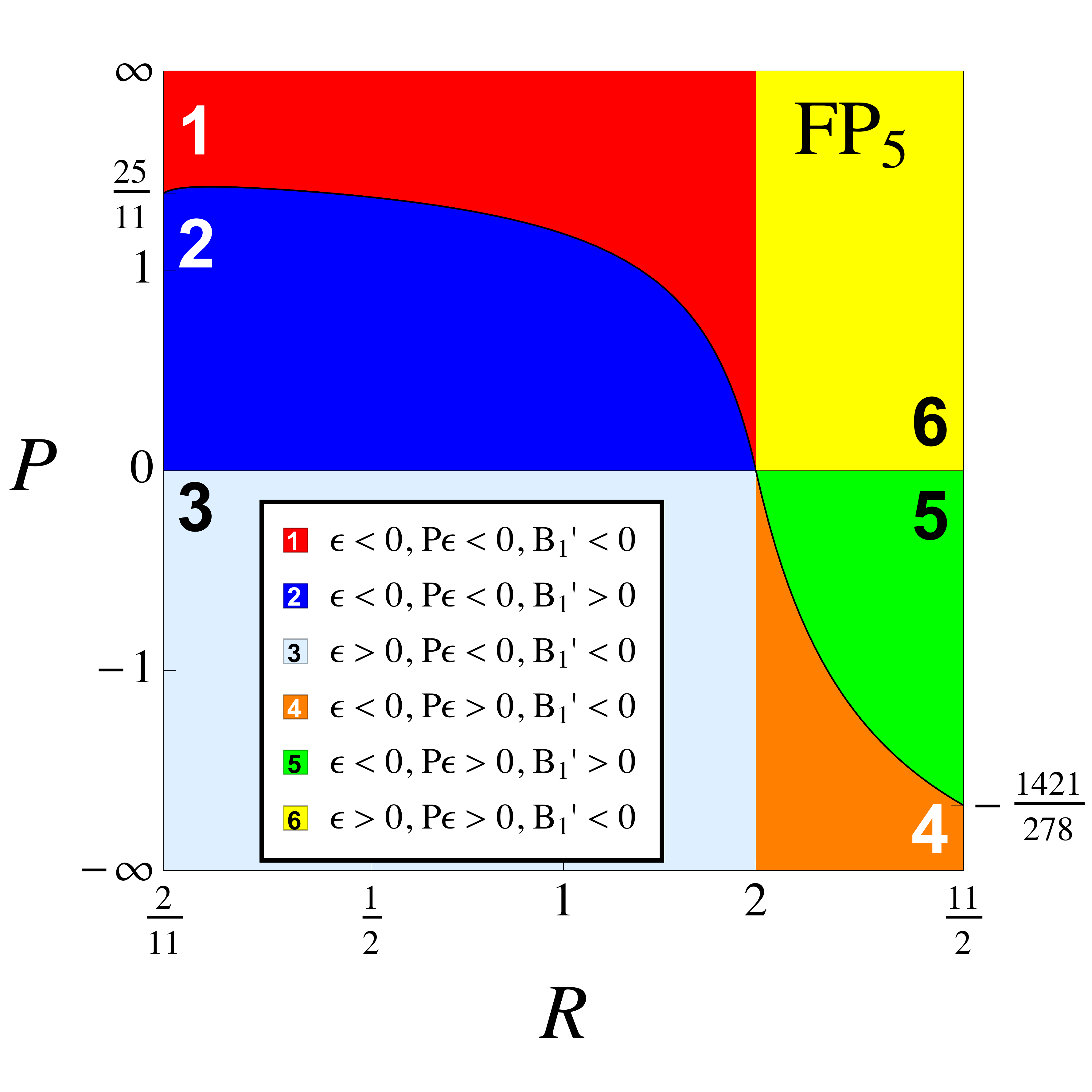}
\caption{The phase space of parameters for the partially interacting fixed points FP${}_4$ and FP${}_5$, where the gauge and Yukawa coupling in one gauge sector take interacting fixed points while those of the other sector remain trivial. The insets indicate the different parameter regions and conditions for existence, 
and whether the non-interacting gauge sector is asymptotically free $(B'>0)$ or infrared free $(B'<0)$, see Tab.~\ref{tBprime},\ref{tPara15}. 
}\label{pFP45} 
\end{center}
\end{figure}

\begin{figure}[t]
\begin{center}
\includegraphics[scale=.95]{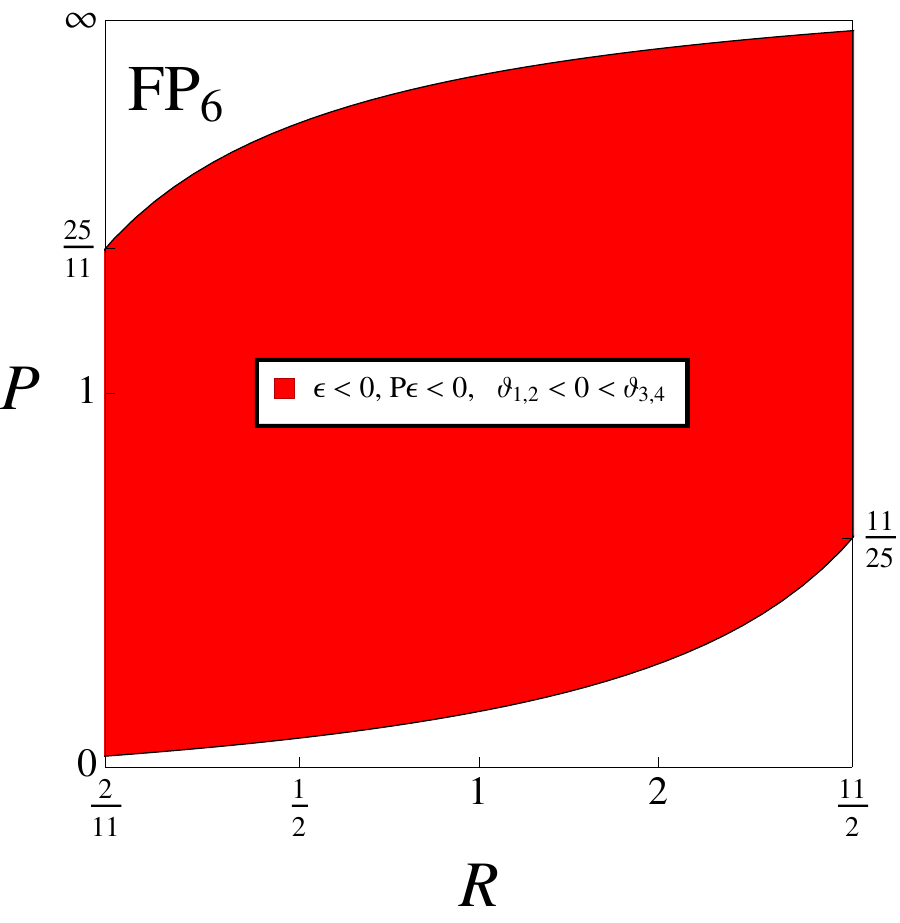}
\caption{The phase space of parameters for  the  interacting  fixed point \fp6 (red) where both gauge sectors take interacting and physical fixed points while all Yukawa couplings vanish.  The eigenvalue spectrum at the fixed point  always displays exactly two relevant eigenvalues of  $O(\epsilon)$ and two irrelevant eigenvalues of order $O(\epsilon^2)$, 
see Tab.~\ref{tPara69}.  Note that this fixed point invariably requires asymptotic freedom for both gauge sectors
 (see main text).
}\label{pFP6} 
\end{center}
\end{figure}

For convenience and better visibility, we  scale the axes in Figs.~\ref{pFP23}~\ref{pFP45},~\ref{pFP6},~\ref{pFP78} and~\ref{pFP9} as
\beq
\label{rescale}
X\to \frac{X}{1+|X|}\quad {\rm where}\quad X=P\ {\rm or}\ R\,, 
\eeq
and within their respective domains of validity $R\in(\s02{11},\s0{11}{2})$ and $P\in (-\infty,\infty)$. The rescaling permits easy display of the entire range of parameters. 

Fig.~\ref{pFP23} shows the results for \fp2 ({BZ$\cdot$G}, upper) and \fp3 ({G$\cdot$BZ}, lower panel), and parameter ranges are given in Tab.~\ref{tPara15}. 
We observe that the Banks-Zaks fixed point always requires an asymptotically free gauge sector. Hence, \fp2 exists for any $R$ as long as $\eps<0$. Provided that $P\eps <0$, the other gauge sector either remain asymptotically free (region 1) or becomes infrared free (region 2). On the other hand, if $P\eps>0$, the  other gauge sector is invariably infrared free. This is a consequence of  \eq{Bprime} which states that the interacting gauge sector can turn asymptotic freedom of the non-interacting gauge sector into infrared freedom (region 2), but not the other way around.  The existence of the parameter region 2 is thus entirely due to the semi-simple character of the theory which cannot arise from a ``direct product''. 

The Banks-Zaks fixed point is invariably attractive in the gauge coupling, and repulsive in the Yukawa coupling. The eigenvalue spectrum in the gauge-Yukawa sector is therefore of the form \eq{3R} or \eq{1R}, depending on whether the free gauge sector is asymptotically free or infrared free, see Tab.~\ref{tPara15}.

Under the exchange of gauge groups we have $(R,P,\eps)\leftrightarrow \left(R^{-1},P^{-1},P\eps\right)$, see \eq{exchange}. On the level of Fig.~\ref{pFP23} this corresponds to a simple rotation by $180$ degree around the  symmetric points $(R,P)=(1,1)$ (for $P>0$) and  $(R,P)=(1,-1)$ (for $P<0$), owing to the rescaling of parameters. Consequently, the results for \fp3 can be deduced from those at \fp2 by a simple rotation, see Fig.~\ref{pFP23}. 
More generally, this  exchange symmetry relates the partially interacting fixed points
\fp2 $\leftrightarrow$ \fp3 (Fig.~\ref{pFP23}), \fp4 $\leftrightarrow$ \fp5 (Fig.~\ref{pFP45}), and the fully interacting fixed points \fp7 $\leftrightarrow$ \fp8 (Fig.~\ref{pFP78}).    
The exchange symmetry is manifest at the fully interacting fixed points \fp6 (Fig.~\ref{pFP6}) and \fp9 (Fig.~\ref{pFP9}).

\subsection{Gauss with Gauge-Yukawa} 

In Fig.~\ref{pFP45} we show the domains of existence for \fp4 ({GY\,$\cdot$\,G}, upper) and \fp5 ({G\,$\cdot$\,GY}, lower panel). We observe that the fixed point exists for any parameter choice though its features vary with matter multiplicities. Specifically, for \fp4, six qualitatively different parameter regions are found. If the interacting gauge coupling is asymptotically free $(\eps<0)$ and provided that $P\eps <0$, the other gauge sector either remains asymptotically free (region 2) or becomes infrared free (region 1), whereas for $P\eps >0$ the other gauge sector invariably remains infrared free (region 3). 
Conversely, if the interacting gauge coupling is infrared free $(\eps>0)$ and provided that $P\eps <0$, the other gauge sector either remains asymptotically free (region 5) or becomes infrared free (region 4), whereas for $P\eps >0$ the other gauge sector invariably remains infrared free (region 6).  Moreover, as explained in Tab.~\ref{tBprime}, the interacting gauge sector can turn asymptotic freedom of the non-interacting gauge sector into infrared freedom (region 1 and 4). 
The eigenvalue spectrum in the gauge-Yukawa sector has therefore no relevant eigendirection \eq{0R} in region 1 and 3, one relevant eigendirection \eq{1R} in region 4 and 6, two relevant eigendirections \eq{2R} in region 2, and  three relevant eigendirections \eq{3R} in region 5, see Tab.~\ref{tPara15}.

We make the following observations. Firstly, we note that \fp4 in region 1 and 3 corresponds to a fully attractive IR fixed point with all RG trajectories terminating in it. The fixed point then acts as an infrared ``sink'' for massless trajectories and all canononicaly marginal couplings of the theory. Once scalar masses are switched on, RG flows may run away from the hypercritical surface of exactly massless theories, leading to massive phases with or without spontaneous breaking of symmetry. The quantum phase transition at \fp4 in region 1 and 3 is of the second order. Notice that in the ``direct product'' limit only models with $P\eps>0>\eps$ and $R>\s012$ (analogous to region 3) would lead to a fully infrared attractive ``sink''. Hence, the availability of region 1 is an entirely new effect, solely due to the $\psi$ fermions and the semi-simple nature of our models. We conclude that the semi-simple structure opens up new types of fixed points which cannot be achieved through a product structure. In region 2, we find that \fp4 has two relevant eigendirections as it would  in ``direct product'' settings. 

Secondly, in regions 4 and 6, \fp4 shows a single relevant eigendirection.  In the ``direct product'' limit, only models with $P,\eps>0$ and $R<\s012$ (analogous to region 6) would lead to a single relevant direction. Again, the availability of region 4 is a novel feature, and solely due to the $\psi$ fermions and thus a consequence of the semi-simple nature of the model. 

In the parameter region 5 the fixed point shows the largest number of UV relevant directions as it would without the $\psi$ fermions.  Moreover, in this parameter regime the Gaussian fixed point has only two relevant directions ($\eps>0, P\eps <0$).  Therefore \fp4 in region 5  qualifies as an asymptotically safe UV fixed point. On the other hand, in region 2,4 and 6, it takes the role of a cross-over fixed point. 
Results for \fp5 (Fig.~\ref{pFP45}, lower panel) follow from those for \fp4 via \eq{exchange}, and the distinct regions stated for \fp5 relate to the same physics as those for \fp4.

\begin{table}[b]
\begin{center}
\begin{tabular}{cccccc}
  \rowcolor{LightGreen}
\toprule
\rowcolor{LightGreen}
\cellcolor{LightBlue}
&\multicolumn{3}{c}{\cellcolor{LightGreen}\bf \ \ \ parameter range\ \ \ }
&&
\\
\rowcolor{LightGreen}
\cellcolor{LightBlue}
\multirow{-2}{*}{$\bm{\#}\ $}	& \ sign$\ \epsilon\ $	& $R$	& $P$	
&\multirow{-2}{*}{\cellcolor{LightGreen}\bf eigenvalue spectrum}
&\multirow{-2}{*}{\cellcolor{LightGreen}\bf info}
\\[.4mm]
\rowcolor{white}
\midrule
\rowcolor{white}
\FP6 
	&$-$ 
	& $\left(\frac{2}{11},\frac{11}{2}\right)$
	& $\left(\frac{2}{25}R, \frac{25}{2}R\right)$	
	& $\vartheta_{1,2} < 0 < \vartheta_{3,4}$
	&\cellcolor{LightRed}
Fig.~\ref{pFP6}\\[.4mm]
\midrule
\rowcolor{LightGray} &&&&& \cellcolor{LightRed} Fig.~\ref{pFP78} (upper panel)
\\
\rowcolor{LightGray}
	 &$-$ 
	 &$\left(\frac{2}{11},\frac{1}{2}\right)$
	 &$ \left(\frac{25}{2}R ,+\infty\right)$	
	 &$\vartheta_{1,2} < 0 < \vartheta_{3,4}$
	&region 1
	\\[.4mm]
\rowcolor{LightGray}
	&$-$
	&  $\left(\frac{1}{2},R_1\right)$
	& $\left(\frac{25}{2}R , X(R)\right)$		
	& $\vartheta_{1,2} < 0 < \vartheta_{3,4}$
	&region 1
\\[.4mm]
\rowcolor{LightGray}
	&$-$ 
	&  $\left(R_1,\frac{11}{2}\right)$ 
	& $\left(X(R) , \frac{25}{2}R\right)$		
	& $\quad\ \vartheta_1 < 0 < \vartheta_{2,3,4}$
	&region 2
\\[.4mm]
\rowcolor{LightGray}
\multirow{-5}{*}{\cellcolor{LightGray}\FP7}		 	
	&$+$ 
	&  $\left(\frac{2}{11},\frac{1}{2}\right)$
	&$ \left(-\infty, X(R)\right)$	
	&$\vartheta_{1,2} < 0 < \vartheta_{3,4}$
	& region 3
\\[.4mm]
\midrule
\rowcolor{white} &&&&& \cellcolor{LightRed} Fig.~\ref{pFP78} (lower panel)
\\
\rowcolor{white}
	&$-$ 
	&  $\left(R_4,\frac{11}{2}\right)$
	&  $\big(\xalt(R),\frac{2}{25}R\big)$	 
	& $\vartheta_{1,2} < 0 < \vartheta_{3,4}$
	&region 1 \& 3 
\\[.4mm]
\rowcolor{white}
\multirow{-3}{*}{\FP8}		
	&$-$ 
	& $\left(\frac{2}{11},R_4\right)$
	&  $\big(\frac{2}{25}R , \xalt(R)\big)$	 
	& $\quad\ \vartheta_1 < 0 < \vartheta_{2,3,4}$
	& region 2
\\[.4mm]

\midrule
\rowcolor{LightGray} &&&&& \cellcolor{LightRed} Fig.~\ref{pFP9}
\\
\rowcolor{LightGray}
	 &$-$ 
	 &  $(\frac{2}{11},\frac{1}{2})$
	 &$  (\xalt(R) ,+\infty) $		
	 &  $\quad\ \vartheta_1 < 0 < \vartheta_{2,3,4}$
	 & region 1
\\[.4mm]
\rowcolor{LightGray}
	&$-$ 
	&  $\left(\frac{1}{2},R_2\right)$
	&$  \big(\xalt(R) , X(R)\big)$		
	& $\quad\ \vartheta_1 < 0 < \vartheta_{2,3,4}$
	&region 1 
\\[.4mm]
\rowcolor{LightGray}
	 &$-$ 
	 &  $\left(R_3,\s0{11}2\right)$
	 &$  \big(\xalt(R) , X(R)\big)$			
	 & $\quad\  \vartheta_1 < 0 < \vartheta_{2,3,4}$
	 & region 1 \& 4
\\[.4mm]
\rowcolor{LightGray}
	&$-$
	&  $\left(R_2,R_3\right)$
	&$\ \ \big(X(R) , \xalt(R)\big)\ \ $			
	& $\quad\quad\quad\quad\  0 < \vartheta_{1,2,3,4}$
	&region 2
\\[.4mm]
\rowcolor{LightGray}
\multirow{-6}{*}{\cellcolor{LightGray}\FP9}		
	&$+$ 
	&  $\left(\frac{2}{11},\frac{1}{2}\right)$
	&$ \left(-\infty, X(R)\right)$	
	& $\quad\ \vartheta_1 < 0 < \vartheta_{2,3,4}$
	& region 3
\\[.4mm]
\bottomrule
\end{tabular}
\end{center}
 \caption{Parameter regions where the fully interacting fixed points \fp6 -- \fp9 exist, along with the eigenvalue spectrum for the various parameter regions.
}\label{tPara69}
\end{table}

\subsection{Banks-Zaks with Banks-Zaks}

Next, we turn to fully interacting fixed points where both gauge couplings are non-vanishing, see Tab.~\ref{tPara69}. In general, the eigenvalue spectrum is determined through \eq{dtheta} with solutions \eq{theta}, with $\eps$ taking the role of the parameter $B$.
In the ``direct product'' limit,  fully interacting fixed points reduce to direct products from each pairing of the possibilities (BZ, GY) in each of the simple gauge sectors.  For $N_\psi\neq 0$, the fermions $\psi$ introduce a direct mixing between the gauge groups and we may then expect to find something close to a product structure, potentially modified by new effects parametrised via $R$ in fixed points not involving Gaussian factors.

The first such fixed point is \fp6 ({BZ$\cdot$BZ}), where each gauge sector achieves a Banks-Zaks fixed point. Yukawa couplings play no role, see  Fig.~\ref{pFP6}.  The fixed point invariably requires $\eps<0$ and $P\eps<0$ and entails an eigenvalue spectrum with two relevant directions of order ${\cal O}(\eps^2)$, and two irrelevant directions of order ${\cal O}(\eps)$ associated to the Yukawas, 
\beq\label{12-34}
\vartheta_{1},\vartheta_{2}<0<\vartheta_{3},\vartheta_{4}\,.
\eeq
The quartics are  marginally irrelevant. 
The Gaussian is necessarily the UV fixed point in these settings which makes \fp6 a cross-over fixed point.  The accessible parameter region, shown in Fig.~\ref{pFP6},  is invariant under the exchange of gauge groups \eq{exchange}. The ``direct product'' limit has qualitatively the same spectrum \eq{12-34}. The main effect due to the semi-simple character of the theory relates to the exclusion of certain parameter regions (white regions). We conclude that the semi-simple nature of the theory leads to parameter restrictions without otherwise changing the overall appearance of the fixed point.

\begin{figure}[t]
\begin{center}
\hskip-.5cm
\includegraphics[scale=0.285]{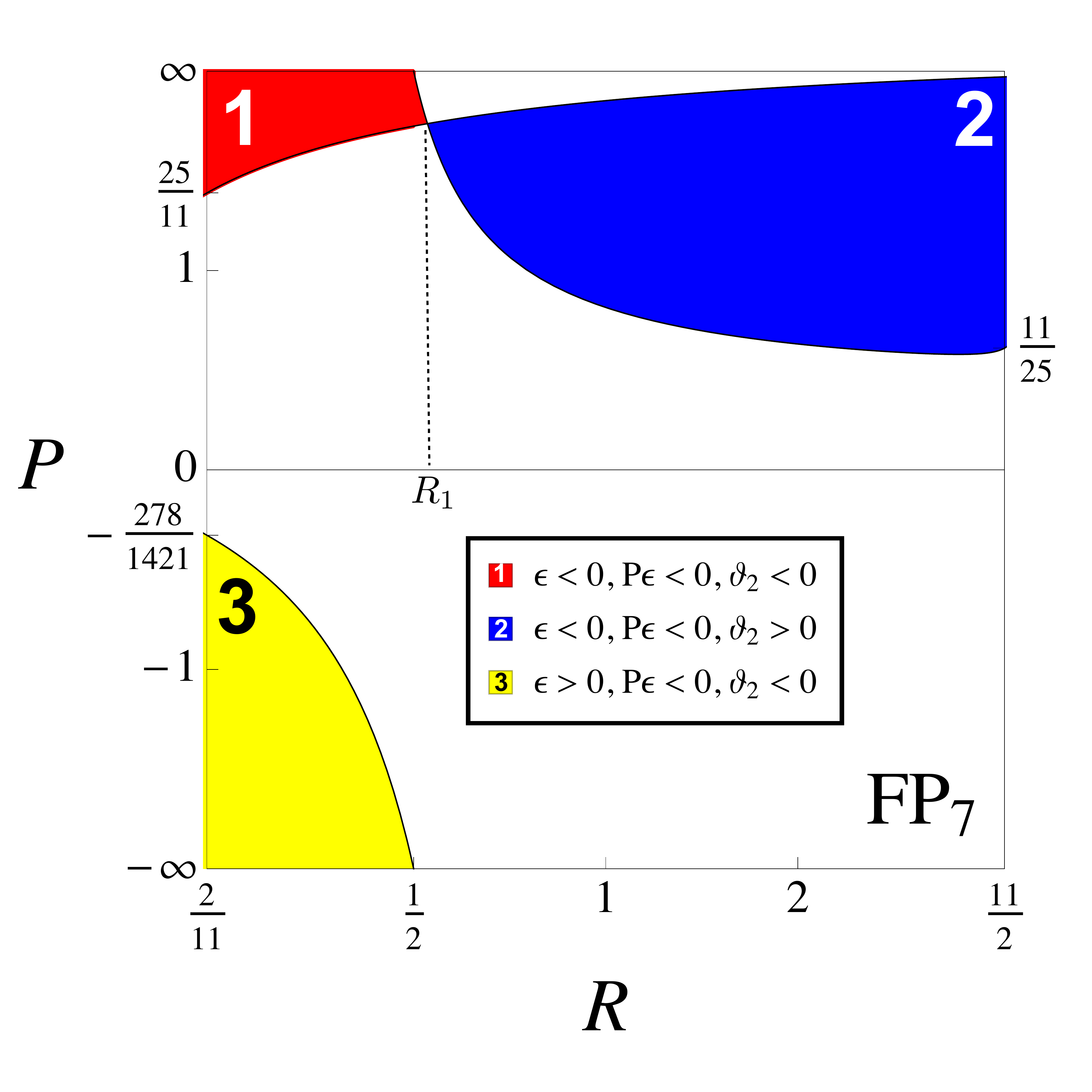}
\\
\vskip-.3cm
\hskip.5cm
\includegraphics[scale=0.285]{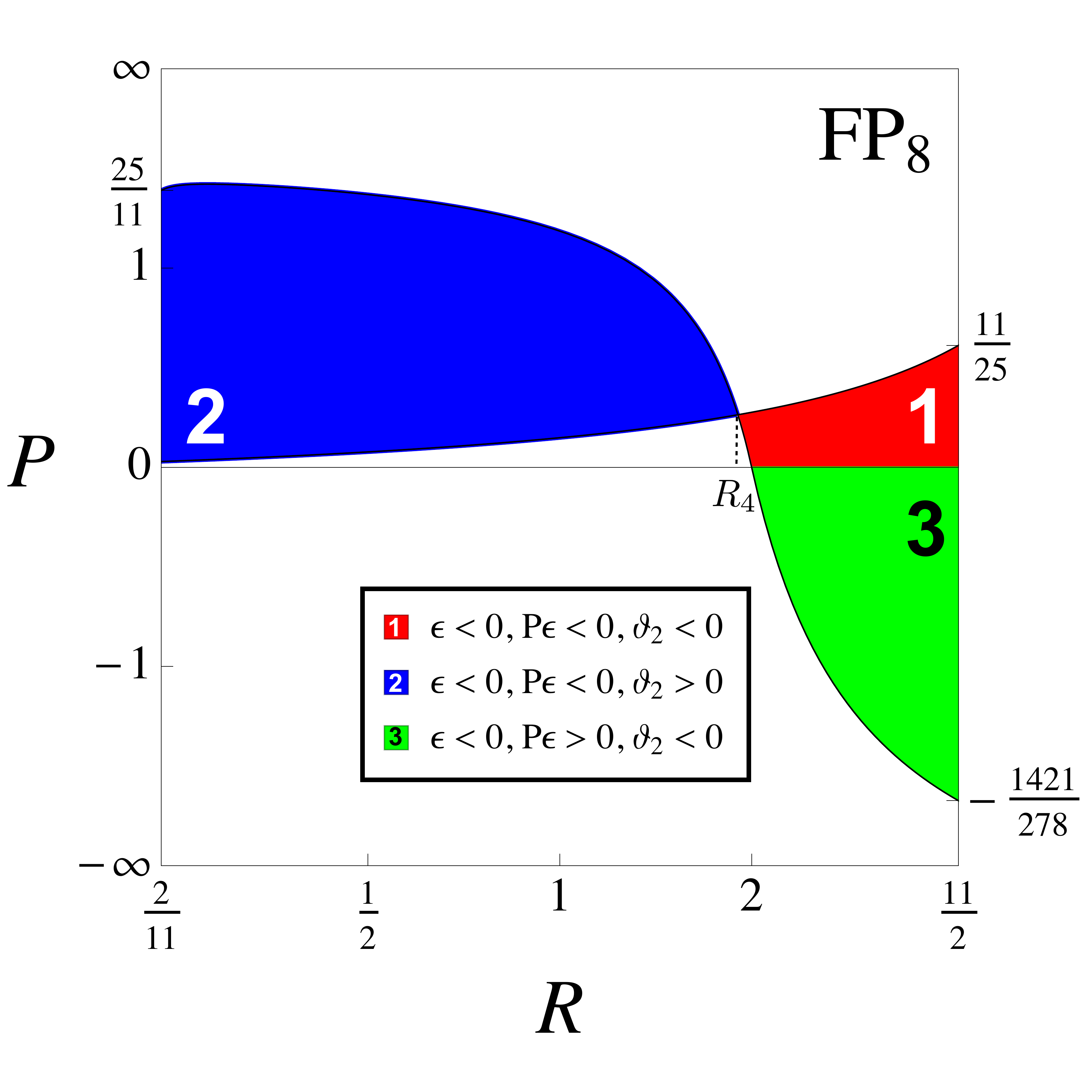}
\caption{The phase space of parameters for the fixed points FP${}_7$ and FP${}_8$ where two gauge and one of the Yukawa couplings take interacting and physical fixed points, while the other Yukawa coupling remains trivial. The inset indicates the signs for $\eps$ and $P\eps$, together with the sign for the eigenvalue $\vartheta_2$, Tab.~\ref{tPara69}  
 (see main text). }
\label{pFP78} 
\end{center}
\end{figure}

\subsection{Banks-Zaks with Gauge-Yukawa} 
At the interacting fixed points \fp7 ({BZ\,$\cdot$\,GY}, upper panel), and \fp8 ({GY\,$\cdot$\,BZ}, lower panel), we have that both gauge and one of the Yukawa couplings are non-trivial. Our results for the condition of existence and the eigenvalue spectra are displayed in Fig.~\ref{pFP78}. By definition, this type of fixed point requires that either $\eps<0$ or $P\eps<0$, or both, meaning that at least one of the gauge sectors is asymptotically free. In Fig.~\ref{pFP78}, this relates to three different parameter regions (see inset for the colour coding). In region 1 and 2, the theory is asymptotically free in both gauge sectors, whereas in region 3 the theory is asymptotically free in only one gauge sector. We observe that large regions of parameter space are excluded. Valid parameter regions are further distinguished by their eigenvalue spectrum which either takes the form \eq{2R} or \eq{1R}, meaning that minimally one and maximally two eigenoperators constructed out of the gauge kinetic terms and the Yukawa interactions are UV relevant, $\vartheta_1 <0<\vartheta_3,\vartheta_4$.
The sign of $\vartheta_2$ depends on the matter field multiplicities. 
In region 1 and 3, and for either of \fp7 and \fp8, we find that 
\beq\label{1-fp78}
\vartheta_1 , \vartheta_2<0<\vartheta_3,\vartheta_4\,.
\eeq
In region 2, conversely, we have
\beq\label{2-fp78}
\vartheta_1 <0< \vartheta_2,\vartheta_3,\vartheta_4\,.
\eeq
Hence, at \fp7 and in the regime where both gauge sectors are asymptotically  free $(P > 0 > \eps)$, two types of valid fixed points are found. For sufficiently low $R<R_1$ \eq{R1234}, and large $P$, the fixed point has two relevant directions (region 1). Increasing $R>R_1$ at fixed $P$ may lead to a second type of IR fixed point with a single relevant direction (region 2). On the other hand, in the regime $\eps>0>P$ only one type of fixed point exists with two relevant directions (region 3). 
It is worth comparing  these results with the ``direct product'' limit. For $P > 0 > \eps$ the latter leads to the eigenvalue spectrum \eq{2-fp78}, as  found in region 2. Also, for $\eps>0>P$ the ``direct product'' fixed point has the eigenvalue spectrum \eq{1-fp78}, which is qualitatively in accord with findings in region 3. 
We conclude that the semi-simple nature of the interactions plays a minor quantitative role in region 2 and 3.
On the other hand, in region 1 where $P > 0 > \eps$, the semi-simple nature of the theory leads to an important qualitative modification: an eigenvalue spectrum with two relevant directions at \fp7  cannot be achieved through a direct product setting; rather, it necessarily requires matter fields charged under both gauge groups. We conclude that the semi-simple nature of interactions play a key qualitative role in region 1.
Analogous results hold true for \fp8 after the substitutions \eq{exchange} and the replacement $R_1\to R_4=1/R_1$, see \eq{R1234}.

\subsection{Gauge-Yukawa with Gauge-Yukawa} 
At the fully interacting fixed point \fp9 ({GY\,$\cdot$\,GY}), we have that both gauge and both Yukawa couplings are non-trivial. We find that the eigenvalue spectrum in the gauge-Yukawa sector reads either \eq{1R} or \eq{0R}, meaning that at least three of the four eigenoperators constructed out of the gauge and fermion fields 
are strictly irrelevant, $0<\vartheta_2,\vartheta_3,\vartheta_4$.
The sign of the  eigenvalue $\vartheta_1$ depends on the matter field multiplicities of the model. 

\begin{figure}[t]
\begin{center}
\includegraphics[scale=0.3]{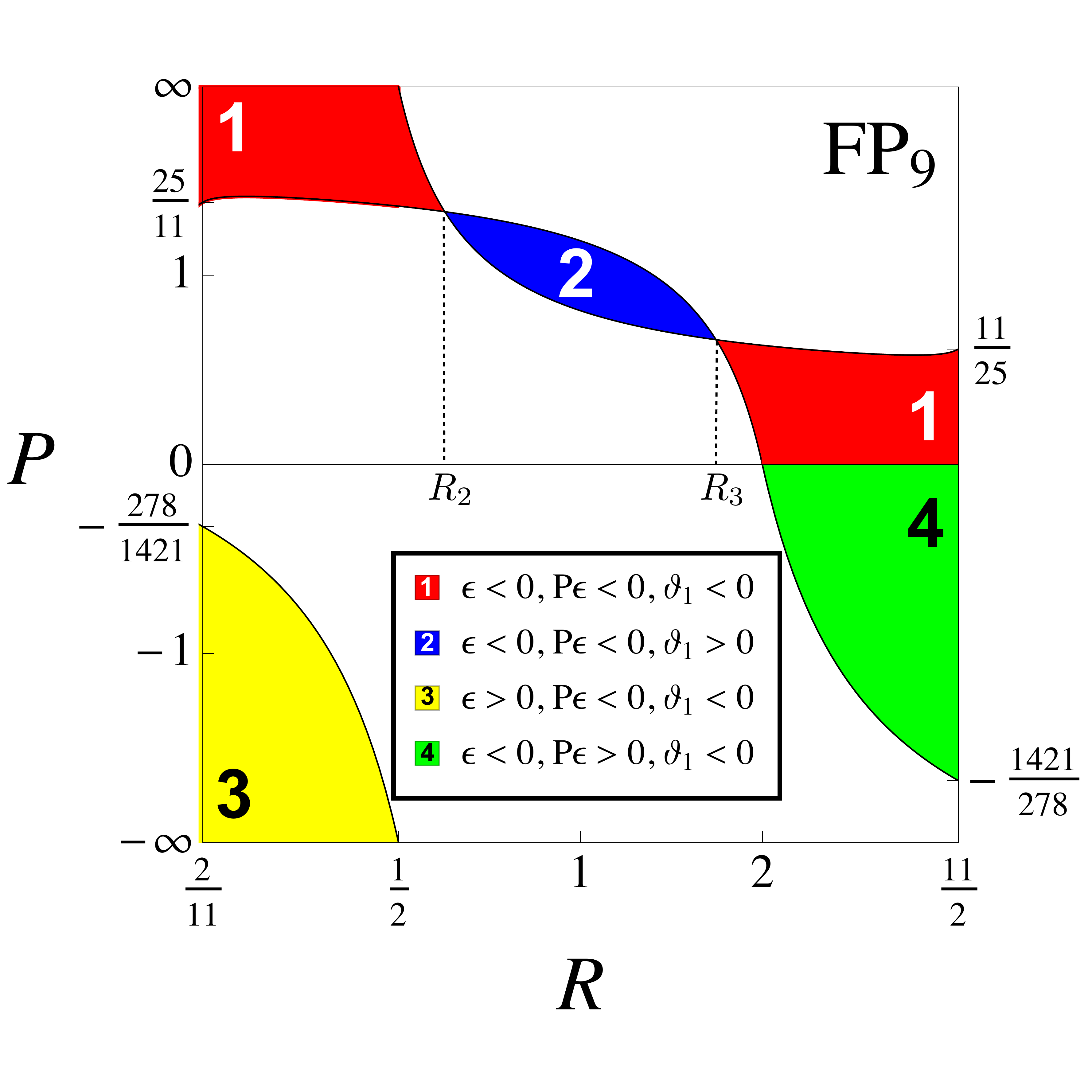}
\vskip-.5cm
\caption{The phase space of parameters for the  fully interacting fixed point FP${}_9$ where all gauge and all Yukawa couplings are non-trivial.
The coloured regions relate to the portions of parameter space where the fully interacting fixed point is physical. The inset provides additional information including the sign for the eigenvalue $\vartheta_4$ (see main text).
}\label{pFP9} 
\end{center}
\end{figure}

Our results for the condition of existence and the eigenvalue spectrum are stated in Fig.~\ref{pFP9}. 
We observe four qualitatively different parameter regions (see inset for the colour coding).
For $P > 0 >\eps$, the theory is asymptotically free in both gauge sectors and we find two types of valid parameter regions, depending on whether $R$ takes values below $R_2$ or  above $R_3$ (region 1), or  in between (region 2); see \eq{R1234}. Moreover, in region 2, we find that the fixed point is strictly IR attractive in all couplings, owing to
\beq\label{2-fp9}
0<\vartheta_1\,,\vartheta_2,\vartheta_3,\vartheta_4\,.\eeq
Hence, the fixed point \fp9 in region 2 corresponds to a fully attractive IR fixed point acting as an infrared ``sink'' for massless trajectories and all canononicaly marginal couplings of the theory. Ultimately it describes a second order quantum phase transition between a symmetric and a symmetry broken phase, characterised by the vacuum expectation value of the scalar field. Qualitatively, the same  type of result is achieved in the ``direct product'' limit.
Hence, the main effect due to semi-simple interactions
is to have generated a boundary in parameter space. 
In region 1 we find
\beq\label{1-fp9}
\vartheta_1<0<\vartheta_2,\vartheta_3,\vartheta_4\,.
\eeq
This type of eigenvalue spectrum cannot be achieved without semi-simple interactions mediated by the $\psi$ fields and is therefore a novel feature, entirely due to the semi-simple nature of the theory. In this regime, \fp9 corresponds to a cross-over fixed point (the Gaussian is the UV fixed point) with a single unstable direction where trajectories escape either towards a weakly coupled IR fixed point, or towards a regime of strong coupling  with (chiral) symmetry breaking, confinement, or infrared conformality.

For $\eps>0>P$ or $P\eps>0>\eps$, the theory is asymptotically free in one and infrared free in the other gauge sector. Valid fixed points then correspond to region 3 or region~4, respectively. In either of these cases, the eigenvalue spectrum shows a single relevant direction, \eq{1-fp9}. This agrees qualitatively with the eigenvalue spectrum in the  ``direct product'' limit. We conclude, once more,  that the main impact of the $\psi$ fields relates to the boundaries in parameter space which restrict the fixed point's domain of availability.

Finally, for $\eps,P>0$, the theory is infrared free in both gauge sectors. We observe that  no such interacting fixed point arises, irrespective of matter multiplicities. Interestingly though, such fixed points do exist in the ``direct product'' limit with spectrum
  \beq\label{thetafp9put}
\vartheta_1,\vartheta_2<0<\vartheta_3,\vartheta_4\,.
\eeq
The reason for their non-existence in our models is the presence of the $\psi$ fermions. The requirement of perturbativity in both gauge couplings then leads to limitations on the parameter $R$ which  cannot be satisfied at \fp9 with eigenvalue spectrum \eq{thetafp9put}. This result provides us with an example where the semi-simple nature of the theory
``disables'' a fixed point. This completes the overview of interacting fixed points in the gauge-Yukawa sector and their key properties.

\section{\bf Scalar fixed points and vacuum stability}\label{stab}
In this section, we analyse the scalar sector and establish conditions for stability of the quantum vacuum. We  also provide results for all scalar couplings  at all interacting fixed points, Tab.~\ref{tsFPeps}. 

\subsection{Yukawa and scalar nullclines}\label{YSnull}

Following  \cite{Bond:2016dvk}, we begin by exploiting the results \eq{EF} to express the Yukawa nullclines in terms of the gauge couplings and the parameter $R$.  To leading order in the small expansion parameters \eq{eps}, and using \eq{EF}  together with \eq{12Yy}, the non-trivial Yukawa nullclines $\beta_Y=0$ and $\beta_y=0$ take the explicit form
\beq\label{Ynull}
\frac{\alpha_Y}{\alpha_1}=\frac{6}{13-2R}\,,\quad
\frac{\alpha_y}{\alpha_2}=\frac{6}{13-2/R}\,.
\eeq
For fixed gauge couplings, we observe that the Yukawa couplings are enhanced over their values in the absence of the fermions $\psi$. The relevance of nullcline solutions \eq{Ynull} is as follows. By their very definition, the Yukawa couplings no longer run with the RG scale when taking the values \eq{Ynull}. If at the same time the gauge couplings take fixed points on their own, the nullcline relations then provide us with the correct fixed point values for the Yukawa couplings.  Evidently, \eq{Ynull} together with \eq{R} guarantees that the Yukawa fixed points are physical as long as the gauge fixed points are. Note also that the slope of the nullcline remains positive and finite for all $R$ within the domain \eq{R}.  Hence strict perturbativity in the Yukawa couplings follows from strict perturbativity in the gauge couplings, in accord with the general discussion  in \cite{Bond:2016dvk} based on dimensional analysis.

Next we turn to the scalar nullclines. Since the beta functions for the two scalar sectors decouple at this order, we may analyse their nullclines individually.\footnote{This simplification solely arises provided the mixing coupling $\alpha_w$ takes its exact Gaussian fixed point \eq{wFP}. For non-trivial $\alpha_w$ the nullclines take more general forms.}   All results for the subsystem $(\alpha_U,\alpha_V)$ can straightforwardly be translated to the subsystem  $(\alpha_u,\alpha_v)$ by  substituting $R\leftrightarrow 1/R$, also using \eq{eps}. Furthermore, since the scalars are uncharged, their one loop beta functions are independent of the gauge coupling. Dimensional analysis then shows that all non-trivial scalar nullclines are proportional to the corresponding Yukawa coupling \cite{Bond:2016dvk}. The scalar nullclines  represent exact fixed points of the theory provided the Yukawa couplings take interacting fixed points. Perturbativity of scalar couplings at an interacting fixed point then follows from the perturbativity of Yukawa couplings which, in turn, follows from perturbativity in the gauge couplings.

Specifically, the  nullclines for the single trace scalar couplings are found from \eq{HVhv} by resolving $\beta_U=0$ for $\alpha_U$. We find two solutions
\beq\label{Hnull}
\frac{\alpha_{U\pm}}{\alpha_Y}=\frac14\left(-1\pm\sqrt{23-4R}\right)\,.
\eeq
Note that the double trace coupling does not couple back into  the running of the single trace coupling. Within the parameter range \eq{R} we observe that $\alpha_{U+}>0>\alpha_{U-}$.
Next, we consider the nullclines for the double-trace quartic coupling $\alpha_V$. Inserting $\alpha_{U+}$ into $\beta_V=0$, we find a pair of nullclines given by
\beq\label{Vnull}
\frac{\alpha_{V\pm}}{\alpha_Y}=\frac14\left(-2\sqrt{23-4R}\pm 
\sqrt{20-4R+6\sqrt{23-4R}}\right)\,.
\eeq
Both nullclines take real values for all $R$ within the range \eq{R}, and we end up with $\alpha_{U+}\ge 0$ together with $0>\alpha_{V+}>\alpha_{V-}$. Analogously, inserting $\alpha_{U-}$ into $\beta_V=0$, we find a second pair of nullclines given by
\beq\label{Vnull2}
\frac{\alpha_{V2\pm}}{\alpha_Y}=\frac14\left(2\sqrt{23-4R}\pm 
\sqrt{20-4R-6\sqrt{23-4R}}\right)\,.
\eeq
In this case, however, the result \eq{Vnull2} comes out complex within the parameter range \eq{R}, meaning that even if $\alpha_Y^*$ takes a real positive fixed point the corresponding scalar fixed point is invariably unphysical. 

The replacement $R\to 1/R$ in \eq{Hnull} and \eq{Vnull}, \eq{Vnull2} allows us to obtain explicit expressions for the nullclines for ${\alpha_{u\pm}}/{\alpha_y}$  and ${\alpha_{v\pm}}/{\alpha_y}$.  The real solutions are given by
\beq\label{hnull}
\frac{\alpha_{u\pm}}{\alpha_y}=\frac14\left(-1\pm\sqrt{23-4/R}\right)
\eeq
with $\alpha_{u+}\ge 0>\alpha_{u-}$. The solution $\alpha_{u+}$ leads to real nullclines for the double-trace coupling $\alpha_{v}$ given by
\beq\label{vnull}
\frac{\alpha_{v\pm}}{\alpha_y}=\frac14\left(-2\sqrt{23-4/R}\pm 
\sqrt{20-4/R+6\sqrt{23-4/R}}\right)\,,
\eeq
and we end up with $\alpha_{u+}\ge 0$ together with  $0>\alpha_{v+}>\alpha_{v-}$. On the other hand, the solution $\alpha_{u-}$ does not lead to real solutions for $\alpha_{v\pm}$. 
This completes the overview of Yukawa and scalar nullcline solutions.

\subsection{Stability of the vacuum}
We are now in a position to reach firm conclusions concerning the stability of the ground state at interacting fixed points. The reason for this is that this information is encoded in the scalar nullclines. The explicit form of the fixed point in the gauge-Yukawa sector is not needed. 
To that end, we recall the stability analysis for potentials of the form
\begin{align}\label{potential}
W \propto \alpha_U\,\Tr\,(H ^\dagger H )^2  +\alpha_V/N_{\rm F}\,(\Tr\,H ^\dagger H )^2 \,,
\end{align}
In the limit where $\alpha_w=0$ the scalar field potential in our models \eq{L} are given by \eq{potential} together with its counterpart $(H,\alpha_U,\alpha_V)\leftrightarrow (h,\alpha_u,\alpha_v)$. For potentials of the form \eq{potential}, the general conditions for vacuum stability read \cite{Paterson:1980fc,Litim:2015iea}
\beq\label{stability}
\begin{array}{rcl}
{\it a)}\quad \alpha^*_{U}>0&{\rm \ \ and\ \ }&\alpha^*_{U}+\alpha^*_{V}> 0\\
{\it b)}\quad \alpha^*_{U}<0&{\rm \ \ and\ \ }&\alpha^*_{U}+\alpha^*_{V}/N_{\rm F}>0
\end{array}
\eeq
and similarly for $(\alpha_U,\alpha_V)\leftrightarrow(\alpha_u,\alpha_v)$. 
In the Veneziano limit, case $b)$ effectively becomes void and cannot be satisfied for any $\alpha_U^*$, irrespective of the sign of   $\alpha_V^*$. 
Inserting the fixed points into \eq{stability} we find 
\beq\label{vac11}
\begin{array}{rcl}
\alpha^*_{U+}+\alpha^*_{V+}&=&
\displaystyle
\frac{\alpha^*_Y}4
\left(+\sqrt{20-4R+6\sqrt{23-4R}}-\sqrt{23-4R} -1\right)\ge 0\,,\\[2ex]
\displaystyle
\alpha^*_{U+}+\alpha^*_{V-}&=&
\displaystyle
\frac{\alpha^*_Y}4
\left(-\sqrt{20-4R+6\sqrt{23-4R}}-\sqrt{23-4R} -1\right)\le -\alpha^*_Y\,,
\end{array}
\eeq
Stability of the quantum vacuum is evidently  achieved at the fixed point $(\alpha^*_{U+},\alpha^*_{V+})$  following case $a)$ and irrespective of the value for the Yukawa fixed point as long as $\alpha_Y^*>0$. 
The potential \eq{potential} becomes exactly flat at the fixed point  iff $R=\s0{11}{2}$. In this case, higher order or radiative corrections must be taken into consideration to guarantee stability in the presence of flat directions. 
Stability is not achieved at the fixed point $(\alpha^*_{U+},\alpha^*_{V-})$, 
for any $R$. 
Turning to the second scalar sector, we find 
\beq\label{vac21}
\begin{array}{rcl}
\alpha^*_{u+}+\alpha^*_{v+}&=&
\displaystyle
\frac{\alpha^*_y}4
\left(+\sqrt{20-4/R+6\sqrt{23-4/R}}-\sqrt{23-4/R} -1\right)\ge 0\,,\\[2ex]
\alpha^*_{u+}+\alpha^*_{v-}&=&
\displaystyle
\frac{\alpha^*_y}4
\left(-\sqrt{20-4/R+6\sqrt{23-4/R}}-\sqrt{23-4/R} -1\right)\le -\alpha^*_y\,,
\end{array}
\eeq
where the bounds refer to $R$ varying within the range \eq{R}. 
This part of the potential becomes exactly flat at the fixed point  iff $R=\s02{11}$. 
The result establishes vacuum stability at the fixed point $(\alpha^*_{u+},\alpha^*_{v+})$.
We also confirm that the fixed point $(\alpha^*_{u+},\alpha^*_{v-})$  
does not lead to a stable ground state. 
We  conclude that vacuum stability is guaranteed at the interacting fixed points $(\alpha^*_{U+},\alpha^*_{V+})$ and $(\alpha^*_{u+},\alpha^*_{v+})$, together with $\alpha_w=0$, irrespective of the fixed points in the gauge Yukawa sector, as long as the later is physical. Out of the a priori  $2^3$ different fixed point candidates in the scalar sector at one loop (half of which lead to real fixed points) the additional requirement of vacuum stability has identified a {\it unique} viable solution. 
In this light, vacuum stability 
dictates that the anomalous dimensions \eq{gammam} are strictly positive at interacting fixed points, \eq{gammaPos}, with
\beq\label{gammam*}
\begin{array}{rcl}
\gamma^*_M&=&\displaystyle\al Y^*\, \sqrt{20-4R+6\sqrt{23-4R}}>0\,,\\[1ex]
 \gamma^*_m&=&\displaystyle\al y^*\, \sqrt{20-4/R+6\sqrt{23-4/R}}>0  \,,
\end{array}
\eeq
and provided that \eq{R} is observed.

\subsection{Portal coupling}
Now we clarify whether the stability of the vacuum is affected by the presence of the ``portal'' coupling   $\alpha_w\neq 0$ which induces a mixing between the scalar sectors. 
In this case the scalar potential is given by $W=-L_{\rm pot}$ in \eq{L},
\bea
	W&=& U\, \Tr(H^\dagger H)^2 + V\, (\Tr H^\dagger H )^2 
		+ u\, \Tr(h^\dagger h)^2 + v\, (\Tr h^\dagger h )^2 + w\, \Tr(H^\dagger H)\Tr(h^\dagger h)\,,\quad
\eea
where $H$  and $h$ are $N_{\rm F} \times N_{\rm F}$ and $N_{\rm f} \times N_{\rm f}$ matrices, respectively. Following the reasoning of  \cite{Paterson:1980fc,Litim:2015iea}, we observe that the potential has a global $U(N_{\rm F})_L \otimes U(N_{\rm F})_R\otimes U(N_{\rm f})_L \otimes U(N_{\rm f})_R$
symmetry which allows us to bring each of the fields into a real diagonal configuration, $H = \operatorname{diag}(\Phi_1,\Phi_2,\dots)$ and $h = \operatorname{diag}(\phi_1,\phi_2,\dots)$. As the potential is homogeneous in either field,  $W(c\,H, c\,h) = c^4 W(H,h)$, 
it suffices to guarantee positivity on a fixed surface $\sum_i \Phi_i^2 =1= \sum_j \phi_j^2$ which is implemented using Lagrange multipliers $\Lambda$ and $\lambda$.
From
\beq
\begin{array}{rl}
	\di\frac{\partial W}{\partial \Phi_i} 
	&= 2 \Phi_i (2U \Phi_i^2 + 2V + w - 2\Lambda)\,,\\[2ex]
	\di \frac{\partial W}{\partial \phi_i} &= 2 \phi_i (2u \phi_i^2 + 2v + w - 2\lambda)\,,
\end{array}
\eeq
it  follows that 
extremal field configurations are those where all non-zero fields take equal values. If we have $M$ non-zero $\Phi$ fields and $m$ non-zero $\phi$ fields, the extremal field values are
$\Phi_i^2 = 0$ or $\Phi_i^2 = \frac{1}{M}$  alongside with
$\phi_i^2 = 0$ or $\phi_i^2= \frac{1}{m}$.
Three non-trivial cases arise. If $m = 0$ the  extremal potential is
$W_e = {U}/{M} + V$. Likewise if $M = 0$ we have $W_e = {u}/{m} + v$.
Lastly, if both $m, M \neq 0$, we have $W_e = {U}/{M} + V + {u}/{m} + v + w$.
The values of $M,m$ for which these extremal potentials are minima depend on the signs of the couplings $U,u$, leaving us with the four possible cases  $U,u>0$, $u>0>U$, $U>0>u$, and $0>U,u$. We thus obtain four distinct sets of conditions for vacuum stability which we summarise as follows:
\beq\label{conditionsW}
\begin{array}{llccccl}
a)\ \ 
	&\alpha_u, \alpha_U\ge 0\,, 
	&\alpha_U + \alpha_V \geq 0\,, 
	&
	&\alpha_u + \alpha_v \geq 0\,, 
	&
	&\di F\,(\alpha_U + \alpha_V) + \frac{\alpha_u 
	+ \alpha_v}{F}+ \alpha_w \geq 0\,, 
\\[2ex]
b)	
	&\alpha_u>0>\alpha_U\,, 
	&\di \alpha_U + \frac{\alpha_V}{N_{\rm F}} \geq 0 \,,
	&
	&\alpha_u + \alpha_v \geq 0\,,
	&
	&\di
	\alpha_U + \frac{\alpha_V}{N_{\rm F}} + \frac{\alpha_u + \alpha_v}{F\,N_{\rm f}} 
	+ \frac{\alpha_w}{N_{\rm f}}\geq 0\,, 
	\\[2ex]
c)	
	& \alpha_U>0>\alpha_u\,, 
	&\alpha_U + \alpha_V \geq 0\,,
	&
	&\di \alpha_u + \frac{\alpha_v}{N_{\rm f}} \geq 0\,,
	&
	&\di 
	\alpha_u + \frac{\alpha_v}{N_{\rm f}} + F\,\frac{\alpha_U + \alpha_V}{N_{\rm F}} 
	+ \frac{\alpha_w}{N_{\rm F}}\geq 0\,, 
	\\[2ex]
d)
	&0\ge\alpha_u,\alpha_U\,, 
	&\di \alpha_U + \frac{\alpha_V}{N_{\rm F}} \geq 0\,,
	&
	&\di \alpha_u + \frac{\alpha_v}{N_{\rm f}} \geq 0\,,
	&
	&\di	 \alpha_u + \frac{\alpha_v}{N_{\rm f}} 
	+ F\left(\alpha_U + \frac{\alpha_V}{N_{\rm F}}\right) 
	+ \frac{\alpha_w}{N_{\rm F}} \geq 0\,.
\end{array}
\eeq
Notice that we have rescaled the couplings as in \eq{couplings} and \eq{w} to make contact with the notation used in this paper. The parameter $F\equiv N_{\rm f}/N_{\rm F}>0$ can be expressed in terms of the parameter $R$ to leading order in $\eps\ll 1$,  see \eq{epsF}.

We make the following observations. In all four cases, the additional condition owing to the mixing coupling \eq{w} takes the form of a lower bound for $\alpha_w$. Furthermore, $\alpha_w$ is allowed to be negative without destroying the stability of the potential, provided it does not become too negative. 
We also note that none of the three cases $b)$, $c)$ or $d)$  in \eq{conditionsW} can have consistent solutions in the Veneziano limit where $N_{\rm F}, N_{\rm f} \rightarrow \infty$. This uniquely leaves the case $a)$ as the only possibility for vacuum stability in the parameter regions considered here. These solutions neatly fall back onto  the solutions discussed previously in the limit $\alpha_w\to 0$. As long as the auxiliary condition 
\beq
\label{aux}
\alpha_w \geq  -[F\left(\alpha_U + \alpha_V\right) + F^{-1}\left(\alpha_u + \alpha_v\right)]
\eeq
is satisfied,   we can  safely conclude that  a non-vanishing $\alpha_w\neq 0$ does not spoil vacuum stability, not even for negative portal coupling $\alpha_w$.

\begin{table*}[t]
\begin{center}
\begin{tabular}{cc}
\toprule
\rowcolor{LightBlue}
&\cellcolor{LightGreen} 
\\[-1mm]
\cellcolor{LightBlue}
\multirow{-2}{*}{
$\bm{\#}$
}
&\cellcolor{LightGreen}  
\multirow{-2}{*}{\bf quartic scalar couplings}
\\
\midrule
\rowcolor{LightGray}
\FP{1-3} 
&
$
\alpha^*_U=0\,,\quad
\alpha^*_V=0\,,\quad 
\alpha^*_u=0\,,\quad
\alpha^*_v=0\,,
$\\[1ex]
& \\[-2.5ex]
\FP4
&
$
\alpha^*_U=\s0{4F_1(R)R\eps}{(2R-1)(3R-19)}\,,\quad
\alpha^*_V=\s0{4F_2(R)R\eps}{(2R-1)(3R-19)}\,,\quad
\alpha^*_u=0\,,\quad 
\alpha^*_v=0\,,
$ 
\\[1ex]
\rowcolor{LightGray}
& \\[-2.5ex]
\rowcolor{LightGray}
\multirow{1}{*}{\bf FP${}_{\bf 5}$}
&$
\alpha^*_U=0\,, \quad
\alpha^*_V=0\,, \quad 
\alpha^*_u=\frac{4\,F_1(1/R)P\eps/R}{(2/R-1)(3/R-19)}\,,\quad
\alpha^*_v=\frac{4\,F_2(1/R)P\eps/R}{(2/R-1)(3/R-19)}\,,
$\\[1ex]
& \\[-2.5ex]
\FP6
&$\alpha^*_U=0\,,
\quad\alpha^*_V=0\,,
\quad \alpha^*_u=0\,,
\quad\alpha^*_v=0\,,
$\\[1ex]
\rowcolor{LightGray}
& \\[-2.5ex]
\rowcolor{LightGray}
\FP7 
&
$\alpha^*_U=\0{4}3 \frac{(25 - 2P/R)F_1(R)}{50R^2-343R+167} R\eps\,,\quad
\alpha^*_V= \0{4}3 \frac{(25 - 2P/R)F_2(R)}{50R^2-343R+167} R\eps\,,\quad
\alpha^*_u=0\,,\,\quad 
\alpha^*_v=0\,,
$\quad\quad
\\[1ex]
& \\[-2.5ex]
\multirow{1}{*}{\FP8}
&$
\alpha^*_U=0\,, \quad
\alpha^*_V=0\,, \quad 
\alpha^*_u=\0{4}3 \frac{(25 - 2R/P)F_1(1/R)}{50/R^2-343/R+167} \frac{P\eps}{R}\,,\quad
\alpha^*_v=\0{4}3 \frac{(25 - 2R/P)F_2(1/R)}{50/R^2-343/R+167} \frac{P\eps}{R}\,,$
\\[1ex]
\rowcolor{LightGray}
& \\[-2.5ex]
\rowcolor{LightGray}
&
$\alpha^*_U=
\0{4}3\0{[(13-2/R)P/R+(2/R-1)(3/R-19)]F_1(R)}{(19R^2 - 43R + 19)(2/R^2 -13/R + 2)}
R\eps,$
$\alpha^*_u=
\0{4}3\0{[(13-2R)R/P+(2R-1)(3R-19)]F_1(1/R)}{(19/R^2 - 43/R + 19)(2R^2 -13R + 2)} \frac{P\eps}{R},$
\\[1ex]
\rowcolor{LightGray}
\multirow{-3}{*}{\FP9}
&
$\alpha^*_V=
\0{4}3\0{[(13-2/R)P/R+(2/R-1)(3/R-19)]F_2(R)}{(19R^2 - 43R + 19)(2/R^2 -13/R + 2)}
R\eps,$
$\alpha^*_v=
\0{4}3\0{[(13-2R)R/P+(2R-1)(3R-19)]F_2(1/R)}{(19/R^2 - 43/R + 19)(2R^2 -13R + 2)} \frac{P\eps}{R}\,. $ \\[1ex]
\bottomrule
\end{tabular}
\caption{Quartic scalar couplings at all weakly interacting fixed points to leading order in $\eps$ following  Tab.~\ref{tFPs} using the auxiliary functions  \eq{F12}.  Same conventions as in Tab.~\ref{tFPeps}. Within the admissible parameter ranges (Tab.~\ref{tPara15},~\ref{tPara69}) we observe vacuum stability.
}
 \label{tsFPeps}
\end{center}
\end{table*}

\subsection{Unique scalar fixed points}
In Tab.~\ref{tsFPeps}, we summarise our results for the quartic scalar couplings at all weakly interacting fixed points to leading order in $\eps$ following  Tab.~\ref{tFPs}, using \eq{PRepsN}. 
We also introduce the auxiliary functions
 \beq\label{F12}
\begin{array}{l}
 F_1(x)=\s014\left(\sqrt{23-4x}-1\right)\,,\\[1.5ex]
 F_2(x)=\s014\left(\sqrt{20-4x+6\sqrt{23-4x}}-2\sqrt{23-4x}\right)
\end{array}
\eeq
which originate from the scalar nullclines. The main result is that vacuum stability together with a physical fixed point in the gauge-Yukawa sector singles out a {\it unique} fixed point in the scalar sector. The scalar fixed points do not offer further parameter constraints other than those already stated in Tabs.~\ref{tPara15} and~\ref{tPara69}. Within the admissible parameter ranges 
we invariably find that the scalar couplings are either strictly irrelevant (at interacting fixed points) or marginally irrelevant (at the Gaussian fixed point).

\section{\bf Ultraviolet  completions}\label{class}
 In this section, we discuss  interacting fixed points and the weak coupling phase structure of minimal models \eq{L} in dependence on matter field multiplicities. Differences  from the viewpoint of their high- and low-energy behaviour are highlighted.

\subsection{Classification}
\begin{figure}[t]
\begin{center}
\includegraphics[scale=0.35]{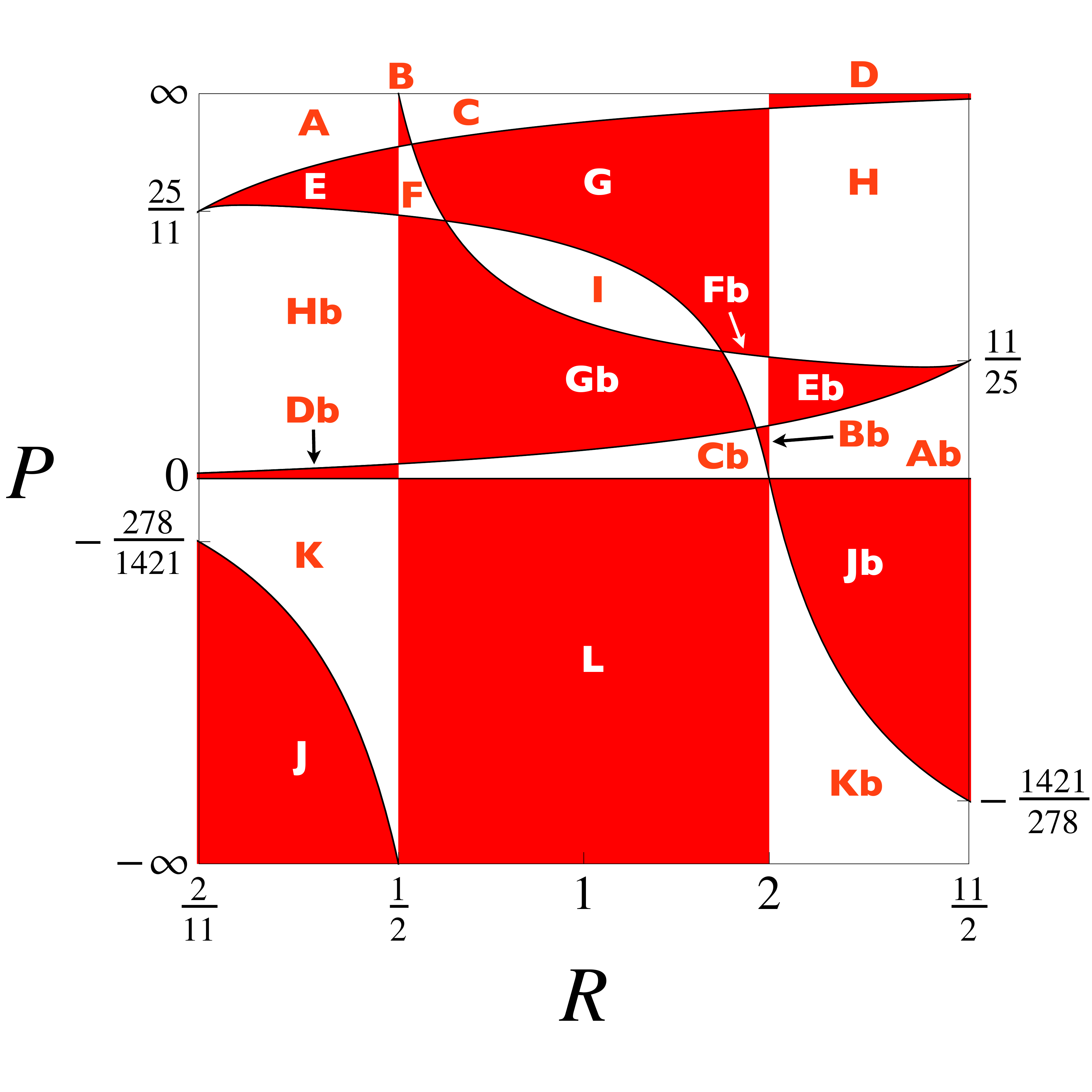}
\vskip-.75cm
\caption{The ``phase space'' of   quantum field theories with fundamental action \eq{L} expressed as a function of field multiplicities and written in terms of $(P,R)$, see \eq{PRepsN}. The  22 different parameter regions are indicated by roman letters.
Theories with parameters in region X are dual to those in region Xb under the exchange of gauge groups following the  map \eq{exchange}. Further details on fixed points and their eigenvalue spectra per parameter region are summarised in Figs.~\ref{pAll_AF},~\ref{pAll_AS} and~\ref{pAll_EFT}.}\label{pFPall} 
\end{center}
\end{figure}

In Figs.~\ref{pFPall},~\ref{pAll_AF},~\ref{pAll_AS} and~\ref{pAll_EFT}   we summarise  results for the qualitatively different types of quantum field theories with Lagrangean \eq{L}  in view of their fixed point structure at weak coupling, together with their behaviour in the deep UV and IR. Theories differ primarily through their matter multiplicities \eq{RST}, which translate to the parameters $(P,R)$ and the sign of $\eps$, \eq{PRepsN}.  As such, the ``phase space'' shown in Fig.~\ref{pFPall} arises as the overlay of Figs.~\ref{pFP23},~\ref{pFP45},~\ref{pFP6},~\ref{pFP78} and~\ref{pFP9}. Distinctive parameter regions are separated from each other by the seven characteristic curves  $P=0\,, X\,, \xalt\,, Y$ or $\yalt$ and $R=\s0{1}{2}$ or $2$. The functions $X(R)\,, \xalt(R)\,, Y(R)$ and $\yalt(R)$ are given explicitly in \eq{P1234}.
Overall, this leads to the 22 distinct regions shown in Fig.~\ref{pFPall} and denoted by capital letters.  
Together with the sign of $\eps$ this leaves us with 44 different cases. Some of these are redundant and related under the exchange of gauge groups, see  \eq{exchange}. In fact, for $P>0$ and for either sign of $\eps$, we find nine fundamentally independent cases corresponding to the parameter regions
\beq\label{++}
\begin{array}{c}
\text{A\,,\ B\,,\ C\,,\ D\,,\ E\,,\ F\,,\ G\,,\ H\,,\ I}
\end{array}
\eeq
given in Fig.~\ref{pFPall}. Theories with parameters in the regime
\beq\label{++b}
\begin{array}{c}
\text{Ab\,,\ Bb\,,\ Cb\,,\ Db\,,\ Eb\,,\ Fb\,,\ Gb\,,\ Hb}\,,
\end{array}
\eeq
are ``dual'' to  those in \eq{++} 
under the exchange of gauge groups (X $\leftrightarrow$ Xb)  and for the same sign of $\eps$, except for the theories within (I, $\eps$), which are ``selfdual'' and mapped onto themselves under \eq{exchange}.
For $P<0$ we find five parameter regions for either sign of $\eps$,
\beq\label{+-}
\text{J\,,\ K\,,\ L\,,\ Kb\,,\ Jb}\,.
\eeq
For these, the manifest ``duality'' under exchange of gauge groups  involves a change of sign for $\eps$ 
with (X, $\eps<0$) being dual to  (Xb, $-\eps>0$) except for the parameter region L  which is selfdual. In total, we end up with $2\times 9+5=23$ fundamentally distinct scenarios underneath the $2\times (9+8+5)=44$ cases tabulated in Figs.~\ref{pAll_AF},~\ref{pAll_AS} and~\ref{pAll_EFT} and discussed more extensively below. 

A comment on the nomenclature: in each row of Figs.~\ref{pAll_AF},~\ref{pAll_AS} and~\ref{pAll_EFT}, we  indicate the parameter region $(P,R)$ as in Fig.~\ref{pFPall} together with the sign of $\eps$ (if required), followed by the set of  fixed points. For each of these, the (marginally) relevant and irrelevant eigenvalues in the gauge-Yukawa sector are indicated by a $-$ and $+$ sign. For the Gaussian fixed point \fp1, the signs relate pairwise to the $SU(N_{\rm C})$  and $SU(N_{\rm c})$  gauge sector, respectively; for all other fixed points eigenvalues are sorted by magnitude. Red shaded slots indicate eigenvalue spectra which uniquely arise due to the semi-simple character of the theory. The column `` UV'' states the UV  fixed point, differentiating between complete asymptotic freedom (AF), asymptotic safety (AS), asymptotic freedom in one  sector without asymptotic safety in the other (pAF), asymptotic safety  in one sector without asymptotic freedom in the other (pAS), or none of the above. The column ``IR'' states the fully attractive IR fixed point (provided it exists), distinguishing the cases where none (0), one (Y) or (y), or  both (Yy)  Yukawa couplings are non-trivial at the fixed point; a hyphen indicates that the IR regime is strongly coupled.

\subsection{Asymptotic freedom}\label{secAF}

\begin{figure}[t]
\begin{center}
\includegraphics[scale=0.24]{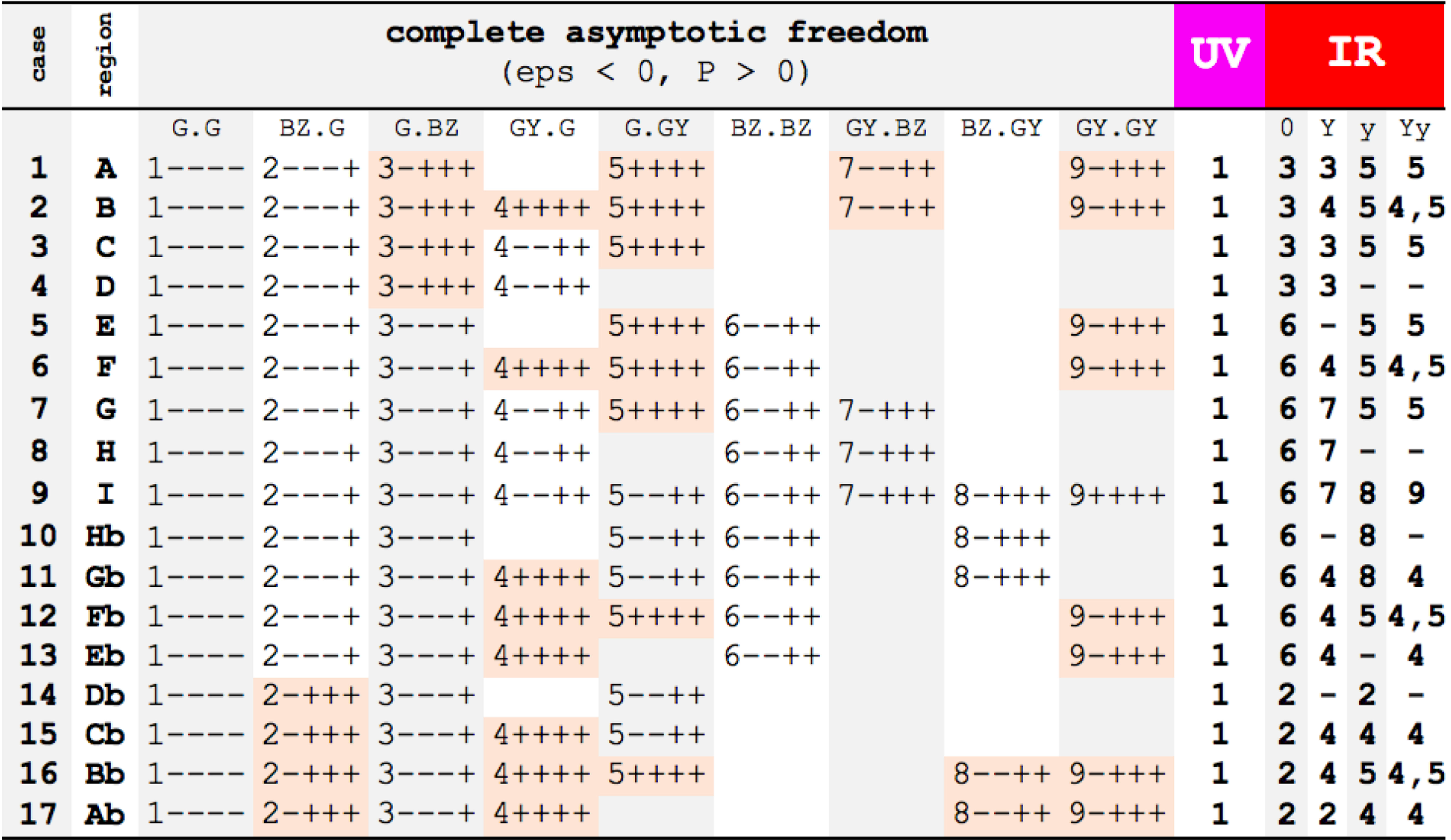}
\caption{
Shown are the fixed points and eigenvalue spectra
of quantum field theories with Lagrangean \eq{L}  for the 17 parameter regions with $\eps<0$ and $P >0$ in Fig.~\ref{pFPall}. Scalar selfinteractions are irrelevant at fixed points. All cases display  complete asymptotic freedom in the UV. Red shaded slots indicate eigenvalue spectra which  arise due to the semi-simple character of the theory.
 In the deep IR, various types of interacting conformal fixed points  are achieved depending on whether both, one, or none of the Yukawa couplings $Y$ and $y$ vanish (from left to right). Regimes with ``strong coupling only'' in the IR are indicated by a hyphen. 
}
\label{pAll_AF} 
\end{center}
\end{figure}

\begin{figure}[t]
\begin{center}
\includegraphics[scale=0.23]{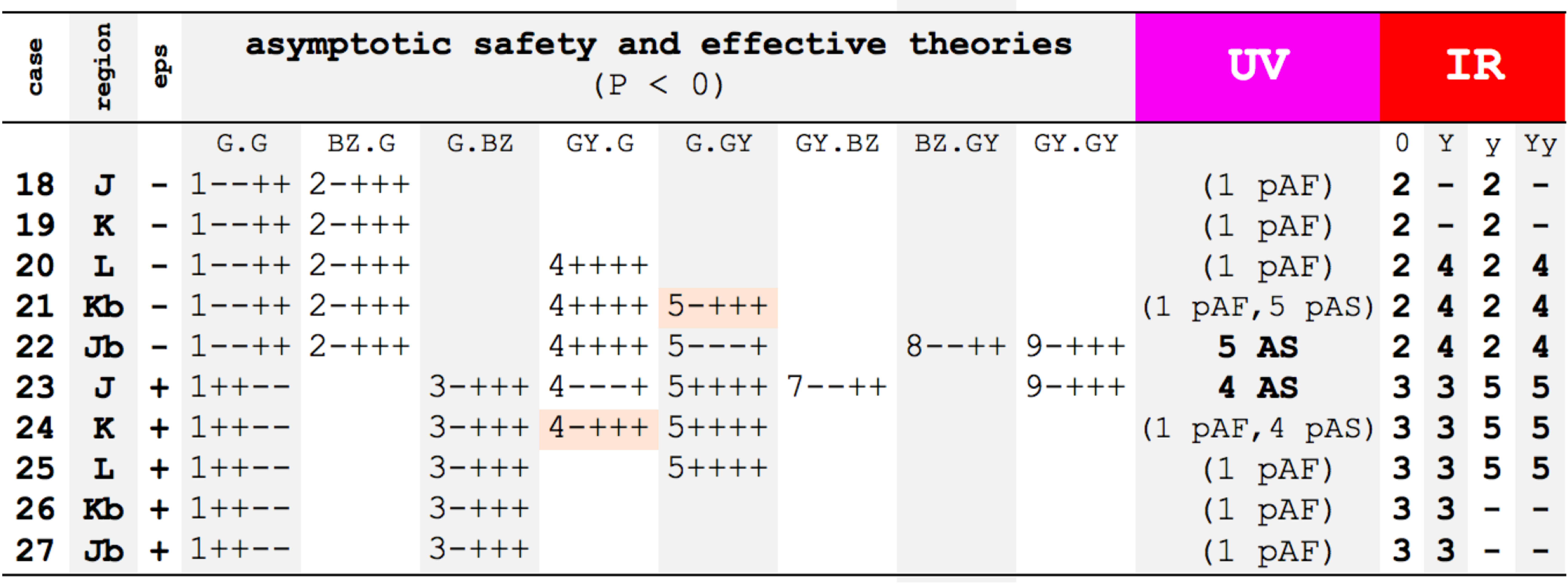}
\caption{
Same as Fig.~\ref{pAll_AF},  covering the 10 parameter regions with $P <0$ of Fig.~\ref{pFPall}. 
 Notice that \fp6 is absent throughout. Exact asymptotic safety  (AS) is realised in the cases 22 and 23. Red shaded slots indicate eigenvalue spectra which  arise due to the semi-simple character of the theory. For the cases $18-21$ and $24-27$, partial asymptotic freedom  (pAF) or partial asymptotic safety (pAS) is observed whereby one gauge sector decouples entirely at all scales. The latter theories are only UV complete in one of the two  gauge sectors and must be viewed as effective rather than fundamental. 
}
\label{pAll_AS} 
\end{center}
\end{figure}

\begin{figure}[t]
\begin{center}
\includegraphics[scale=0.225]{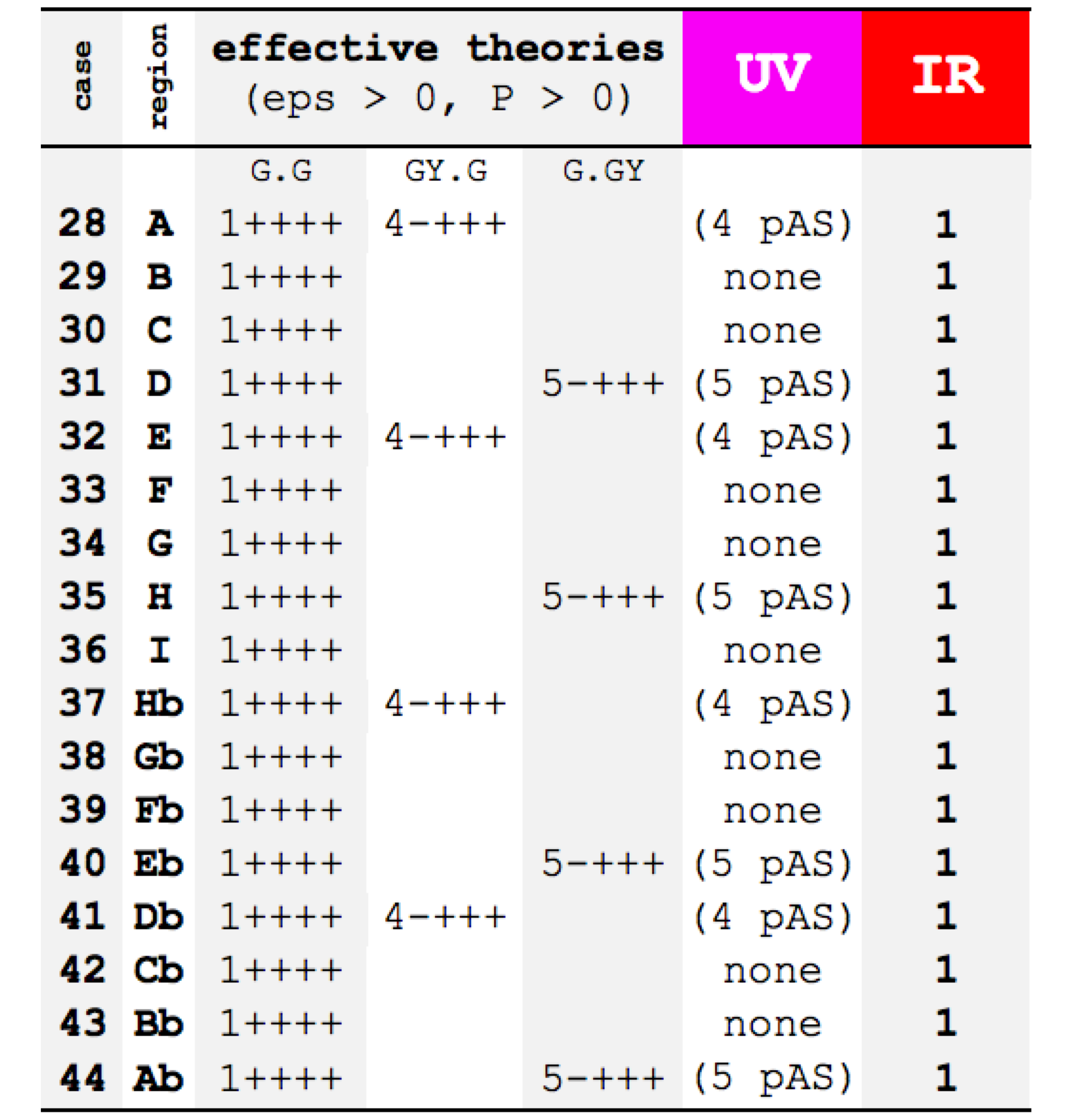}
\caption{Same as Figs~\ref{pAll_AF} and~\ref{pAll_AS},
 covering the 17  parameter regions where
$\eps>0$ and $P>0$ in Fig.~\ref{pFPall}. Asymptotic freedom is absent in both gauge sectors implying that \fp2, \fp3, \fp6, \fp7 and \fp8 cannot arise. 
Partial asymptotic safety (in one gauge sector) is observed in case $28, 31, 32, 35, 37, 40, 41$ and $44$, whereby the other gauge sector remains free at all scales  (pAS). All models must be viewed as effective rather than fundamental theories. All theories become trivial in the IR. 
}
\label{pAll_EFT} 
\end{center}
\vskip-.5cm
\end{figure}

We  discuss main features of the different quantum field theories \eq{L} starting with those where  each gauge sector is asymptotically free from the outset $(P > 0 > \eps)$, corresponding to the cases $1-17$ in Fig.~\ref{pAll_AF}. The Gaussian fixed point \fp1 is always the UV fixed point. Any other weakly interacting fixed point displays a lower number of relevant directions. All weakly interacting fixed points can be reached from the Gaussian.  Another point in common is that all theories are completely asymptotically free meaning that-- besides the gauge and the Yukawa couplings --  all quartics  reach the Gaussian UV fixed point. 

Differences arise as to the set of interacting fixed points, summarised in  Fig.~\ref{pAll_AF}. Overall, theories display between three and eight distinct weakly interacting fixed points. The partial Banks-Zaks fixed points (\fp2, \fp3) are invariably present in all 17 cases. This is a consequence of a general theorem established in \cite{Bond:2016dvk}, which states that the two loop gauge coefficient is strictly positive for any gauge theory in the limit where the one-loop coefficient vanishes. This guarantees the existence of a partial Banks-Zaks fixed point in either gauge sector. At least one of the partial gauge-Yukawa fixed points (\fp4, \fp5) also arises in all cases. Moreover, the fully interacting Banks-Zaks (\fp6) as well as the fully interacting gauge-Yukawa fixed points (\fp7, \fp8, \fp9) are present in many, though not all, cases. All nine distinct fixed points are  available in the ``most symmetric'' parameter region I (case 9).

It is  noteworthy that many theories
display a fully IR attractive ``sink'', invariably given by an IR gauge-Yukawa fixed point in one  (\fp4, \fp5) or both gauge sectors (\fp9). In Fig.~\ref{pFPall}, this happens for matter field multiplicities in the regions A, B, C, E, F, G, I and their duals (cases 1, 2, 3,  5, 6, 7, 9, 11, 12, 13, 15, 16 and 17 of Fig.~\ref{pAll_AF}).
 
 At \fp9, the fully IR attractive fixed point is largely a consequence of IR attractive fixed points in each gauge sector individually. This is not altered qualitatively by the semi-simple nature of the model. As such, a fully IR attractive fixed point  \fp9 also arises in the ``direct product'' limit where the $\psi$ fermions are removed. 
 
 At \fp4 and \fp5, in contrast, the IR sink is  a direct consequence of the semi~simple nature of the theory in that it would be strictly absent as soon as the messenger fermions $\psi$ are removed. Most importantly, the IR gauge~Yukawa fixed point in one gauge sector changes the sign of the effective one loop coefficient in the other, mediated via the $\psi$ fermions. This secondary effect means that one gauge sector becomes IR free dynamically, rather than remaining UV free. Overall, the  fixed point becomes  IR attractive in all canonically marginal couplings (including the quartic couplings). 
In most cases the IR sink is unique except in parameter regions B and F (case 2, 6, 12 and 16) where we find two competing and inequivalent IR sinks (\fp4 versus \fp5). 

Provided that one or both Yukawa couplings take Gaussian values, other fixed points may take over the role of IR ``sinks''. In these settings, one or both of the elementary  ``meson'' fields remain free for all scales and decouple from the outset. Specifically, the IR sink is given by \fp6 provided that  $y=0=Y$ (cases 5 -- 13); by \fp2 or \fp7 provided that $y=0$  (cases 14 or  7 -- 9, respectively); and by \fp3 or \fp8 provided that $Y=0$  (cases 4  or  9 -- 11). We note that \fp6, \fp7 and \fp8  are natural IR sinks, with or without $\psi$ fermions, provided that all Yukawa couplings of those fermions which interact with the Banks-Zaks fixed point(s) vanish. On the other hand, the result that \fp2 and \fp3 may become IR sinks is a strict consequence of the $\psi$ fermions and would not arise otherwise. Once more, one of the gauge sectors becomes IR free owing to the BZ fixed point in the other, an effect which is mediated via the $\psi$ fermions.  
In the presence of non-trivial Yukawa couplings, no fully IR stable fixed point arises for theories with field multiplicities in the parameter  regions D and H  (case 4, 8, 10 and 14). Generically, trajectories will then run towards strong coupling with $e.g.$~confinement or strongly-coupled IR conformality.  Analogous conclusions hold true in settings with fully attractive IR fixed points provided their basins of attraction do not include the Gaussian.

Finally, another interesting feature which is entirely due to the semi~simple nature of the theory are models where \fp9 has a single relevant direction (cases 1, 2, 5, 6, 12, 13, 16 and 17). Whenever this arises, the theory also always displays a fully IR attractive fixed point (\fp4, \fp5, or both).

\subsection{Asymptotic safety}

We now turn to quantum field theories with \eq{L} where asymptotic safety is realised. Asymptotic safety relates to settings where some or all couplings take non-zero values in the UV \cite{Bond:2016dvk}. A prerequisite for this is the absence of asymptotic freedom in at least one of the gauge sectors. We find two such examples  provided $P<0$ (cases 22 and 23 in Fig.~\ref{pAll_AS}), corresponding precisely to settings where one gauge sector is QCD-like whereas the other is QED-like. For these theories, we furthermore find that all other interacting fixed points are also present, except those of the Banks-Zaks type involving the QED-like gauge sector. More specifically, in case 22 the role of the asymptotically safe UV fixed point is now taken by \fp5. The UV critical surface is three-dimensional, in distinction to asymptotically free settings where it is four-dimensional. This reduction, ultimately a consequence of an interacting fixed point in one of the Yukawa couplings, leads to enhanced predictivity of the theory. The Gaussian necessarily becomes a cross-over fixed point with both attractive and repulsive directions, similar to the interacting \fp8. Also, \fp2 and \fp9 are realised with a one-dimensional critical surface. The fully IR attractive \fp4 -- the counterpart of the UV fixed point  \fp5 -- takes the role of an IR ``sink''. 
In the low energy limit, the theory displays free $SU(N_c)$ ``gluons''  in one gauge sector and weakly interacting $SU(N_C)$ ``gluons'' in the other. Moreover, the spectrum includes both free and weakly interacting mesons related to the former and the latter sectors, as well as free and weakly interacting fermions. Qualitatively, a similar result arises in the ``direct product'' limit, showing that the semi-simple nature of \eq{L} is not crucial for this scenario.  

A noteworthy feature of semi-simple theories with asymptotic safety is that they connect an interacting UV fixed point with an interacting IR fixed point. Hence, our models offer  examples of quantum field theories with exact UV and IR conformality, strictly controlled by perturbation theory for all scales. In the massless limit, the phase diagram has trajectories connecting the interacting UV fixed point with the interacting IR fixed point.  Some trajectories may escape towards the regime of strong coupling where the theory is expected to display confinement, possibly infrared conformality. The same picture arises in case 23 after exchange of gauge groups.

No asymptotically safe fixed point arises if both gauge sectors are IR free $(P, \eps > 0)$. This result is in marked contrast to findings in the ``direct product'' limit where models with an interacting UV fixed points exist -- simply because it exists for the simple gauge factors \eq{Simple} and \eq{Simple2}, given suitable matter field multiplicities. We conclude that it is precisely the semi-simple nature of the speific set of theories \eq{L} which disallows asymptotic safety for settings with $P,\eps>0$, see \eq{PRepsN}.

\subsection{Effective field theories}\label{eff}

We now turn to quantum field theories with \eq{L} which are not UV complete semi-simple gauge theories and, as such, must  be seen as effective field theories. We find three different types of these. Firstly, we find models with partial asymptotic freedom (pAF), where one gauge sector remains asymptotically free whereas the other stays infrared free. These models always realise a Banks-Zaks fixed point (as they must), and some also realise an IR gauge-Yukawa fixed point.  When viewed as a fundamental theory, the IR free  sector decouples exactly, for all RG scales, and the theory becomes a simple asymptotically free gauge theory (which is UV complete). The IR-free sector can be interacting  when viewed as an effective theory, very much like the $U(1)_Y$ sector of the Standard Model. This setting requires $P<0$ and is realised in cases 18 -- 21 and 23 -- 27. 

Secondly, we find models with partial asymptotic safety (pAS), where one gauge sector becomes asymptotically safe whereas the other remains free at all scales. All such models display a UV gauge-Yukawa fixed point. 
When viewed as a fundamental theory, these semi-simple gauge theories in fact  reduce to a simple asymptotically safe gauge theory (which is UV complete). The IR-free sector can be interacting when viewed as a non-UV complete effective theory. 
This setting mostly requires $P,\eps>0$ and is realised in cases 28, 31, 32, 35, 37, 40, 41 and 44. Curiously, pAS is also realised in cases 21 and 24 where $P<0$ alongside pAF in the other gauge sector --- such models have two disconnected UV scenarios, where we can choose to have either asymptotic freedom in one sector, or asymptotic safety in the other, in each case with the remaining sector decoupling at all scales. Once more, if both gauge sectors are interacting these models must be viewed as (non-UV complete) effective theories.

Finally, we find models with none of the above. In these settings (cases 29, 30, 33, 34, 36, 38, 39, 42 and 43), both gauge sectors are IR free and no other weakly coupled fixed points are realised, leaving us with no perturbative UV completion. 
In the cases 28 -- 44, the Gaussian acts as in IR ``sink'' for RG trajectories. Along these, the long-distance behaviour is trivial, characterised by free massless non-Abelian gauge fields, quarks, and elementary mesons.

In summary,  the semi-simple gauge Yukawa theories \eq{L} have a well-defined UV limit with either asymptotic freedom or asymptotic safety in 9+1=10 cases out of the 23 fundamentally distinct parameter settings covered in  Fig.~\ref{pFPall}.  The remaining 4+9=13 parameter settings do not offer  a well-defined UV limit at weak coupling. This completes the classification of the models with \eq{L}.

 \begin{figure}[t]
\begin{center}
\vskip-1cm
\includegraphics[scale=.4]{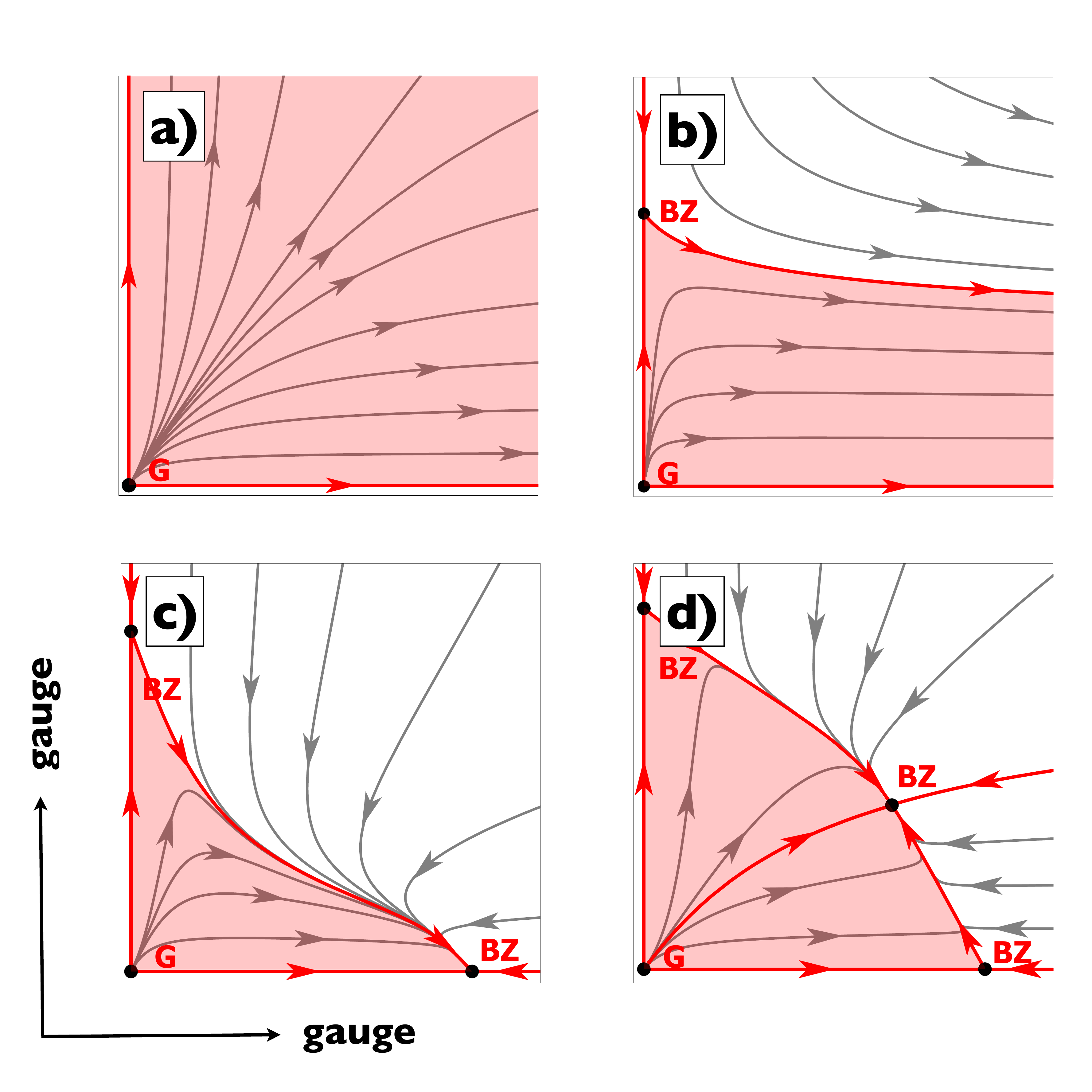}
\caption{Phase diagrams of asymptotically free  semi-simple gauge theories (two gauge groups) coupled to matter without Yukawas, covering  $a)$ asymptotic freedom and the Gaussian (G) without interacting fixed points and trajectories running towards strong coupling and confinement, $b)$ the same, with an additional  Banks-Zaks (BZ) fixed point, $c)$  two BZ fixed points, one of which turned into an IR sink for all trajectories, or $d)$ three  BZ fixed points, the fully interacting one now becoming the IR sink. Axes show the running gauge couplings, fixed points (black) are connected by separatrices (red), and red-shaded areas cover all UV free trajectories with arrows pointing from the UV to the IR. 
}\label{pSemiSimpleBZs} 
\end{center}
\end{figure}

  \begin{figure}[t]
\begin{center}
\vskip-1cm
\includegraphics[scale=.4]{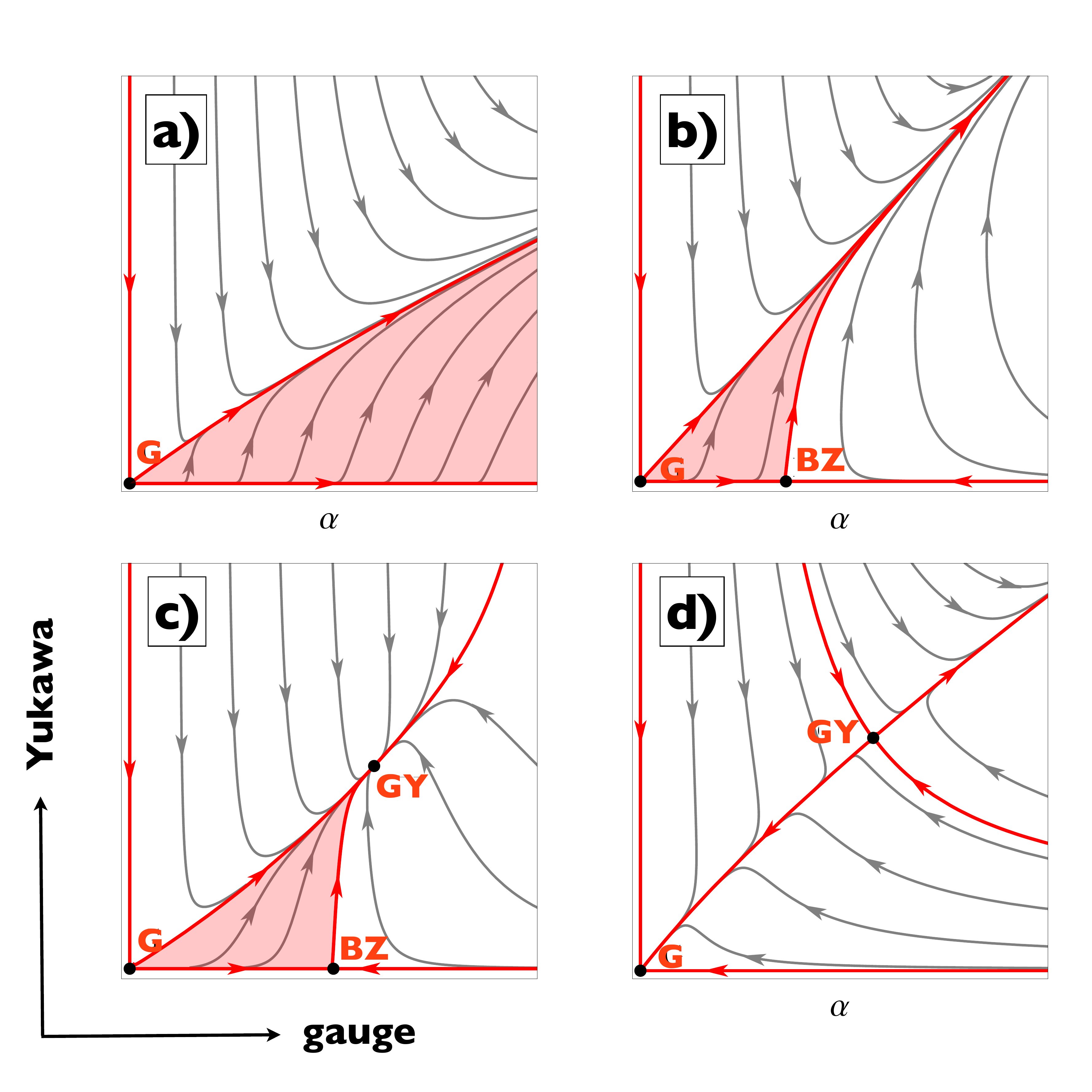}
\caption{Phase diagrams of UV complete and weakly interacting  simple gauge theories coupled to matter with a single Yukawa coupling, covering  $a)$ asymptotic freedom with the Gaussian UV fixed point and no other weakly interacting fixed point, $b)$ asymptotic freedom with a Banks-Zaks (BZ) fixed point, $c)$ asymptotic freedom with a Banks-Zaks and an IR gauge-Yukawa (GY) fixed point, and $d)$ asymptotic safety with an UV gauge-Yukawa fixed point. Axes display the running gauge and Yukawa couplings, fixed points (black) are connected by separatrices (red), and red-shaded areas cover all UV free trajectories with arrows pointing from the UV to the IR \cite{Bond:2016dvk,Bond:2017sem}. Examples are given by  \eq{Simple}, \eq{Simple2} (see main text).
}\label{pSimpleCasesPD} 
\end{center}
\end{figure}

\begin{figure}[t]
\hskip1cm
\includegraphics[scale=.33]{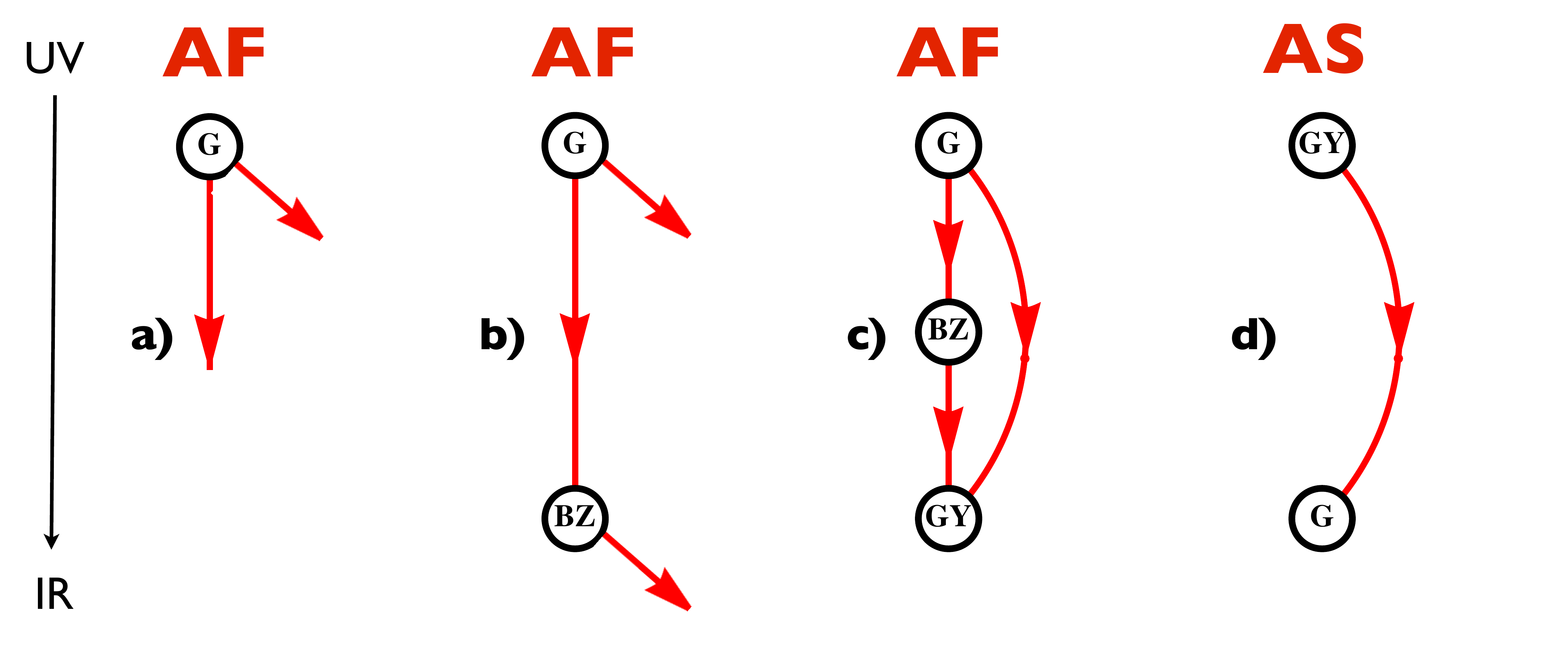}
\caption{``Primitives'' for  phase diagrams of  simple gauge-Yukawa theories with asymptotic freedom (AF) or asymptotic safety (AS), corresponding to the different setting shown in Fig.~\ref{pSimpleCasesPD}. Arrows   point from the UV to the IR and connect the different fixed points. Open arrows point towards strong coupling in the IR. 
The number of outgoing red arrows gives  the dimensionality of the UV critical surface. The separate UV safe trajectory towards strong coupling in case $d)$ is not indicated. Yukawa-induced IR unstable directions in $a,b)$ or gauge Yukawa fixed points in $c,d)$ 
are absent as soon as Yukawa interactions are switched off from the outset. }\label{pPrimitivesSimple} 
\end{figure}

\section{\bf Phase diagrams of gauge theories}\label{PD}
In this section, we discuss the phase diagrams of UV complete theories of the type \eq{L}, particularly in view of theories with asymptotic freedom or asymptotic safety.

\subsection{Semi-simple gauge theories without Yukawas}

We begin with settings where Yukawa couplings are switched off.  In these cases, interacting fixed points can only arise for asymptotically free gauge sectors, and fixed points are of the Banks-Zaks type or products thereof  \cite{Bond:2016dvk,Bond:2017sem}. Qualitatively different cases realised amongst the theories \eq{L} are summarised in
Fig.~\ref{pSemiSimpleBZs} for semi-simple gauge theories with two gauge groups $G_1\times G_2$. Results generalise to more gauge groups in an obvious manner. 

Specifically, Fig.~\ref{pSemiSimpleBZs}$a)$ shows theories with asymptotic freedom but without any BZ fixed points. UV free trajectories  emanate out of the Gaussian fixed point and invariably escape towards strong coupling where the theory is expected to display confinement, or IR conformality.  Similarily, Fig.~\ref{pSemiSimpleBZs}$b)$ shows theories with asymptotic freedom and a BZ fixed point in one of the gauge sectors. The other gauge coupling remains an IR relevant perturbation even at the BZ. Therefore UV free trajectories will again escape towards strong coupling in the IR. 
 
Fig.~\ref{pSemiSimpleBZs}$c)$ shows asymptotic freedom with a BZ fixed point in both  gauge sectors individually. Here, and much unlike  Fig.~\ref{pSemiSimpleBZs}$b)$, one of the BZ fixed points has turned into an exact IR ``sink'', and both BZ fixed points are connected by a separatrix. As we have already noticed in Sect.~\ref{secAF}, the presence of an interacting fixed point in one gauge sector can turn the other gauge sector from UV free to IR free. 
This new type of phenomenon has become possibe owing to the $\psi$ fermions and is once again due to the semi-simple nature of the theory.  Therefore, all UV free trajectories invariably are attracted into the IR sink. In the deep IR, the theory approaches a conformal fixed point with massless and unconfined free and weakly coupled  gluons and quarks. Regimes of strong coupling cannot be reached.
 
Fig.~\ref{pSemiSimpleBZs}$d)$ shows asymptotic freedom with a (partial) BZ fixed point in either gauge sector individually, as well as a fully interacting BZ fixed point. Most notably, all UV free trajectories are attracted by the later, which acts as an IR sink. No trajectories can escape towards strong coupling. The long distance physics is characterised by an interacting conformal field theory with massless weakly coupled gauge fields and fermions. Here, and unlike in Fig.~\ref{pSemiSimpleBZs}$c)$, all fields remain weakly coupled in the IR.

In the scenarios of Fig.~\ref{pSemiSimpleBZs}$a)$ and $b)$ UV free trajectoires run towards strong coupling and confinement in the IR, in one or both gauge sectors. In contrast,  the scenarios in Fig.~\ref{pSemiSimpleBZs}$c)$ and $d)$ show that all UV free trajectories are attracted by an  IR-stable conformal fixed point. These theories remain unconfined and perturbative at all scales.  All four scenarios in Fig.~\ref{pSemiSimpleBZs} are realised for our template of semi-simple gauge theories with Lagrangean \eq{L}. Explicit examples are given for models without Yukawa couplings  $(Y=0=y)$ and for field multiplicities in the parameter regions 
 $a)$ $\eps_1, \eps_2<-75/26$, 
 $b)$ $\eps_1<-75/26$ and $-75/26<\eps_2<0$, or $(\eps_1\leftrightarrow \eps_2)$,
 $c)$  the cases 1 -- 4 and 14 -- 17 of Fig.~\ref{pAll_AF}, and 
 $d)$  the cases 5 -- 13 of Fig.~\ref{pAll_AF}.

\subsection{Simple gauge theories with Yukawas}
We  continue the discussion of phase diagrams with simple gauge theories with gauge group $G$ and a single Yukawa coupling. Four distinct cases can arise  \cite{Bond:2016dvk,Bond:2017sem}, 
summarised in Fig.~\ref{pSimpleCasesPD}. For asymptotically free settings, the theory either shows $a)$ only the Gaussian UV fixed point, $b)$ the Gaussian together with the Banks-Zaks,  or $c)$ the Gaussian together with the Banks-Zaks and an IR gauge-Yukawa fixed point. Simple gauge theories can also become asymptotically safe, in which case $d)$ a UV gauge-Yukawa fixed point arises. Trajectories are directed towards the IR. The red-shaded areas indicate the set of UV complete trajectories emanating out of the UV fixed point. We genuinely observe a two-dimensional area of trajectories for asymptotically free settings, which is reduced to a one-dimensional set in the asymptotically safe scenario. The IR regime is  characterised by either strong interactions and confinement such as in Fig.~\ref{pSimpleCasesPD}$a,b,d)$, or by an interacting conformal field theory with weakly coupled gluons and fermions alongside free or interacting scalar mesons ---corresponding to the BZ fixed points in Fig.~\ref{pSimpleCasesPD}$b)$ and $c)$, or the IR GY fixed point in Fig.~\ref{pSimpleCasesPD}$c)$, respectively---,  or by Gaussian scaling, Fig.~\ref{pSimpleCasesPD}$d)$. 

All four scenarios in Fig.~\ref{pSimpleCasesPD} are realised for simple gauge theories with \eq{Simple} corresponding to the parameter regions 
 $a)$ $\eps_1<-75/26$, $b)$ $-75/26<\eps_1<0$ and $R>\s012$, $c)$ $-75/26<\eps_1<0$ and $R<\s012$, or $d)$  $\eps_1>0$ and $R<\s012$, respectively, with $R$ additionally bounded by \eq{Rrelaxed}.

\begin{figure}[t]
\begin{center}
\includegraphics[scale=.45]{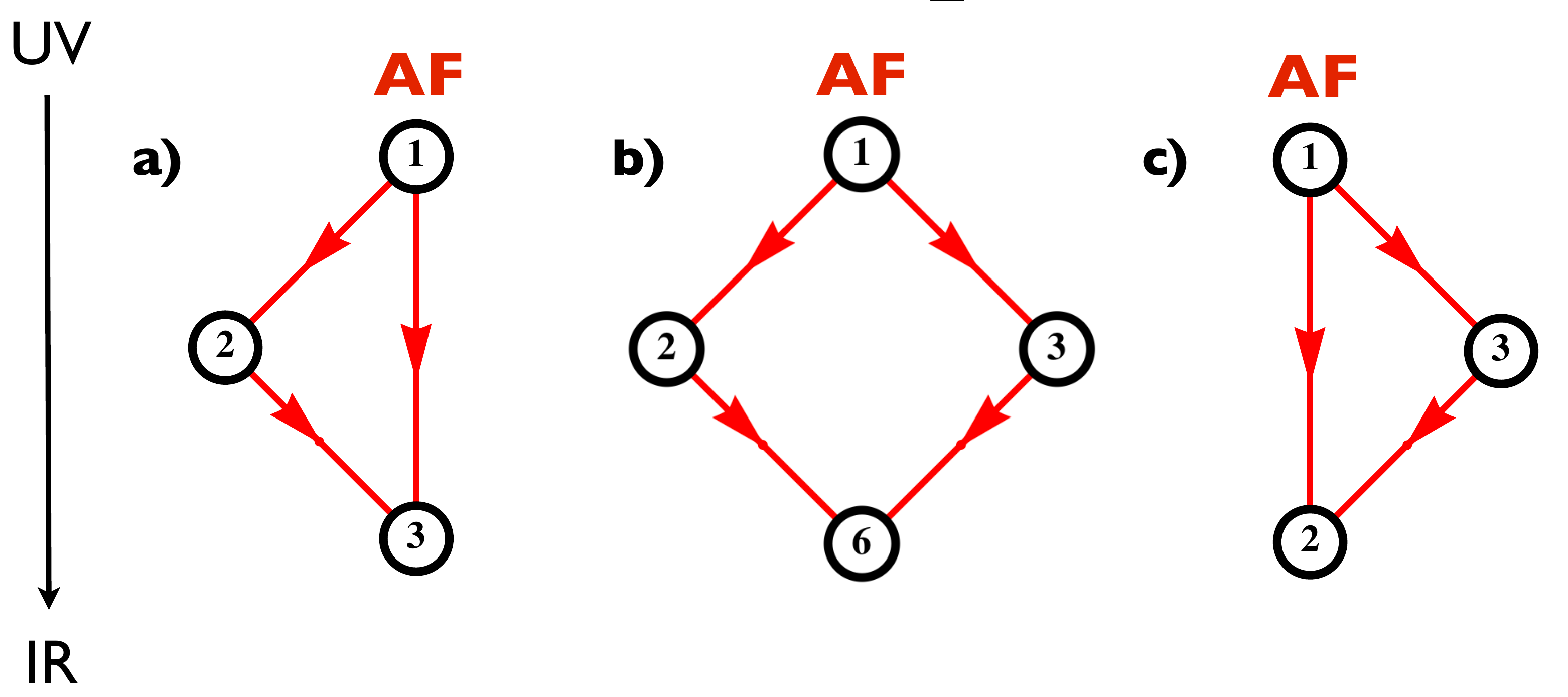}
\caption{
Schematic phase diagram for asymptotically free semi-simple  gauge theories \eq{L}  with Banks-Zaks type fixed points without Yukawas and exact IR conformality. Field multiplicities correspond to the cases  $a)$  1 -- 4, $b)$ 5 -- 13, and $c)$ 14 -- 17 of Fig.~\ref{pAll_AF}, respectively, with scalars decoupled.
RG flows point from the UV to the IR (top to bottom). At each fixed point, the dimensionality of the UV critical surface is  given by the number of outgoing red arrows. All UV free trajectories terminate at  \fp2, \fp6 and \fp3, respectively, which act as fully attractive IR ``sinks''.
The topology of the phase diagram $b)$ is the ``square'' of Fig.~\ref{pPrimitivesSimple}$b)$, representing Fig.~\ref{pSemiSimpleBZs}$d)$. The phase diagrams $a)$ and $c)$, representing Fig.~\ref{pSemiSimpleBZs}$c)$, cannot be constructed from the primitives in Fig.~\ref{pPrimitivesSimple}.
}\label{pPD_BZ_IR} 
\end{center}
\end{figure}

An economic way to display phase diagrams for semi-simple theories with or without Yukawas is achieved by introducing a schematic diagrammatic language, see Fig.~\ref{pPrimitivesSimple}. Each of the four  basic phase diagrams in Fig.~\ref{pSimpleCasesPD} are represented by a ``primitive'' diagram,  Fig.~\ref{pPrimitivesSimple}, where full dots indicate (free or interacting) fixed points, red arrows indicate the outgoing trajectories, and RG flows schematically run ``top-down'' from the UV to the IR. Also, at each fixed point the number of outgoing arrows indicates the dimensionality of the fixed point's ``UV critical surface''. Fixed points are connected by separatrices. We use straight lines to indicate separatrices involving the BZ fixed point, curved lines to indicate separatrices connecting  GY fixed points with the Gaussian, and open-ended lines to denote RG trajectories running towards strong coupling without reaching any weakly coupled fixed points. 

Specifically, in case $a)$,  a two-dimensional array of RG flows are running out of the Gaussian UV fixed point towards strong coupling, with no weakly interacting fixed points. In case $b)$, we additionally observe a Banks-Zaks fixed point. It is connected with the Gaussian by a separatrix shown in red. Arrows invariably point towards the IR. Yukawa couplings act as an unstable direction at both fixed points. 
In case $c)$, we additionally observe a gauge-Yukawa fixed point besides the Gaussian and the BZ. All three fixed points are connected by separatrices. Note that two lines emanate from the Gaussian, reflecting that the UV critical surface is two dimensional. The GY fixed point arises as an IR sink, which attracts all UV-free trajectories emanating out of the Gaussian. 
In case $d)$, the model is asymptotically safe and the GY fixed point has become the interacting UV fixed point. A Banks-Zaks fixed point can no longer arise \cite{Bond:2016dvk}. The theory has a one-dimensional UV critical surface connecting the  GY fixed point with the IR Gaussian fixed point via a separatrix. A second UV safe trajectory which leaves the GY fixed point towards strong coupling is not depicted. Finally, we note that the Yukawa-induced IR unstable directions in $a)$ and $b)$ or gauge Yukawa fixed points in $c)$ and $d)$ are absent as soon as Yukawa interactions are switched off from the outset.

\begin{figure}[t]
\begin{center}
\includegraphics[scale=.35]{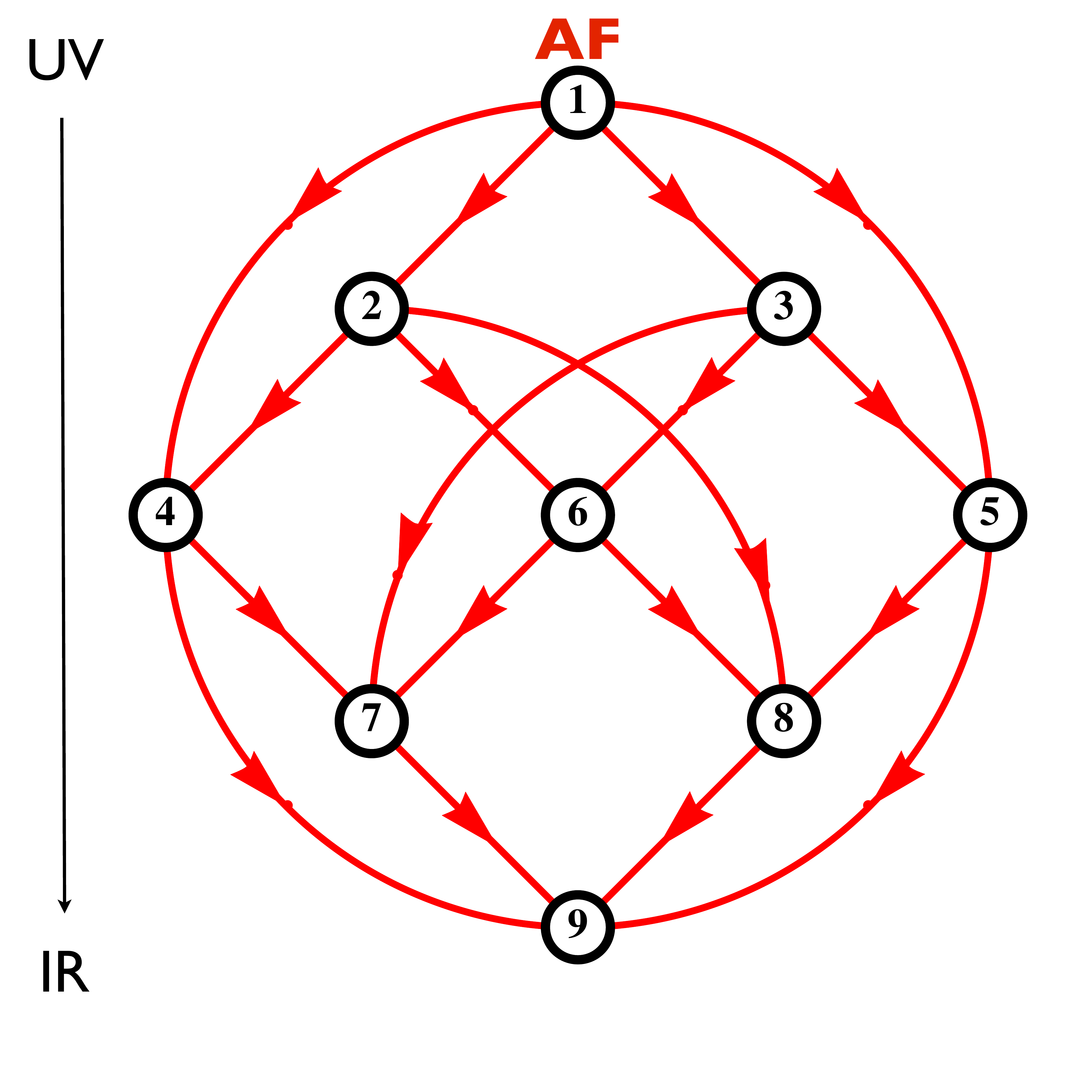}
\caption{
Asymptotic freedom and 
schematic phase diagram for semi simple  gauge-Yukawa theories with field multiplicities as in
case 9 of Fig.~\ref{pAll_AF}.
RG flows point from the UV to the IR (top to bottom). Besides the Gaussian UV fixed point (FP${}_1$), the theory displays all eight weakly interacting fixed points, see Tab.~\ref{tFPs}.
At each fixed point, the dimensionality of the UV critical surface is  given by the number of outgoing red arrows.  FP${}_9$ is fully attractive and acts as an IR ``sink''. The topology of the phase diagram is the ``square'' of Figs.~\ref{pSimpleCasesPD},\ref{pPrimitivesSimple}$c)$; see main text.
}\label{pPD_AF} 
\end{center}
\end{figure}

\begin{figure}[t]
\begin{center}
\includegraphics[scale=.35]{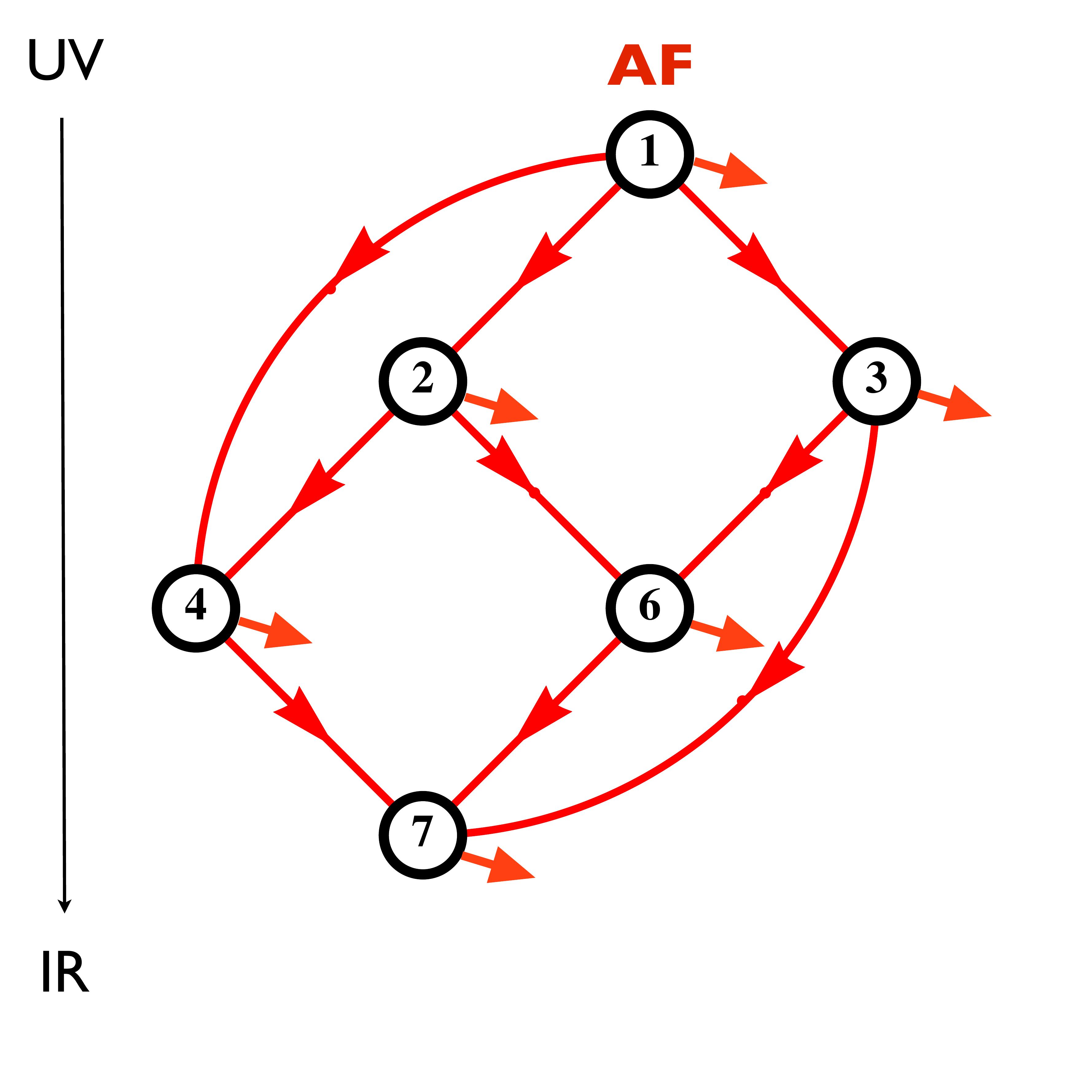}
\caption{
Asymptotic freedom and schematic  phase diagrams for semi-simple  gauge-Yukawa theories 
with field multiplicities as in case 8 of Fig.~\ref{pAll_AF}.
Flows point from the UV to the IR (top to bottom). The theories display   five weakly interacting fixed points besides the Gaussian UV fixed point (FP${}_1$). The unavailability of \fp5, \fp8 and \fp9 implies that some trajectories escape towards
strong coupling (short arrows), and none of the fixed points acts as a complete IR attractor.
The topology of the phase diagram is the ``direct product'' of Fig.~\ref{pSimpleCasesPD},~\ref{pPrimitivesSimple}$c)$ with  Fig.~\ref{pSimpleCasesPD},~\ref{pPrimitivesSimple}$b)$; see main text. 
The IR unstable direction is removed provided that the Yukawa coupling $y\equiv 0$, in which case the singlet mesons $h$ decouple.
}\label{pPD_AF_pt2} 
\end{center}
\end{figure}

\subsection{Semi-simple gauge theories with asymptotic freedom}
 
We consider phase diagrams for semi-simple theories \eq{L}  with complete asymptotic freedom, exemplified by all models in Fig.~\ref{pAll_AF}.  When Yukawa couplings are absent, the meson-like scalar degrees of freedom remain free  at all scales and decouple from the theory. In the regime with asymptotic freedom solely Banks-Zaks fixed points can arise in the IR. Fig.~\ref{pSemiSimpleBZs}$d)$ and  Fig.~\ref{pPD_BZ_IR}$b)$ shows settings where all Banks-Zaks fixed points are present, corresponding to 
the cases 5 -- 13 of Fig.~\ref{pAll_AF}. 
RG flows point from the UV to the IR (top to bottom) and connect the Gaussian UV fixed point (FP${}_1$) with either of the partially (\fp2 and \fp3) and the fully interacting (\fp6) Banks-Zaks fixed points. The latter  is fully attractive and acts as an IR sink. The topology of the phase diagram is the ``square'' of Figs.~\ref{pSimpleCasesPD},\ref{pPrimitivesSimple}$b)$. In the deep IR the theory is unconfined yet weakly interacting, and the elementary gauge fields $A, a$ and fermions $Q, q$ and $\psi$ appear as massless particles at the IR conformal fixed point.
The phase diagrams in Figs.~\ref{pPD_BZ_IR}$a)$ and $c)$ 
cannot be constructed out of the simple primitives, Fig.~\ref{pPrimitivesSimple}. The reason for this is that the eigenvalue spectrum at one of the fixed points deviates from the ``direct product'' spectrum due to interactions.

Next we include Yukawa interactions. 
We have already concluded from Fig.~\ref{pAll_AF}  that the eigenvalue spectrum in the cases 8, 9 and 10 agrees qualitatively, for all fixed points, with the eigenvalue spectrum in the  corresponding ``direct product'' limit. In these settings, we may then use the primitives in Fig.~\ref{pPrimitivesSimple} to find the semi-simple phase diagrams. 
We consider the case where the parameters  \eq{PRepsN} take values within the range I of Fig.~\ref{pFPall} and for  $\eps<0$, corresponding to case 9 of Fig.~\ref{pAll_AF}. This family of theories includes the ``symmetric'' setup $(R,P)=(1,1)$ where symmetry under the exchange of gauge groups is manifest. The UV fixed point is given by the Gaussian  (FP${}_1$), and  the UV critical surface at the Gaussian is four-dimensional, owing to the marginal UV relevancy of the two gauge and the two Yukawa couplings. All scalar couplings are irrelevant in the UV and can be expressed in terms of the gauge and the Yukawa couplings along UV-free trajectories. 
Moreover, each gauge sector displays the Banks-Zaks and a gauge-Yukawa fixed point individually, and all nine fixed points are realised in the full theory. 

Since the sign pattern of the eigenvalue spectra at all fixed points
is equivalent to the ``direct product'' limit, the topology of the semi-simple phase diagram is the ``square'' of Fig.~\ref{pPrimitivesSimple}$c)$ -- shown in Fig.~\ref{pPD_AF}. Fixed points are connected by separatrices (red lines), and arrows always point towards the IR.
From top to bottom, the fixed points FP${}_1$ (FP${}_{2,3})$ [FP${}_{4,5,6}$] (FP${}_{7,8})$ and FP${}_9$ have a 4 (3) [2] (1) and 0-dimensional UV critical surface, respectively, corresponding to the number of outgoing red arrows.   FP${}_9$ acts as an IR attractor for all trajectories within its basin of attraction. Consequently, the elementary quarks and gluons are not confined and  the theory corresponds to a conformal field theory of weakly interacting massless gluons, fermions and mesons in the deep IR.  For certain fine-tuned settings, the IR limit would, instead, correspond to one of the other interacting fixed points FP${}_{2}$ -- FP${}_{8}$, relating to different conformal field theories.  Also, while all other fixed points can be reached from the Gaussian FP${}_1$ (whose UV critical surface has the largest dimensionality), it is not true in general that a fixed point with a smaller UV critical dimension can be reached from a fixed point with a larger one. Fixed points are also not connected ``horizontally''.

As a further example we consider a less symmetrical setting given by models with \eq{PRepsN} in the parameter range H (or Hb) of Fig.~\ref{pFPall}, and for  $\eps<0$. In these theories, only one of the two gauge sectors can achieve a gauge-Yukawa fixed point. Consequently,  six different types of fixed points are realised. The sign pattern of the eigenvalue spectrum (cases 8 or 10, Fig.~\ref{pAll_AF})  ensures 
that the topology of the semi-simple phase diagram  obtains as the direct product of Fig.~\ref{pPrimitivesSimple}$b)$ with Fig.~\ref{pPrimitivesSimple}$c)$, shown in Fig.~\ref{pPD_AF_pt2}. From top to bottom, the fixed points FP${}_1$ (FP${}_{2,3})$ [FP${}_{4,6}$] and FP${}_7$ have a 4 (3) [2] and 1-dimensional UV critical surface, respectively.
Fixed points are connected by separatrices. The absence of \fp5, \fp8 and \fp9 implies that some trajectories escape towards
strong coupling, indicated by short arrows, from each of the fixed points. The unstable direction relates to the Yukawa coupling $y$ in \eq{L}. Provided it is switched off, \fp7 would become the fully attractive  IR ``sink''. In this case,  the elementary mesons $h$ are spectators and remain free at all scales. Also, the elementary quarks and gluons remain unconfined. In the deep IR, the theory corresponds to a conformal field theory of weakly interacting massless gluons $A$, fermions $Q, \psi$ and mesons $H$, together with free and massless gluons $a$,  fermions $q$ and  mesons $h$, see Tab.~\ref{tCharges}.  For certain fine-tuned settings, the IR limit would, instead, correspond to one of the other interacting fixed points FP${}_{2}$ -- FP${}_{8}$, relating to different conformal field theories.

The phase diagrams of asymptotically free theories in the cases 1 -- 7 and 11 -- 17 of Fig.~\ref{pAll_AF} cannot be constructed out of the simple primitives, Fig.~\ref{pPrimitivesSimple}. The reason for this is that their eigenvalue spectrum at some of the interacting fixed points deviates from the ``direct product'' spectrum. Once again this effect is due to the semi-simple nature of the theory. A more detailed study of these cases is left for future work.

\begin{figure}[t]
\includegraphics[scale=0.35]{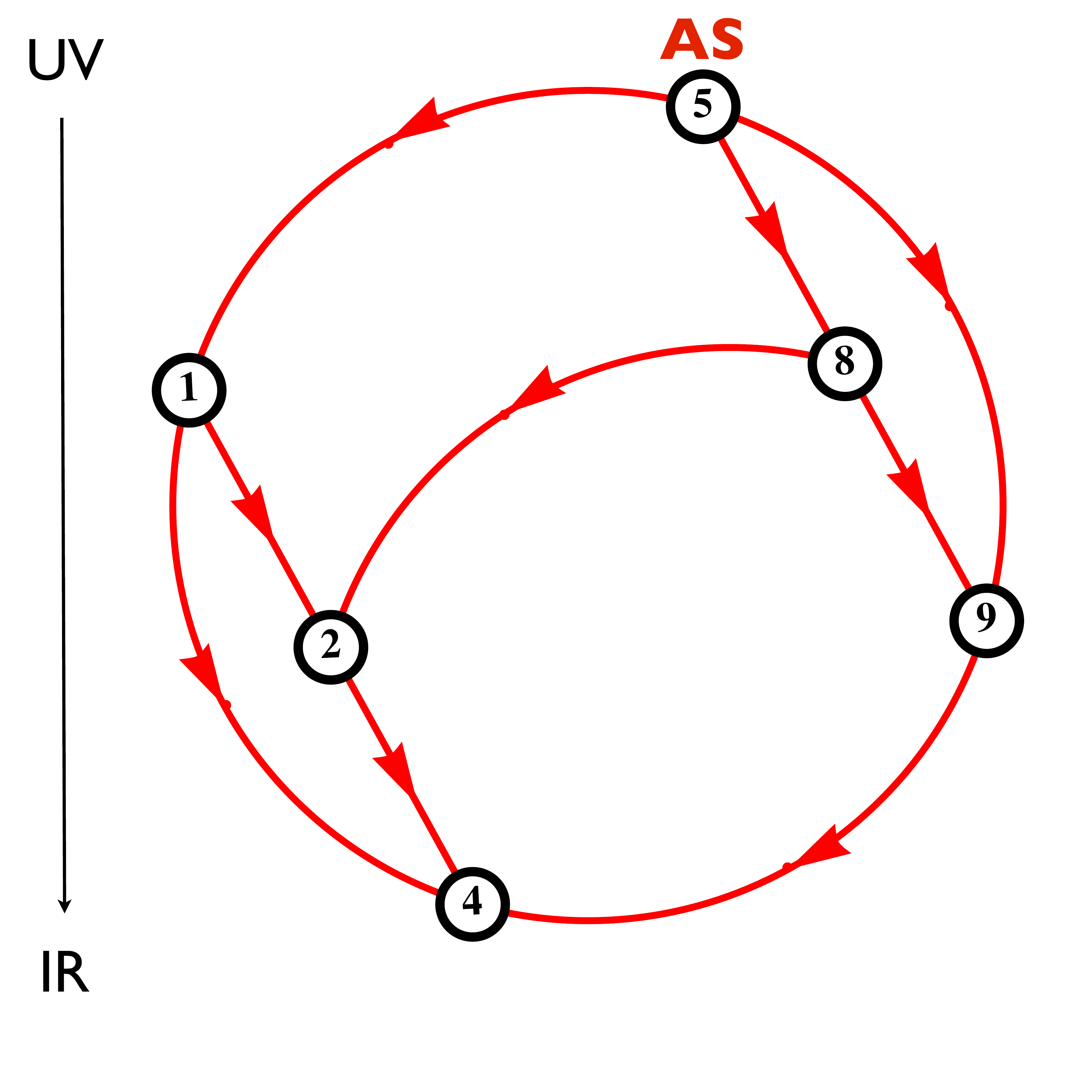}
\begin{center}
\caption{
Asymptotic safety and schematic phase diagram of semi simple  gauge-Yukawa theories with field multiplicities as in case 22 of Fig.~\ref{pAll_AS}).
Besides the partially interacting UV fixed point (FP${}_5$), the theory displays five weakly interacting fixed points. The Gaussian (FP${}_1$) takes the role of a crossover fixed point and 
FP${}_4$ takes the role of an IR sink. The topology of the phase diagram is the ``direct product'' of Fig.~\ref{pSimpleCasesPD},~\ref{pPrimitivesSimple}$c)$ with  Fig.~\ref{pSimpleCasesPD},~\ref{pPrimitivesSimple}$d)$; see main text.}
\label{pPD_AS}
\end{center}
\end{figure}

\subsection{Semi-simple gauge theories with asymptotic safety}

We finally turn to the phase diagram of semi-simple gauge theories with exact asymptotic safety.  From Figs.~\ref{pAll_AS} and~\ref{pAll_EFT} we conclude that asymptotic safety arises through a partially interacting UV fixed point where one gauge sector is interacting whereas the other gauge sector is free. This is achieved for matter field multiplicities \eq{PRepsN} taking values within the range J or Jb  of Fig.~\ref{pFPall}, corresponding to cases 22 or 23 of Fig.~\ref{pAll_AS}. Once more, the eigenvalue spectra at all fixed points are  equivalent to the ones in the direct product limit, implying that the phase diagram arises as the direct product 
of the corresponding ``simple factors''  Fig.~\ref{pSimpleCasesPD},~\ref{pPrimitivesSimple}$c)$ and  Fig.~\ref{pSimpleCasesPD},~\ref{pPrimitivesSimple}$d)$.

Fig.~\ref{pPD_AS} shows the schematic phase diagram for case 22,
where the asymptotically safe UV fixed point FP${}_5$ is of the {G$\,\cdot\,$GY}~type (see Tab.~\ref{tFPs} and~\ref{tFPeps}). Unlike the cases with asymptotic freedom, here, the UV hypercritical surface  is  three rather than four dimensional. The reason for this is that one of the Yukawa couplings is taking an interacting UV fixed point. 
At each fixed point, the number of outgoing directions indicate the dimensionality of the fixed point's critical hypersurface. From top to bottom, the fixed points FP${}_5$ (FP${}_{1,8})$ [FP${}_{2,9}$] and FP${}_4$ have a 3 (2) [1] and 0-dimensional UV critical surface.
UV finite trajectories connect  FP${}_5$ via  intermediate cross-over fixed points with the fully IR attractive fixed point FP${}_4$, which is of the {GY$\,\cdot\,$G} type. At weak coupling, all UV-IR connecting trajectories  proceed either via the Gaussian FP${}_1$ ({G$\,\cdot\,$G}) and FP${}_2$ ({BZ$\,\cdot\,$G}), or via  FP${}_8$ ({BZ$\,\cdot\,$GY}) and FP${}_2$ or FP${}_9$ ({GY$\,\cdot\,$GY}). The Gaussian fixed point is IR free in one of the gauge couplings meaning that is necessarily arises as a cross-over fixed point. There are no trajectories connecting the fixed points  FP${}_1$ with FP${}_9$ because the sole relevant direction at the latter is an irrelevant direction at the former. \fp4 acts as an IR ``sink'' for RG trajectories.  While all other fixed points can be reached from the interacting UV fixed point  FP${}_5$ (whose UV critical surface has the largest dimensionality), it is not true in general that a fixed point with a smaller UV critical dimension can be reached from a fixed point with a larger one (e.g.~\fp9 cannot be reached from \fp1). Fixed points are also not connected ``horizontally''.

An intriguing novelty of our models with asymptotic safety is that both the deep UV and the deep IR limits are characterised by weakly interacting conformal field theories. For example, in the deep UV the  theories of case 22  correspond to  conformal field theories of weakly interacting massless gluons $a$, fermions $q, \psi$ and mesons $h$, together with free and massless gluons $A$,  fermions $Q$ and  mesons $H$. Along the UV -- IR transition,  the fields $(A,Q,H)$ and $(a,q,h)$ effectively ``interchange'' their roles, ultimately approaching  conformal field theories of weakly interacting massless gluons $A$, fermions $Q, \psi$, and mesons $H$, together with free and massless gluons $a$,  fermions $q$ and  mesons $h$ in the IR.  Hence, one may say that IR conformality in the $SU(N_{\rm c})$  gauge sector arises  from UV conformality in the $SU(N_{\rm C})$  gauge sector through a  ``see-saw'' mechanism transmitted via the $\psi$  fermions, $i.e.$ the only fields which are interacting at all scales including  the UV and the IR limits. For certain fine-tuned settings, the IR limit would, instead, correspond to one of the other interacting fixed points FP${}_{1}$, \fp2, FP${}_{8}$ or \fp9, relating to different conformal field theories. Also, for certain UV parameters, theories may escape towards strong coupling in the IR.

\subsection{Mass deformations and phase transitions}
In the vicinity of fixed points phase transitions between different phases arise once mass terms are switched on. At weak coupling mass anomalous dimensions are perturbatively small (Sec.~\ref{anomalous}). The running of scalar or fermion mass terms, once switched on, will then be dominated by their canonical mass dimensions -- modulo small quantum corrections. Consequently, mass terms add additional relevant directions at all fixed points (e.g.~Figs.~\ref{pPD_AF} -- \ref{pPD_AS}). Each of the eight interacting UV fixed points relates to a quantum phase transition between phases with and without spontaneous breaking of symmetry where the vacuum expectation value of the scalar fields serves as an order parameter. 
In particular, fixed points which act as IR sinks for the canonically marginal interactions (such as \fp9 in  Fig.~\ref{pPD_AF} and \fp4 in Fig.~\ref{pPD_AS}) develop new unstable directions driven by the mass. Scalar fields may or may not develop vacuum expectation values leading to symmetric and symmetry broken phases, respectively. Also, fermions may  acquire masses spontaneously. Thereby a variety of different phases may arise, connected by first and higher order quantum phase transitions.  Close to interacting fixed points, phase transitions are continuous and, in some cases, of the Wilson-Fisher type with a single relevant parameter.  We leave a more detailed investigation of  phase transitions for a future study.

\section{\bf Discussion}\label{disc}
In this section, we address  further aspects of interacting fixed points covering
universality and operator ordering, triviality bounds, perturbativity in and beyond the Veneziano limit, conformal symmetry, and conformal windows.

\subsection{Gap, universality, and operator ordering}

At partially or fully interacting fixed points, the degeneracy of the nine classically marginal couplings \eq{couplings}, \eq{w} is partly or fully lifted.  We have computed scaling exponents to the leading non-trivial order in $\eps$. Interacting fixed points have  non-trivial exponents of order $\sim \eps$, except if a gauge coupling is involved in which case one of  the exponents is parametrically smaller $\sim\eps^2$. Hence, the eigenvalue spectrum opens up $\sim \eps$ because eigenvalues of order $\eps$ are invariably present at any of the interacting fixed points. It is convenient to denote the difference between the smallest negative eigenvalue and the smallest positive eigenvalue as the ``gap'' in the eigenvalue spectrum, which serves as an indicator for interaction strength \cite{Falls:2013bv,Falls:2014tra}. Simple  $SU(N)$ gauge theories in the Veneziano limit such as \eq{Simple} display a gap of order $\sim \eps$ $(\eps^2)$ at the Banks-Zaks or the UV gauge Yukawa (IR gauge Yukawa) fixed point, respectively \cite{Litim:2014uca}.  
In semi-simple theories, and depending on the specifics of the fixed point, we again find that the gap is either of order $\eps$ or of order $\eps^2$. (The gap trivially vanishes if one of the gauge sectors is asymptotically free and takes Gaussian values.) The gap still depends on the remaining free parameters $(P,R)$. 

Also, all results for fixed points and scaling exponents are universal and independent of the RG scheme, although we have used a specific scheme (MS bar) throughout. This is obviously correct for dimensionless couplings at one loop where divergences are logarithmic. We have checked that it also holds at two loop level both for the gauge sectors, and for the Yukawa contributions to the running of the gauge coupling(s) \cite{Litim:2014uca}. The field strengths and the Yukawa couplings are marginally relevant operators at asymptotically free Gaussian UV fixed points (case 1 -- 17 of Fig.~\ref{pAll_AF}). At asymptotically safe UV fixed points, one of the field strengths becomes relevant and the corresponding Yukawa coupling irrelevant  (case 22, 23 of Fig.~\ref{pAll_AS}). There is no UV fixed point where both gauge sectors remain interacting. The scalar selfinteractions are (marginally) irrelevant at any fixed point.

\subsection{Elementary gauge fields and scalars}

Triviality bounds relate to perturbative UV Landau poles of infrared free interactions. They limit the predictivity of theories to a maximal UV extension \cite{Callaway:1988ya}.  
For theories with action \eq{L}, perturbative UV Landau poles can arise for gauge couplings
in the absence of asymptotic freedom or asymptotic safety.
Examples for this are given in  cases 18 -- 21 and 24 -- 27 of Fig.~\ref{pAll_AS} where one gauge sector is IR free, as well as in cases 28 -- 44 of Fig.~\ref{pAll_EFT}  
where both gauge sectors are IR free. In these cases the theories can at best be treated as effective rather than fundamental  (see Sect.~\ref{eff}).  Conversely, triviality in gauge sectors is trivially avoided in settings with asymptotic freedom (such as in cases 1 -- 17), and non-trivially in settings with asymptotic safety (case 22 and 23). In the latter cases, the loss of asymptotic freedom is compensated through an interacting fixed point in the Yukawa and scalar couplings, which enabled a fixed point for the gauge coupling \cite{Litim:2014uca}. We stress that scalar fields and Yukawa interactions play a key role. Without them, triviality of any QED-like gauge theories cannot be avoided \cite{Bond:2016dvk,Bond:2017sem}.

Triviality also  relates to the  difficulty of defining elementary self-interacting scalar quantum fields in four dimensions  \cite{Wilson:1973jj,Luscher:1987ek,Hasenfratz:1987eh}. It is interesting to notice that the quartic scalar couplings always take a unique physical  fixed points as soon as the gauge and Yukawa coupling take weakly coupled fixed points. Hence, in theories with \eq{L} scalar fields can be viewed as elementary and  triviality is evaded in all settings with asymptotic freedom and asymptotic safety.
In either case gauge fields play an important role, albeit for different reasons \cite{Litim:2014uca}. For gauge interactions with asymptotic freedom, the running of gauge couplings dictates  the running for Yukawa and scalar couplings, and conditions for complete asymptotic freedom have been derived  \cite{Coleman:1973sx} which ensure that gauge theories coupled to matter reach the free UV fixed point  \cite{Chang:1974bv}. For theories with asymptotic safety, scalars are required to help generate a combined fixed point in the gauge, Yukawa, and quartic scalar couplings. This leads invariably to a ``reduction of couplings'' and enhanced predictivity over models with asymptotic freedom through a reduced UV critical surface.

\subsection{Veneziano limit and beyond}
Our findings, throughout, rely on the existence of exact small parameters $\eps_1\ll 1$ and $\eps_2\ll 1$ \eq{eps12} [or $\eps\ll1 $ see \eq{P}] in the Veneziano limit, which relate to  the gauge one loop coefficients.
Consequently, an iterative solution of perturbative beta functions becomes exact and interacting fixed points arise as exact power series in the small parameters.   More specifically, the leading non-trivial approximation which is  NLO${}^\prime$  (Tab.~\ref{tNLO})  retains the gauge beta functions up to two loop, and the Yukawa and scalar beta functions up to one loop. The parametric smallness of the gauge one-loop coefficients allows an exact cancellation of one and two loop terms implying that interacting fixed points for the gauge couplings must be of the order of the one loop coefficient  $\sim\eps$. The Yukawa nullclines at one loop imply that Yukawa couplings are necessarily proportional to the gauge couplings, and 
the scalar nullcline impose that scalar couplings are proportional to the Yukawas (see Sect.~\ref{YSnull}); hence either of these come out  $\sim\eps$.  Higher order loop approximations $n$NLO${}^\prime$  starting with $n= 2$ then correspond to retaining $n+1$ loops in the gauge, and $n$ loops in the Yukawa and scalar beta functions respectively, see Tab.~\ref{tNLO}.  Hence, solving the beta functions for interacting fixed points  order-by-order in perturbation theory $(n\to n+1)$ we have that
\beq\label{nNLO}
 \alpha_i^*=\alpha_i^*\Big|_{n{\rm NLO'}}
+{\cal O}(\eps^{n+1})
 \eeq
for all couplings \eq{couplings}, \eq{w} and all fixed points, with corrections from the $(n+1)$NLO${}^\prime$ level being at least one power in $\eps$ smaller than those from the preceding level. 
We conclude that the expressions for the interacting fixed points $\alpha_i^*|_{n{\rm NLO'}}$ are accurate polynomials in $\eps$ up to including terms of order $\eps^{n}$, for all $n$.

Beyond the Veneziano limit,  the parametrically small control parameter $\eps$ is no longer available. Instead, $\eps$ will take finite, possibly large, values dictated by the (finite) field multiplicities. Still, for sufficiently large matter field multiplicities, $\eps$ remains sufficiently small  and perturbativity remains in reach  \cite{Bond:2017wut}. It is then conceivable that the fixed points found in the Veneziano limit persist even for finite $N$.\footnote{An example for a conformal window with asymptotic safety  is given in \cite{Bond:2017tbw} for the model introduced in  \cite{Litim:2014uca}.}
At finite $N$, however, we stress that the $n$NLO${}^\prime$ approximations and \eq{nNLO} are  no longer exact order-by-order. It then becomes important to check numerical convergence of higher loop approximations, including non-perturbative resummations.   In this context it would be particularly useful to know the radius of convergence of beta functions (in $\eps$)  in the Veneziano limit. A finite radius of convergence  has been established rigorously in certain large-$N_F$ limits of 
gauge theories without Yukawa interactions \cite{PalanquesMestre:1983zy,Holdom:2010qs} which makes it conceivable that the radius of convergence might be finite here as well.\footnote{Results for resummed beta functions of large-$N$ gauge theories with Yukawa couplings are presently not available.} If so, this would offer additional indications for the existence of interacting fixed points beyond the Veneziano limit.

\subsection{Conformal symmetry and conformal windows}

By their very definition, the gauge-Yukawa theories investigated here are scale-invariant at (interacting) fixed points. Conditions under which scale invariance entails exact conformal invariance have been discussed by Polchinski  \cite{Polchinski:1987dy} (see also \cite{Luty:2012ww}). Applied to the theories \eq{L} at weak coupling, it implies that exact conformal invariance is realised at all interacting fixed points discovered here. 
It would then be interesting to find the full conformally invariant effective action beyond the classically marginal invariants retained in \eq{L}.
First steps into these directions have been reported in \cite{Buyukbese:2017ehm}.
Moreover,  for a quantum theory to be compatible with unitarity, scaling dimension of (primary) scalar fields must be larger than unity. This is  confirmed for all fixed points by using the results of Sect.~\ref{anomalous} for the anomalous dimensions of fields and composite scalar operators, together with the results for fixed points at NLO${}'$ accuracy  (Tab.~\ref{tPara15} and~\ref{tPara69}). We conclude that the residual interactions are compatible with unitarity.

Away from the Veneziano limit, findings for the various interacting conformal fixed points persist once $\eps$ is finite. One may then think of keeping the parameters in the gauge sectors $(N_{\rm C}, N_{\rm c}$) fixed  and finite while varying the matter field content $(N_{\rm F}, N_{\rm f}, N_\psi)$. Then, the domain of existence for each of the interacting fixed points  (Tab.~\ref{tPara15},~\ref{tPara69}) turns into a ``conformal window''  as a function of the matter field multiplicities. The fixed point ceases to exist outside the conformal window. The conformal window for asymptotic safety with a simple $SU(N)$ gauge factor has been determined in \cite{Bond:2017tbw}. Boundaries of conformal windows can be estimated within perturbation theory though more accurate results invariably require non-perturbative tools.\footnote{See \cite{DelDebbio:2010zz} for lattice studies of conformal windows in QCD with fermionic matter  (Banks-Zaks fixed points).}

\section{\bf Summary}\label{sum}

We have used perturbation theory and large-$N$ techniques for a 
rigorous and 
comprehensive investigation of weakly  interacting fixed points 
of gauge theories coupled to fermionic and scalar matter. 
For  concrete families of simple and semi-simple
gauge theories  with action \eq{L} and following the classification of fixed points put forward in \cite{Bond:2016dvk,Bond:2017sem},  we have discovered a large variety of exact high- and low-energy fixed points (Tab.~\ref{tFPs},~\ref{tPara15},~\ref{tPara69}).
These include partially interacting ones (Tab.~\ref{tPara15}) where one gauge sector remains free, and 
fully interacting ones (Tab.~\ref{tPara69}) where both gauge sectors are interacting. 
We have determined the domains of existence for all of them
(Fig.~\ref{pFP23}--~\ref{pFP9}). Interestingly, we also find that the requirement of vacuum stability always singles out a unique viable fixed point in the scalar sector. 
 
As a function of field multiplicities, the phase space  of distinct quantum field theories (Fig.~\ref{pFPall}) includes models with asymptotic safety and asymptotic freedom, and effective theories without UV completion (Figs.~\ref{pAll_AF},~\ref{pAll_AS} and~\ref{pAll_EFT}).  In the IR,  theories display either strong coupling and confinement, or weakly coupled  fixed points where the elementary gauge fields and fermions are unconfined and appear as massless particles.  Many features are a consequence of the semi-simple nature and would not arise in simple (or ``direct products'' of simple) gauge theories. Highlights include  massless semi-simple gauge-matter theories where one gauge sector can be both UV free and IR free owing to a fixed point in the other, Fig.~\ref{pSemiSimpleBZs} $c)$, and theories with inequivalent scaling limits in the IR. Semi-simple  effects are particularly pronounced for asymptotically free theories where they enhance the diversity of different IR scaling regimes  (Fig.~\ref{pAll_AF}). 

Another central outcome of our study is the first explicit ``proof of existence'' for asymptotic safety in semi-simple quantum field theories with elementary gauge fields, scalars and fermions. It establishes the important result  that asymptotic safety is not limited to simple gauge factors \cite{Litim:2014uca}, fully in line with general theorems and structural results  \cite{Bond:2016dvk}.  Our findings, together with their supersymmetric counterparts in \cite{Bond:2017suy},  make it conceivable that semi-simple theories display interacting UV fixed points even beyond the Veneziano limit,  thus further paving the way  for asymptotic safety beyond the Standard Model \cite{Bond:2017wut}.  
The stability of the vacuum (Sect.~\ref{stab}) in all models studied here suggests
that the near-criticality  of the Standard Model Higgs \cite{EliasMiro:2011aa,Buttazzo:2013uya} can very well expand into full criticality at an interacting UV fixed point  \cite{Bond:2017wut}.

In addition, we have investigated phase diagrams for simple and semi-simple gauge theories
with and without Yukawa interactions, continuing an analysis initiated in \cite{Bond:2016dvk,Bond:2017sem}.
We find that transitions from the UV to the IR can proceed from  free or interacting fixed points to confinement and strong coupling.  We also find transitions from free to interacting (Figs.~\ref{pSemiSimpleBZs},~\ref{pSimpleCasesPD}, \ref{pPrimitivesSimple}, \ref{pPD_BZ_IR}, \ref{pPD_AF}, \ref{pPD_AF_pt2}) 
 or  from interacting to other interacting conformal fixed points  (Fig.~\ref{pPD_AS}). 
  In the latter cases, 
  theories display a variety of exact 
``IR sinks'', meaning free or interacting IR conformal fixed points which are fully attractive in all classically marginal interactions. Once more, many new features have come to light  beyond those observed in simple gauge theories \cite{Bond:2016dvk,Bond:2017sem}. 
  
  Our study used minimal models with a low number of Yukawa and gauge couplings. Already at this basic level,  an intriguing diversity of fixed points and scaling regimes has emerged, with many novel characteristics both at high and low energies. 
We believe that these findings  warrant more extensive studies in view of rigorous results \cite{Bond:2016dvk,Bond:2017suy}, extensions towards strong coupling \cite{Buyukbese:2017ehm}, and its exciting potential for physics beyond the Standard Model \cite{Bond:2017wut}.\\[3ex]

\centerline{\bf Acknowledgements}

${}$\\ Some of the results have been presented at the workshops {\it The Exact Renormalisation Group}, Trieste (Sept 2016), and {\it Understanding the LHC}, Bad Honnef (Feb 2017). This work is supported by the Science and Technology Research Council (STFC) under the Consolidated Grant [ST/G000573/1] and by an STFC studentship.

\appendix

\renewcommand{\thesection}{{\bf \Alph{section}}}

\section{\bf General expressions for fixed points}
\label{AppN}
Most results in the main text relate to the choice $N_\psi=1$. For completeness,  we summarize fixed point results  for general $N_\psi$ species of fermions in the fundamental of both gauge groups $SU(N_C)$ and $SU(N_c)$. We observe that $N_\psi$ is restricted within the
range \beq\label{rangePsi}
0\le N_\psi \leq \0{11}2\,. 
\eeq
Outside of this range, exact perturbativity is lost. Substituting $N_\psi$ into the RG coefficients and solving for fixed points, we find the following expressions at the partially interacting Banks-Zaks  fixed points \fp2 and \fp3,
\bea
{\Fp2}:\quad\alpha_{1}&=&-\04{75}R\eps\\
{\Fp3}:\quad\alpha_{2}&=&-\04{75}\,\0{P\eps}{R}
\eea
At the partially interacting fixed points \fp4 and \fp5 we have
\bea
{\Fp4}:\quad&
\left\{\begin{array}{rcl}
\alpha_{1}&=&\di\023 \frac{13-2 N_\psi R}{(2N_\psi R - 1)(3N_\psi R - 19)}R\eps\\[2ex]
\alpha_{Y}&=&\di \frac{4}{(2N_\psi R - 1)(3N_\psi R - 19)}R\eps
\end{array}
\right.
\\[2ex]
{\Fp5}:\quad&
\left\{\begin{array}{rcl}
\alpha_{2}&=&\di
\023 \frac{13-2 N_\psi/R}{(2N_\psi/ R - 1)(3N_\psi/ R - 19)}\0{P\eps}{R}\\[2ex]
\alpha_{y}&=&\di
\frac{4}{(2N_\psi/ R - 1)(3N_\psi/ R - 19)} \0{P\eps}{R}
\end{array}
\right.\eea
For the Banks-Zaks times Banks-Zaks-type  fixed point \fp6 we find
\bea
{\Fp6}:\quad
\left\{\begin{array}{rcl}
\alpha_{1}&=&\di
-\043 \left(\0{25 - 2 N_\psi P/R}{625 - 4N_\psi^2}\right)R\eps\\[2ex]
\alpha_{2}&=&\di
-\043 \left(\0{25 - 2 N_\psi R/P}{625 - 4N_\psi^2}\right)\,\0{P\eps}{R}
\end{array}
\right.
\eea
For the interacting fixed points \fp7 and \fp8 we find
\beq
\label{fp7n}
{\Fp7}:\quad
\left\{\begin{array}{rcl}
\alpha_{1}&=&\di
\023\left(\frac{(13 - 2N_\psi R)(25 - 2N_\psi P/R)}{150 N_\psi^2 R^2 - (4N_\psi^2 + 1025)N_\psi R+26 N_\psi^2 + 475}\right)R\eps\\[2ex]
\alpha_{2}&=&\di
-\043\left(\frac{(13 - 2N_\psi R)N_\psi R/P +(2N_\psi R - 1)(3N_\psi R - 19)}{150 N_\psi^2 R^2 - (4N_\psi^2 + 1025)N_\psi R+26 N_\psi^2 + 475}\right)\,\0{P\eps}{R}\\[2ex]
\alpha_{Y}&=&\di
\frac{4(25 - 2N_\psi P/R)}{150 N_\psi^2 R^2 - (4N_\psi^2 + 1025)N_\psi R+26 N_\psi^2 + 475}R\eps
\end{array}
\right.
\eeq
\beq
\label{fp8n}
{\Fp8}:\quad
\left\{\begin{array}{rcl}
\alpha_{1}&=&\di
-\043\left(\frac{(13 - 2N_\psi/ R)N_\psi P/R +(2N_\psi /R - 1)(3N_\psi /R - 19)}{150 N_\psi^2/ R^2 - (4N_\psi^2 + 1025)N_\psi/ R+26 N_\psi^2 + 475}\right)R\eps\\[2ex]
\alpha_{2}&=&\di
\023\left(\frac{(13 - 2N_\psi /R)(25 - 2N_\psi R/P)}{150 N_\psi^2 /R^2 - (4N_\psi^2 + 1025)N_\psi /R+26 N_\psi^2 + 475}\right)\,\0{P\eps}{R}\\[2ex]
\alpha_{y}&=&\di
\frac{4(25 - 2N_\psi R/P)}{150 N_\psi^2/ R^2 - (4N_\psi^2 + 1025)N_\psi /R+26 N_\psi^2 + 475}\,\0{P\eps}{R}
\end{array}
\right.
\eeq
Finally, at the fully interacting fixed point \fp9 we have
\beq\label{fp9n}
\Fp9
:\ \left\{
\begin{array}{rcl}
\alpha_{1}&=&\di
\023
\0{(13-2N_\psi R)\left[(13-2N_\psi R)N_\psi P/R +(2N_\psi/R-1)(3N_\psi/R -19)\right]\,R\eps}
	{114N_\psi^2(R^2 + 1/R^2) + (32N_\psi^4 + 1512N_\psi^2 + 361) - (220N_\psi^2 + 779)(R + 1/R)}
	\\[2ex]
\alpha_{2}&=&\di
\023
\0{(13-2N_\psi /R)\left[(13-2N_\psi /R)N_\psi R/P +(2N_\psi R-1)(3N_\psi R -19)\right]\,{P\eps}/{R}}
	{114N_\psi^2(R^2 + 1/R^2) + (32N_\psi^4 + 1512N_\psi^2 + 361) - (220N_\psi^2 + 779)(R + 1/R)}
	\\[2ex]
\alpha_{Y}&=&\di
\0{4\left[(13-2N_\psi R)N_\psi P/R +(2N_\psi/R-1)(3N_\psi/R -19)\right]\,R\eps}
	{114N_\psi^2(R^2 + 1/R^2) + (32N_\psi^4 + 1512N_\psi^2 + 361) - (220N_\psi^2 + 779)(R + 1/R)}
	\\[2ex]
\alpha_{y}&=&\di
\0{4\left[(13-2N_\psi /R)N_\psi R/P +(2N_\psi R-1)(3N_\psi R -19)\right]\,{P\eps}/{R}}
	{114N_\psi^2(R^2 + 1/R^2) + (32N_\psi^4 + 1512N_\psi^2 + 361) - (220N_\psi^2 + 779)(R + 1/R)}\,.
\end{array}
\right.	
\eeq
All expressions reduce to those given in the main body in the limit $N_\psi=1$. We note that the parameter range in which fixed points exist changes both qualitatively and quantitatively when varying $N_\psi$ within the range \eq{rangePsi}. Moreover, we also observe that the characteristic boundaries in paramater space depend on $N_\psi$, indicating that domains of existence and eigenvalue spectra depend on $N_\psi$. It is straightforward, if tedious, to investigate regions of validity and scaling exponents for the general case, and to find the analogues of  Figs.~\ref{pFP23},~\ref{pFP45},~\ref{pFP6},~\ref{pFP78} and~\ref{pFP9}  and of Tabs.~\ref{tBprime}~\ref{tPara15},~\ref{tPara69} for general $N_\psi$.

\section{\bf Boundaries}\label{AppX}
We find that the existence and relevancy of fixed points in the parameter space $(P,R)$, see \eq{PRepsN}, is controlled by characteristic curves  $P=X(R), Y(R), \xalt(R)$ or $\yalt(R)$ with the functions
\beq\label{P1234}
\begin{array}{rl}
	X(R) &=\displaystyle \frac{(2R - 13)R}{(2R-1)(3R-19)}\,,\\[2ex]
	Y(R) &=\displaystyle \frac{25}{2}R\,,\\[2ex]
	\xalt(R)  &= \displaystyle\frac{(2/R-1)(3/R-19)}{(2/R - 13)/R}\,,\\[1.5ex]	
	\yalt(R) &=\displaystyle \frac{2}{25}R\,.
\end{array}
\eeq
These appear as boundaries of the ``phase space'' of parameters $(R,P)$ characterising valid fixed points.
Note that the functions $(X,\xalt)$ and $(Y,\yalt)$ in \eq{P1234} are ``dual'' to each other,
\beq
X(R) \cdot \xalt(R^{-1})=1=Y(R) \cdot \yalt(R^{-1})\,.
\eeq
A further set of boundaries is given by the straight lines $R=R_{\rm low} $ or $R_{\rm high}$, with
\beq
\label{R12}
\begin{array}{rl}
R_{\rm low}&=\displaystyle
\frac12\\[1.5ex]
R_{\rm high}&=2\,.
\end{array}
\eeq
The boundaries $P=X(R), Y(R), \xalt(R)$ or $\yalt(R)$  with \eq{P1234} together with \eq{R12} delimit the qualitatively different  quantum field theories in the ``phase space'' shown in Fig.~\ref{pFPall}. 

Certain characteristic values for the parameter $R$ arise  in its domain of vailidity $\s02{11}<R<\s0{11}2$ at points where  the boundaries \eq{P1234} cross. We find four of these $R_{1,\cdots ,4}$ with 
\beq\label{Rvalues}
\02{11}<R_{\rm low}<R_1 < R_2 < 1<R_3 < R_4<R_{\rm high}<\0{11}2\,,
\eeq 
with $R_1$ and $R_2$ arising from
\beq
\begin{array}{rcl}
	X(R_1) &=& Y(R_1)\,,\\[1ex]
	X(R_{2}) &=& \xalt(R_{2}) 
\end{array}
\eeq
together with  $R_3=1/R_2$ and $R_4=1/R_1$. 
Quantitatively we have \eq{R1234} for $R_{1,\cdots ,4}$ as stated in the main text.
The expressions \eq{P1234}, \eq{R12} for the boundaries are modified once $N_\psi\neq 1$.

\bibliographystyle{JHEP}
\bibliography{bib_semi2}

\end{document}